%% file: main.tex
\documentclass[fleqn,nofootinbib,aps,reprint,amsmath,amssymb,10pt,superscriptaddress,floatfix,a4paper,letterpaper,fleqn]{revtex4-1}

\input{package_list}
\input{math_definitions}
\input{preamble}

\begin{document}

\title{
\texorpdfstring{\qquad \\ \qquad \\ \qquad \\  \qquad \\  \qquad \\ \qquad \\}{}
Conception and software implementation of a nuclear data evaluation pipeline
}
\author{G.\thinspace Schnabel}
\email{g.schnabel@iaea.org}
\email{previously affiliated to Uppsala University}
\affiliation{NAPC--Nuclear Data Section, International Atomic Energy Agency, Vienna, Austria}

\author{H.\thinspace Sjöstrand}
\affiliation{Uppsala University, Department of Physics and Astronomy, Uppsala, Sweden}

\author{J.\thinspace Hansson}
\affiliation{Uppsala University, Department of Physics and Astronomy, Uppsala, Sweden}

\author{D.\thinspace Rochman}
\affiliation{Paul Scherrer Institut, 5232 Villigen, Switzerland}

\author{A.\thinspace Koning}
\affiliation{NAPC--Nuclear Data Section, International Atomic Energy Agency, Vienna, Austria}

\author{R. \thinspace Capote}
\affiliation{NAPC--Nuclear Data Section, International Atomic Energy Agency, Vienna, Austria}

\pacs{}

\begin{abstract}
We discuss the design and software implementation of a nuclear data evaluation pipeline applied for a fully reproducible evaluation of neutron-induced cross sections of $^{56}$Fe above the resolved resonance region using the nuclear model code TALYS combined with relevant experimental data.
The emphasis of this paper is on the mathematical and technical aspects of the pipeline and not on the evaluation of $^{56}$Fe, which is tentative.
The mathematical building blocks combined and employed in the pipeline are discussed in detail.
In particular, an intuitive and unified representation of experimental data, systematic and statistical errors, model parameters and defects enables the application of the Generalized Least Squares (GLS) and its natural extension, the Levenberg-Marquardt (LM) algorithm, on a large collection of experimental data without the need for data reduction techniques as a preparatory step.
The LM algorithm tailored to nuclear data evaluation takes into account the exact non-linear physics model to determine best estimates of nuclear quantities.
Associated uncertainty information is derived from a second-order Taylor expansion at the maximum of the posterior distribution.
We also discuss the pipeline in terms of its IT (=information technology) building blocks, such as those to efficiently manage and retrieve experimental data of the EXFOR library, which facilitates their appropriate correction, and to distribute computations on a scientific cluster.
Relying on the mathematical and IT building blocks, we elaborate on the sequence of steps in the pipeline to perform the evaluation, such as the retrieval of experimental data, the correction of experimental uncertainties using marginal likelihood optimization (MLO) and after a screening of thousand TALYS parameters---including Gaussian process priors on energy dependent parameters---the fitting of about 150 parameters using the LM algorithm.
The code of the pipeline including a manual and a Dockerfile for a simplified installation is available at \href{http://www.nucleardata.com}{www.nucleardata.com}. 
\end{abstract}
\maketitle
\lhead{...Nuclear Data Evaluation Pipeline}
\chead{}
\rhead{G. Schnabel \etals}
\lfoot{} \rfoot{}
\renewcommand{\headrulewidth}{0.4pt} \renewcommand{\footrulewidth}{0.4pt}
\tableofcontents
\vfill

\section{INTRODUCTION}
 %
 The ability to reproduce complex scientific workflows becomes more and more an important concern.
 Increasing awareness that many past studies are impossible to reproduce has now even led to the term \textit{reproducibility crisis}~\citep{sep-scientific-reproducibility}.
 Nuclear data evaluation is not exempt of such concerns.
 Many choices related to the selection of the experiments, corrections applied to the experimental data, and the choice of evaluation method are often not well documented enough to enable the reproduction of an evaluation.
 The lack of traceability regarding decisions taken by human experts to produce an evaluation is a serious obstacle to building on the efforts from the past.
 Time and money need to be spent on re-analyzing data and information in articles, which could have been better spent otherwise.
 However, we also want to stress that available knowledge in the form of evaluations, even if they are not reproducible, can still be beneficially used as prior knowledge in future evaluations.
 
 Reproducibility and transparency is the standard practice in everyday neutronics simulations performed in the nuclear industry. It may also be regarded as especially important in the evaluation of standard reactions, such as the neutron standards~\citep{carlson_evaluation_2018,carlson_corrigendum_2020,carlson_international_2009}.
 These reactions serve as a reference in many nuclear experiments to translate relative cross section measurements to absolute ones.
 For this reason, much effort coordinated by the IAEA and various nuclear data evaluation projects has been put into the evaluation of these reactions.
 The documentation of these efforts is important, not only in the form of reports but also as scripts to enable the reproduction of these evaluations.
 The availability of the evaluations as a well-structured sequence of scripts facilitates studying the impact of specific assumptions on the evaluation and provides a solid basis for improved evaluations, such as those of neutron data standards, in the future.
 
 Awareness about the need of storing nuclear data in an accessible and transparent way and the need for automation of evaluation and verification processes is rapidly increasing as indicated by the ongoing efforts of the WPEC subgroup 49 on reproducibility in nuclear data evaluation and the recently approved WPEC subgroup 50 on a curated experimental database with a focus on programmatic readability.
 Another encouraging effort is undertaken by WPEC subgroup 45 to improve the data management and processes to validate nuclear data libraries.
 
 Also established solutions exist that can facilitate the creation of transparent and reproducible evaluations, such as the correction system attached to the EXFOR web retrieval system~\citep{zerkin_experimental_2018} and the \textit{EXFOR-CINDA for application} package available on the IAEA-NDS website enabling programmatic access to information in the EXFOR library~\citep{otuka_towards_2014}.
 Yet another effort to facilitate the retrieval and management of experimental data from the EXFOR library is the conversion of the EXFOR library to a NoSQL JSON database \cite{schnabel_computational_2020}. 
 An example of the benefits of reproducible evaluations is the TENDL library project~\citep{koning_tendl_2019}.
 Its focus on automation and reproducibility from the beginning has been a catalyst in the creation of a comprehensive nuclear data library, setting an example for future evaluations.
 
 Recognizing the need for reproducibility, automation and transparency in nuclear data evaluation and motivated by the benefits of existing software and projects supporting those needs, we present a prototype of a pipeline for nuclear data evaluation, which has been employed for an example evaluation of neutron-induced cross sections of $^{56}$Fe above the resolved resonance region.

 Even though the pipeline (and therefore also its products) has still to be regarded as a prototype, it combines several advancements of evaluation methodology achieved during the last years.
 Examples are Gaussian process priors on energy-dependent model parameters to address model defects~\citep{helgesson_treating_2018}, the treatment of inconsistent experimental data by marginal likelihood optimization~\citep{schnabel_fitting_2018} and the adjustment of model parameters via a prior-aware Levenberg-Marquardt algorithm~\citep{helgesson_fitting_2017} to take into account non-linearities of the nuclear physics model.
 All of these approaches are mathematically related and they can be used in different ways.
 Therefore it was important to implement these essential mathematical concepts and algorithms in a generic way so that at first glance seemingly unrelated task settings can be dealt with in a unified manner.
 For instance, the adjustment of uncertainties of inconsistent experimental data and the adjustment of so-called hyperparameters of Gaussian processes imposed on energy-dependent model parameters rely on the same mathematical approach---marginal likelihood optimization.
 
 An additional important design aspect was to enable the evaluation of a large collection of experimental data.
 By adopting an intuitive perspective on experimental data, model parameters and their relations in the form of a simple Bayesian network~\citep{pearl_causality_2000}, we are able to apply the mathematical techniques on a large collection of experimental data without the need for data reduction techniques as a preparatory step.
 We refer to the mathematical concepts and techniques used in the pipeline as \textit{mathematical building blocks} and discuss them first.
 
 Besides the mathematical aspects, modern nuclear data evaluation is also about the management of information, either in the form of experimental data or scientific knowledge distilled to nuclear models.
 Therefore aspects associated with information management need to be addressed as well, such as the handling of experimental data and the parallelization of nuclear model calculations on a cluster.
 We refer to these needs as IT (=information technology) building blocks and they are discussed after the mathematical building blocks.
 
 After the presentation of the mathematical and IT building blocks, we discuss the pipeline itself, which is a sequence of scripts employing the building blocks to implement the evaluation.
 We go through the individual steps, starting from the retrieval of experimental data from the EXFOR library, over the construction of experimental uncertainty information using a rule-based approach, subsequent automated adjustment of systematic uncertainties associated with
 inconsistent experimental data, and ending with the adjustment of TALYS model parameters and the generation of covariance and random files using a modified version of the TASMAN code.
 TALYS and TASMAN are part of the evaluation code system called T6~\citep{koning_modern_2012}.
 Throughout the description of the individual steps, we discuss features of the pipeline and possible directions for improvement in the future.
 Practicalities, such as computation time, and results are presented in a separate section after the discussion of the steps in the pipeline.
 
 Importantly, the pipeline is open-source and has been made available online~\citep{gschnabel-eval-fe56} and a Dockerfile is also available~\citep{gschnabel-compEXFOR-docker} to facilitate the installation of all components of the pipeline as one bundle.
 The availability of the pipeline and the ability to launch it at the push of a button enable the incremental implementation of more sophisticated algorithms in the future and the systematic study of the impact of assumptions codified in the pipeline.
 
 The pipeline can therefore be regarded as a skeleton for future evaluations, which can be extended according to the scope of the evaluation.
 Going towards more global and automated evaluations, manual choices regarding data selection and uncertainty specifications need to be removed and statistical algorithms made more robust.
 Going towards high-quality evaluations of specific isotopes, the selection of experimental data needs to be more carefully undertaken and assumptions regarding uncertainties and prior assumptions more scrutinized.
 
 In the actual implementation of the pipeline, we had to make many specific choices concerning selection of experimental data, algorithms, and nuclear models.
 We stress that many of the choices in our current implementation of the pipeline, such as the use of the nuclear models code TALYS, are not of fundamental significance for the concept of the pipeline.
 Due to the modular design of the pipeline, specific algorithms or the model for fitting may be replaced by something more pertinent in a certain evaluation context.
 For this reason, we stress that this paper is as much about the concept of a pipeline and general design considerations as it is about the specific implementation details, which are not crucially linked to the conceptual aspects of the pipeline.
 
 The paper can be read selectively depending on the reader.
 A reader interested in the mathematical framework and how its algorithms are applied in the evaluation pipeline may only read~\cref{sec:mathematical_building_blocks} on mathematical building blocks and \cref{sec:evaluation_pipeline} on the sequence of steps in the pipeline.
 A reader who is interested in the mathematical framework and its translation to data structures to enable large scale computations may start with \cref{sec:mathematical_building_blocks} and \cref{subsec:handling_info_bayesian_context}.
 Finally, a reader who is foremostly interested in our ideas on the efficient management of experimental data and model calculations in terms of information technology may directly jump to \cref{sec:IT_building_blocks}.
 
\section{MATHEMATICAL BUILDING BLOCKS}
\label{sec:mathematical_building_blocks}
Concerning the representation of information and algorithms, the pipeline relies on the functionality of the R package \textit{nucdataBaynet} published on GitHub~\citep{gschnabel-nucdataBaynet}.
In this section we review the multivariate normal distribution as it is the mathematical core concept of the pipeline, elaborate on the representation of information to enable large scale evaluations, explain the integration of Gaussian processes into this framework, elaborate on a prior-aware Levenberg-Marquardt algorithm for the optimization of non-linear nuclear physics models, and finally discuss an approximation to the posterior distribution based on the multivariate normal distribution.
All of these different concepts and algorithms have a clear link to the multivariate normal distribution, which we strive to emphasize in the following exposition.

\subsection{Multivariate normal model}
\label{subsec:mvn_model}

The mathematical core concept used at various places throughout the pipeline is the multivariate normal (MVN) distribution.
If $d$ denotes the number of variables, the MVN distribution is characterized by a center vector $\vec{\mu}$ of size~$d$ and 
a covariance matrix $\sigmamat$ of dimension $d \times d$.
The probability density function (pdf) of a MVN distribution is given by
\begin{multline}
    \mathcal{N}(\vec{x}\,|\,\muvec,\sigmamat) =
    \frac{1}{\sqrt{(2\pi)^d \det\sigmamat}}
    \\
    \times
    \exp\left[
        -\frac{1}{2}
        \left( \vec{x} - \muvec \right)^T
        \sigmamat^{-1}
        \left( \vec{x} - \muvec \right)
    \right] \,.
    \label{eq:mvn_pdf}
\end{multline}
Noteworthy, a MVN distribution can exactly capture linear relationships as well as uncertainties, which are both reflected in the specification of the covariance matrix.

The variables associated with elements in $\vec{x}$ can be among other things model parameters, measured values in an experiment, and statistical and systematic errors associated with measured values.
Some of these variables, such as measured values are evidently observable, whereas others, such as model parameters, can only be indirectly inferred from variables whose values were observed.

Before we move on, we stress the important distinction between errors and uncertainties here: An error is a quantity contributing to the difference between the truth and the measurement.
As the value of an error cannot be measured, the likelihood of potential values is typically modeled by a normal distribution and the standard deviation of this distribution is referred to as uncertainty.
In the pipeline, we model measurement errors explicitly because it introduces the possibility to check results for consistency and enables a mathematical representation of information to speed up computations.

To discuss how the values of unobservable variables can be inferred from variables whose values were observed,
we partition both the center vector and the covariance matrix into blocks corresponding to observable variables denoted by the index '$\textrm{obs}$` and unobservable variables denoted by the index '$\textrm{hid}$` (for hidden):
\begin{equation}
    \muvec = \begin{pmatrix}
        \muobs \\ \muhid
    \end{pmatrix},
    \sigmamat = \begin{pmatrix}
        \covobs      & \covobshid \\
        \covobshid^T & \covhid 
    \end{pmatrix}
    \label{eq:mvn_partitioned}
\end{equation}
In the covariance matrix, the off-diagonal blocks $\covobshid$ encode linear relationships between observed and unobserved variables.

Given that values of a subset of the variables have been observed, i.e., $\vec{x}_\textrm{obs} = \vec{o}$, the pdf of the unobserved variables is modified due to the propagation of information gained about the observed variables.
The mathematical operation to propagate this information is called conditioning where we go from the joint distribution $\pi(\vec{x}_\textrm{obs}, \vec{x}_\textrm{hid})$ to the conditional distribution $\pi(\vec{x}_\textrm{hid} \,|\, \vec{x}_\textrm{obs}=\vec{o})$.
For a MVN distribution using the partitioned form of the center vector and covariance matrix in \cref{eq:mvn_partitioned}, the conditional pdf is again multivariate normal.
The center vector and covariance matrix are given by
\begin{align}
    \muhid' &= \muhid + \covobshid^T \covobs^{-1} (\vec{o} - \muobs)
    \label{eq:cond_mvn_hidmean} \\
    \covhid' &= \covhid - \covobshid^T \covobs^{-1} \covobshid \,.
    \label{eq:cond_mvn_hidcov}
\end{align}
These formulas (or mathematical equivalent ones) are known as the Generalized Least Squares (GLS) method in the field of nuclear data evaluation, e.g., \citep{muir_evaluation_1989,smith_least-squares_1993}, implemented in various codes, e.g.,~\citep{gluc1980,muir_global_2007,muir_global_2007_website}.
Usually, unobserved variables are model parameters or the values of cross sections on a predefined energy mesh and observed variables correspond to measurements made in experiments.
However, we stress that any other type of variable can be added as well.
For instance, model parameters and systematic errors can both be present at the same time in the set of unobserved variables.
We further elaborate on this possibility in the next section, as it is used in the evaluation pipeline.

In \cref{eq:cond_mvn_hidmean,eq:cond_mvn_hidcov}, the covariance matrix is assumed to be perfectly known, which may not always be the case in practice.
As an example, if the calibration uncertainty of a detector is not reported, only a reasonable range of this uncertainty may be available by taking into account experiences from similar experiments.
In such cases, we can set the values of missing uncertainties in a way that is consistent to the information from other experiments, the model for fitting, and respecting reasonable limits.

Regarding the consistency with the model for fitting, we should be aware that by our decision of employing a certain nuclear physics model in a conventional statistical inference procedure, such as the Generalized Least Squares (GLS) method, e.g., \citep{muir_evaluation_1989}, we express our absolute confidence in the ability of the model to represent the truth given that values of model parameters are set correctly.
Due to the complexity of nuclear processes, we usually do not believe that nuclear physics models can \textit{perfectly} reproduce the truth.
Therefore, simpler and very flexible mathematical models, such as a piecewise linear function, are sometimes used instead to avoid the distortion of results due to an imperfect model.
This path was taken in the evaluation of neutron data standards~\citep{carlson_evaluation_2018,carlson_corrigendum_2020,carlson_international_2009}.
An alternative is to endow a nuclear physics model with more flexibility by adding a simpler mathematical model on top that captures the trends in the experimental data the nuclear physics model cannot reproduce.
This approach is associated with the idea of \textit{model defects} in the field of nuclear data evaluation, e.g., \citep{leeb_consistent_2008,neudecker_impact_2013,schnabel_differential_2016,helgesson_fitting_2017}.

One approach to fit missing or uncertain elements of the covariance matrix to enforce consistency between experimental datasets and the model for fitting is marginal likelihood optimization (MLO), e.g., \citep[sec. 5.4.1]{rasmussen_gaussian_2006}.
It relies on the marginal distribution of the observed values $\pi(\vec{x}_\textrm{obs})$ obtained from $\pi(\vec{x}_\textrm{obs}, \vec{x}_\textrm{hid})$ by integrating over the unobservable variables $\vec{x}_\textrm{hid}$.
Marginal distributions of a MVN distribution are again multivariate normal.
We obtain them by removing the blocks from the center vector and covariance matrix in \cref{eq:mvn_partitioned} corresponding to the unobserved variables, hence the marginal pdf is given by $\mathcal{N}(\xobs\,|\,\muobs,\covobs)$.

Marginal likelihood maximization seeks to find values for the unknown uncertainties that maximize the probability density for the observed values $\xobs=\vec{o}$.
More technically, missing or uncertain elements in $\covobs$ are adjusted so that $\mathcal{N}(\vec{o}\,|\,\muobs,\covobs)$ attains the largest possible value.
Writing out the logarithmized pdf helps in understanding what the maximization actually means:
\begin{multline}
    2 \ln\mathcal{N}(\vec{x}_\textrm{obs} \,|\,\muobs,\covobs) =
    -d \ln (2\pi) \\ 
    - \ln \det \covobs -  
    \left( \vec{x}_\textrm{obs} - \muobs \right)^T
    \covobs^{-1}
    \left( \vec{x}_\textrm{obs} - \muobs \right)
    \,.
    \label{eq:logmarlike}
\end{multline}
The second term on the right-hand side is proportional to the differential entropy of the multivariate normal distribution.
The third term is the generalized $\chi^2$-value, which takes in comparison to the conventional $\chi^2$-value the full covariance matrix into account.
Maximizing the marginal likelihood therefore amounts to simultaneously minimizing the information entropy and the generalized $\chi^2$-value.
Because the information entropy is a measure of model complexity, higher entropy means a more complex model, the maximization process aims to increase the quality of the fit while avoiding too complex models.

The availability of the partial derivatives of the logarithmized determinant and the inverse covariance matrix with respect to uncertainties in analytic form, see \cref{apx:derivative_logdet,apx:derivative_invmat}, lead to analytic expressions of partial derivatives of \cref{eq:logmarlike}.
Quasi-newton methods, such as the L-BFGS algorithm~\citep{byrd_limited_1995}, can leverage this information in the optimization process.

\subsection{Representation of covariance matrices}
\label{subsec:info_representation}

The computation of the distribution parameters of the conditional pdf in \cref{eq:cond_mvn_hidmean,eq:cond_mvn_hidcov} and the evaluation of the value of the marginal pdf in \cref{eq:logmarlike} is problematic for a large number of observable variables due to the need to invert the associated covariance matrix $\covobs$ and the calculation of its determinant.
Making the pipeline applicable to a large number of observables was a design goal from the very beginning.
In this section we outline the representation of covariance matrices to overcome computational obstacles, which was already suggested by~\citep{larson_concise_2008}. 

For the following we assume a linear model of the form
\begin{equation}
    \mathcal{M}_\textrm{lin}(\modparvec) =
    \modrefvec + \modparmap(\modparvec-\modparrefvec) \,.
    \label{eq:linapprox_model}
\end{equation}
The term `model' is intentionally left open in the current discussion.
It can be a linear approximation of a nuclear physics model or a piecewise linear function.
Both options mentioned are employed at different stages in the pipeline.
Other possibilities include wavelets, Fourier series, smoothing splines and (linear approximations of) neutron transport models.
Also combinations of such models are of interest, e.g., when both differential and integral data are considered in an evaluation.
The treatment of non-linear models, such as implemented in the nuclear model code TALYS is discussed in \cref{subsec:LM_with_prior}.

The relationship between observable variables $\obsvec$ and unobservable variables can be expressed as
\begin{equation}
    \obsvec = 
     \mathcal{M}_\textrm{lin}(\modparvec)
    + \syserrmap\syserrvec
    + \staterrvec
\label{eq:obs_hid_relation}
\end{equation}
where $\syserrmap$ denotes the mapping responsible to distribute the systematic errors $\syserrvec$ to the affected observable variables.
The statistical errors $\staterrvec$ are mutually independent and therefore directly added to the sum.
In principle, the term associated with systematic errors could be split into several terms reflecting different types of errors, such as those related to the detector calibration or to the sample thickness.

The notational distinction between model parameters and systematic errors can be considered irrelevant because they can be dealt with in a unified way.
Therefore we combine the parameter vector and the systematic errors to one vector $\totunobs$ and do the same for the associated mapping matrices:
\begin{equation}
    \totmap = \begin{pmatrix}
        \modparmap, \syserrmap
    \end{pmatrix}
    ,
    \totunobs = \begin{pmatrix}
        \modparvec \\
        \syserrvec
    \end{pmatrix}
    ,
    \totunobsref = \begin{pmatrix}
        \modparrefvec \\
        \vec{0}
    \end{pmatrix}
    \label{eq:Su_abbreviation}
\end{equation}
Using these combined variables, \cref{eq:obs_hid_relation} can be rewritten as
\begin{equation}
    \obsvec = \modrefvec + \totmap (\totunobs - \totunobsref) + \staterrvec \,.
\end{equation}
Going one step further, we denote the sum of the constants $\modrefvec$ and $-\totmap\totunobsref$ as $\totrefvec$ and write
\begin{equation}
    \label{eq:basic_stat_model}
    \obsvec = \totrefvec + \totmap \totunobs + \staterrvec \,.
\end{equation}

This equation serves as the basis to determine the covariance matrix blocks in \cref{eq:mvn_partitioned}.
To that end, we exploit the bilinearity of the covariance operator $\Cov{\vec{x}}{\vec{y}}$ and therefrom induced properties of the variance operator $\Var{\vec{x}}:=\Cov{\vec{x}}{\vec{x}}$. The properties of these operators are briefly reviewed in~\cref{apx:covariance_properties}.
Noteworthy, $\Var{\vec{x}}$ yields the covariance matrix associated with $\vec{x}$.
Thus the covariance matrix $\covobs$ associated with $\obsvec$ is given by
\begin{equation}
\covobs = \Var{\obsvec} = \totmap \Var{\totunobs} \totmap^T + \Var{\staterrvec} \,.
\end{equation}
Please note that this covariance matrix is different from what is typical understood under an experimental covariance matrix.
An experimental covariance matrix $\covexp$ reflects potential deviations of the experimental measurements from an assumed `truth', i.e., $\mathcal{N}(\obsvec \,|\, \mutrue, \covexp)$.
In contrast to that, the covariance matrix $\covobs$ also incorporates the covariance matrix of the unknown model parameters, i.e., it describes fluctuations of potential measurements with respect to the prior expectation $\Exp{\obsvec} = \totrefvec + \totmap \Exp{\vec{u}}$.
Therefore this covariance matrix captures the fluctuations of hypothetical measurements obtained by the following sampling procedure:
Model parameters are sampled from the prior distribution, corresponding predictions are calculated, and finally a perturbation is added to the predictions, obtained by drawing a sample from the experimental covariance matrix $\covexp$. 

The covariance matrix between observable and unobservable variables is given by
\begin{equation}
    \covobshid = \Cov{\obsvec}{\totunobs} = \totmap \Var{\totunobs} \,.
\end{equation}
Both $\covobs$ and $\covobshid$ are thus completely determined by the mapping matrix $\totmap$,
the prior covariance matrix $\covtotunobs=\Var{\totunobs}$ associated with model parameters and systematic errors, and the prior covariance matrix $\covstaterr=\Var{\staterrvec}$ associated with statistical errors.

The covariance matrix associated with observations is therefore amenable to the decomposition
\begin{equation}
    \covobs = \totmap \covtotunobs \totmap^T + \covstaterr
    \label{eq:covobs_factorized} \,.
\end{equation}

We exploit this representation of $\covobs$ in the computation of the inverse and the determinant.
To express the following formulas concisely, we introduce the abbreviation
\begin{equation}
    \mat{Z} =
        \covtotunobs^{-1} + \totmap^T \covstaterr^{-1} \totmap \,.
\end{equation}
Using the Woodbury matrix identity, see \cref{apx:Woodbury_identity}, the inverse appearing in \cref{eq:cond_mvn_hidmean,eq:cond_mvn_hidcov,eq:logmarlike} can be written as
\begin{equation}
    \covobs^{-1} =
    \covstaterr^{-1}
    -\covstaterr^{-1} \totmap
    \mat{Z}^{-1}
    \totmap^T \covstaterr^{-1} \,.
    \label{eq:invcovobs_woodbury}
\end{equation}
We usually do not need to compute  $\covobs$ nor its inverse but rather products of these matrices with other vectors or matrices.
For instance, a matrix-vector product $\covobs^{-1} \vec{x}$ is performed sequentally from right to left starting with the evaluation of $\covstaterr^{-1}\vec{x}$ yielding a vector.
More information on the efficient computation of matrix products involving inverse covariance matrices is in~\cref{apx:prod_inv_mat}.

For the computation of the logarithmized determinant required in \cref{eq:logmarlike}, a similar relationship holds,
\begin{equation}
    \ln\det(\covobs) =
    \ln\det(\covstaterr) + 
    \ln\det(\covtotunobs) +
    \ln\det (\mat{Z}) \,.
    \label{eq:logdetcovobs_detlemma}
\end{equation}
A derivation of this identity is provided in \cref{apx:matrix_determinant_lemma}.
Working with the logarithm is essential for the computation of determinants as they become easily so large or small that the representational capabilities of double precision floats are exceeded.
The logarithm of a determinant can be computed efficiently based on a Cholesky decomposition, i.e., $\mat{Z} = \mat{L}\mat{L}^T$ with $\mat{L}$ being a triagonal matrix, see, e.g.,~\citep{harville_matrix_1997}.
The logarithm of the determinant is then given by $\ln\det \mat{Z} = 2\sum_{i=1}^{N} \ln L_{ii}$.

At first glance, the evaluation of the inverse and determinant of $\covobs$ according to the prescriptions in \cref{eq:invcovobs_woodbury,eq:logdetcovobs_detlemma} seems to be more involved than the direct computation.
The advantage of these formulas is owed to the structure of the matrices.
The covariance matrix $\covstaterr$ associated with statistical errors is diagonal thanks to the mutual independence of statistical errors, hence its inversion is trivial.
Its diagonal contains the squared statistical uncertainties.
The number of model parameters and systematic uncertainties is typically much lower than the number of observable variables, hence the inversion of the smaller dimensional matrix $\mat{Z}$ can be performed faster than of $\covobs$ in \cref{eq:covobs_factorized}.
Moreover, model parameters are almost always assumed to be independent of systematic errors, and systematic errors are mutually independent of each other.
In effect, this leads to $\covtotunobs$ being sparse and block diagonal.
The matrix $\totmap$ is very rectangular, i.e., the number of rows equalling the number of observable variables is usually much larger than the number of model parameters and systematic error components in typical evaluation scenarios.
The blocks in $\totmap$ corresponding to systematic experimental errors, such as normalization errors, contain only a very small number of non-zero elements, hence $\totmap$ exhibits a large degree of sparseness, which can be exploited in matrix multiplications and the calculation of inverses and determinants.
More information on how the sparsity is exploited in computations can be found in~\cref{apx:prod_inv_mat}.

\subsection{Gaussian processes}
A MVN distribution is a model for the joint distribution of a finite number of variables.
A Gaussian process is the generalization of a MVN distribution to an infinite number of variables, i.e., functions.
Gaussian processes appear in the modeling of systematic errors that vary as a function of incident energy, such as the calibration error of a detector, and they can be used to incorporate the notion of model defects into nuclear data evaluations \citep{leeb_consistent_2008,neudecker_impact_2013,schnabel_large_2015,schnabel_differential_2016,helgesson_fitting_2017} and more generally in the calibration of surrogate models for computer experiments, e.g.~\citep{kennedy_bayesian_2001}.

In the pipeline they are used to inject more flexibility in energy-dependent model parameters of TALYS to address model defects.
This approach was discussed and explored in~\citep{helgesson_treating_2018}.
The available functionality in the \textit{nucdataBaynet} package~\citep{gschnabel-nucdataBaynet} goes beyond the assignment of Gaussian processes to model parameters.
Any group of elements present in the vector $\totunobs$ in \cref{eq:Su_abbreviation} can be associated with a Gaussian process.
A good and comprehensive introduction to Gaussian processes with pointers to relevant literature can be found in~\cite{rasmussen_gaussian_2006}.
In the following we provide a brief discussion tailored to their application in the pipeline.

A Gaussian process is defined by a mean function $m(E)$, which provides the a priori expectation of the function values $f(E)$ for all energies $E$ in the admissible domain, and a covariance function $\kappa(E,E')$, which provides the covariance between function values $f(E)$ and $f(E')$ for all admissible pairs of energies $E$ and $E'$: 
\begin{align}
    m(E) &= \Exp{f(E)} \;\;\textrm{and}\;\; \\
    \kappa(E,E') &= \Cov{f(E)}{f(E')} \,.
\end{align}
A defining property of a Gaussian process is that for any finite set of energies $\{E_i\}_{i=1..N}$, the function values $\{f(E_i)\}_{i=1..N}$ are governed by a MVN distribution.
Therefore, if we combine those function values to a vector $\vec{f}$, where $f_i=f(E_i)$, and construct the associated center vector $\vec{m}$, where $m_i=m(E_i)$, and covariance matrix $\mat{K}$, where $K_{ij} = \kappa(E_i,E_j)$, the pdf of the function values is given by $\mathcal{N}\left(\vec{f}\,|\,\vec{m},\mat{K}\right)$.
A typical choice for the covariance function is the so-called squared exponential,
\begin{equation}
    \kappa(E,E') = \delta^2 \exp\left[
        -\frac{1}{2\lambda^2} \left( E - E' \right)^2
    \right] \,,
    \label{eq:covfun_sqrexp}
\end{equation}
which is employed in the implementation of the pipeline to effect the evaluation of $^{56}$Fe.
The amplitude $\delta$ represents the a priori standard deviation of all function values $f(E)$ and is therefore a measure of the range function values are expected to cover.
The length-scale $\lambda$ determines the correlation of function values at different energies, hence is a measure of the expected similarity of function values associated with similar energies.
Other choices of the covariance function are possible. The thesis of Duvenaud~\citep[chap. 2]{duvenaud_automatic_2014} contains a pertinent catalogue of them and discusses their properties.
The possibility of energy-dependent length-scales and amplitudes and their determination on the basis of a collection of neighboring isotopes is explored in~\citep{schnabel_first_2018}.

To predict the function values at energies of interest $\{E_i^\textrm{pred}\}_{i=1..M}$ based on the function values observed at energies $\{E_i^\textrm{obs}\}_{i=1..N}$, we use \cref{eq:cond_mvn_hidmean,eq:cond_mvn_hidcov} to get the conditional pdf.
The elements of required vectors and matrices are computed as follows:
\begin{alignat}{2}
    \left(\muhid\right)_i &= m(E_i^\textrm{pred}) \,,
    &\left(\covhid\right)_{ij} &= \kappa(E_i^\textrm{pred},E_j^\textrm{pred}) \\
    \left(\muobs\right)_i &= m(E_i^\textrm{obs}) \,,
    &\left(\covobs\right)_{ij} &= \kappa(E_i^\textrm{obs},E_j^\textrm{obs}) \\
    (\vec{o})_i &= f(E_i^\textrm{obs}) \,, 
    &\left(\covobshid\right)_{ij} &= \kappa(E_i^\textrm{obs},E_j^\textrm{hid}) 
\end{alignat}

However, there are two issues with the use of Gaussian processes in our situation.
First, the inference based on Gaussian processes can become prohibitively computationally expensive with an increasing number $N$ of observed function values due to the needed inversion of $\covobs$.
The time required for an inversion scales with the cube of the number of observed function values.
Second, typical choices of covariance functions define a Gaussian process that corresponds to a fitting function with an infinite number of parameters.
Thus, imposing a Gaussian process on any block of variables in $\totunobs$ in \cref{eq:Su_abbreviation}, e.g., energy-dependent nuclear model parameters that are propagated to observations via the mapping matrix $\totmap$, would be impossible because an infinite number of energy dependent variables needed to be stored in $\totunobs$.

Both issues can be dealt with by introducing a finite dimensional approximation to Gaussian processes.
There is a variety of sparse Gaussian process approximations, such as in~\citep{snelson_sparse_2006}, which was also explored for the determination of model defects in nuclear data evaluation in~\cite{schnabel_estimating_2018}.
A review of various sparse GP approximations and unified framework for comparison is presented in~\cite{quinonero-candela_unifying_2005}.

Instead of using any of those approximations, we decided to construct a low-dimensional approximation to a GP based on linear interpolation.
The reason being that the sparse approximations referenced in the previous paragraph are associated with dense mapping matrices $\totmap$ whereas this matrix is very sparse if using linear interpolation.
Consult, e.g.,~\citep[sec. 2.2]{schnabel_estimating_2018}, for a brief discussion of sparse GPs in terms of the associated mapping matrix.
In the remainder of this section, we introduce the essential formulas for a sparse GP approximation based on linear interpolation.

Given two mesh points associated with energies $E_{i}$ and $E_{i+1}$, where $E_{i} < E_{i+1}$ and function values $y_i$ and $y_{i+1}$, respectively, values at intermediate energies can be determined by linear interpolation:
\begin{multline}
    g(E) = 
    \left(
        \frac{E_{i+1}-E}{E_{i+1}-E_i}
    \right) y_i
    +
    \left(
        \frac{E-E_i}{E_{i+1}-E_i} 
    \right) y_{i+1} \\
    \textrm{if}\;\; E_i \leq E < E_{i+1}
    \label{eq:basic_linearinterpol}
\end{multline}
To state the formulas in a general way for a complete mesh of energies, we introduce the abbreviation
\begin{equation}
    c_i(E) = \begin{cases}
        \frac{E_{i+1} - E}{E_{i+1}-E_i} & \textrm{if}\;\; E_i \leq E < E_{i+1} \\
        0 & \textrm{otherwise}
    \end{cases}
\end{equation}
and
\begin{equation}
    d_i(E) = \begin{cases}
        \frac{E - E_{i-1}}{E_{i}-E_{i-1}} & \textrm{if}\;\; E_{i-1} \leq E < E_{i} \\
        0 & \textrm{otherwise}
    \end{cases}
\end{equation}
as well as their sum,
\begin{equation}
    f_i(E) = c_i(E) + d_i(E) \,.
\end{equation}
Piecewise linear interpolation, i.e., locating for an energy of interest $E$ the enclosing energies on the mesh and then performing linear interpolation according to \cref{eq:basic_linearinterpol}, can now be concisely written as
\begin{equation}
    g(E) =
    \sum_{i=1}^{M} f_i(E) y_{i} \,.
    \label{eq:pwlinint}
\end{equation}
Thanks to the bilinearity of the covariance operator, the covariance between function values at arbitrary energies can be computed by
\begin{multline}
    \Cov{g(E)}{g(E')} = \\
    \sum_{i=1}^{M} \sum_{j=1}^{M}
    f_i(E) f_j(E') \Cov{y_i}{y_j}.
    \label{eq:covpwlinint}
\end{multline}
This formula shows that the covariance matrix $\tilde{\mat{K}}$ with $\tilde{K}_{ij}=\Cov{y_i}{y_j}$ associated with a \textit{finite} number of variables $M$, in combination with linear interpolation enables the computation of covariances for arbitrary pairs of energies $E, E'$.
In other words, it is a valid specification of a covariance function $\kappa(E,E')$ and together with a mean function $m(E)$ completely characterizes a Gaussian process.

Furthermore, we can identify a mapping matrix $\totmap$ which allows us to incorporate Gaussian processes in a principled way into the representation presented in \cref{eq:covobs_factorized}.
For example, let us assume that we have an energy-dependent systematic error assigned to a certain experimental dataset that contains measured values at energies $\{E^\textrm{obs}_i\}_{i=1..N}$.
Now using a reduced energy mesh containing the energies $\{E^\textrm{hid}_j\}_{j=1..M}$ as a basis for linear interpolation, the partial derivatives of \cref{eq:pwlinint} with respect to the $y_i$, i.e., the values at the knot points, constitute the elements of the mapping matrix:
\begin{equation}
    S_{ij} = \frac{\partial g(E^{\textrm{obs}}_i)}{\partial y_j} = f_j(E^{\textrm{obs}}_i) \,.
    \label{eq:Sgplinint}
\end{equation}
Noteworthy, for a given energy $E^{\textrm{obs}}_i$, there are only two non-zero values of $f_j(E^{\textrm{obs}}_i)$, hence the mapping matrix is very sparse with only two non-zero elements in each row.
For the sake of an example, if $\covobs$ were only made up of an energy-dependent systematic uncertainty given as a GP and uncorrelated statistical uncertainties reflected in $\covstaterr$, we would have
\begin{align}
    \covobs &= \totmap \tilde{\mat{K}} \totmap^T + \covstaterr \, \\
    \covobshid &= \totmap \tilde{\mat{K}} \, \\
    \covhid &= \tilde{\mat{K}}
\end{align}
with the elements of $\tilde{\mat{K}}$ defined below \cref{eq:covpwlinint} and those of $\totmap$ in \cref{eq:Sgplinint}.

\subsection{Levenberg-Marquardt algorithm with prior}
\label{subsec:LM_with_prior}
We assumed a linear link between model parameters and observables in the discussion of \cref{subsec:info_representation}.
However, the model parameters of a nuclear physics model are typically linked to observables in a non-linear way.
In some evaluations, the model parameters are cross sections and adjusted according to measurements of ratios of these cross sections.
This would also result in a non-linear link between parameters and observables.
Irrespective of the cause of the non-linearity, the Jacobian matrix $\modparmap$ becomes a function of the parameter vector, $\modparmap(\modparrefvec)$.
For simplicity, we refer from now on to a nuclear model but it should be understood that the discussion is valid for any non-linear link between parameters and observations.

One possible solution is to apply the GLS formulas iteratively and to use an updated linear approximation of the nuclear model in each iteration.
For instance, this approach is one option offered by the CONRAD evaluation code~\citep{jean_uncertainty_2011,archier_conrad_2014} and SAMMY~\citep{larson_updated_2008} and briefly discussed in~\citep{jean_uncertainty_2011} and in more detail in \citep[sec. IV.A.3]{larson_updated_2008}.
The modified Levenberg-Marquardt algorithm outlined in this section improves upon this approach by the introduction of an adaptive damping term.

In the following we abbreviate $\mat{J} = \modparmap$.
A linear approximation of a nuclear model as in \cref{eq:linapprox_model}, which we reiterate here for convenience,
\begin{equation}
    \mathcal{M}_\textrm{lin}(\modparvec) =
    \modrefvec + \mat{J}(\modparrefvec) \, (\modparvec-\modparrefvec) \,,
    \label{eq:linapprox_model2}
\end{equation}
is only reliable in vicinity of the expansion point $\modparrefvec$.
The choice to pick the a priori best guess $\vec{p}_0$ as reference point $\modparrefvec$ is plausible but an updated best guess $\vec{p}_1$ obtained by the conditioning formula in \cref{eq:cond_mvn_hidmean} can be misleading due to the fact that the linear map $\mat{J}(\vec{p}_0)$ is not representative anymore for the region of the parameter space where $\vec{p}_1$ resides.

The Levenberg-Marquardt algorithm~\citep{levenberg_method_1944,marquardt_algorithm_1963} extends the idea of the linear least squares method to a non-linear link between model parameters and predictions.
We use a modified version of the Levenberg-Marquardt algorithm~\citep{helgesson_fitting_2017} to take into account an experimental covariance matrix and a prior parameter covariance matrix.

To prepare the discussion of the modified Levenberg-Marquardt algorithm, we write the pdf of $\modparvec$ conditioned on the observable variables $\obsvec$ in the factorized form
\begin{equation}
    \pi(\modparvec \,|\, \obsvec) = \frac{1}{\pi(\obsvec)} \ell(\obsvec \,|\, \modparvec) \pi(\modparvec) \,.
    \label{eq:bayesupdate}
\end{equation}
The first factor $1/\pi(\obsvec)$ is independent of $\modparvec$ and hence represents a normalization constant to ensure the integral of the conditional pdf being unity.
For a non-linear model link to map from $\modparvec$ to the observables in $\obsvec$, this normalization constant usually cannot be obtained analytically and one either needs (1) to take recourse to a Monte Carlo approach to estimate it or (2) approximate the conditional pdf by a simpler functional form for which it can be calculated analytically.
Later in the discussion, we take the second approach, but for the moment it is not necessary to have the conditional pdf $\pi(\modparvec \,|\, \obsvec)$ correctly normalized.

We continue by specifying for both pdfs, i.e., the so-called likelihood $\ell(\obsvec \,|\, \modparvec)$ and prior $\pi(\modparvec)$, multivariate normal distributions,
\begin{align}
    \ell(\obsvec \,|\, \modparvec) &= \mathcal{N}(\obsvec \,|\, \mathcal{M}(\modparvec); \covexp) \,,
    \label{eq:mvn_likelihood_exactmodel} \\
    \pi(\modparvec) &= \mathcal{N}(\modparvec \,|\, \vec{p}_0; \mat{P}_0) \,.
\end{align}
If the model were linear, i.e., could be written in the form of \cref{eq:linapprox_model2}, the resulting conditional pdf $\pi(\modparvec \,|\, \obsvec)$ in \cref{eq:bayesupdate} would be multivariate normal. Its maximum would then be associated with the updated vector $\vec{p}_1$ obtained by the formula for the conditional mean vector in \cref{eq:cond_mvn_hidmean} if making the identifications:
\begin{align}
    \muhid &= \vec{p}_0 \,,
    &\covhid &= \mat{P}_0 \\
    \muobs &= \mathcal{M}_\textrm{lin}(\vec{p}_0) \,,
    &\covobs &= \mat{J}\mat{P}_0\mat{J}^T + \covexp \\
    \vec{o} &= \obsvec \,,
    &\covobshid &= \mat{J} \mat{P}_0
\label{eq:LMcondsubst}
\end{align}

However, the nuclear model represented by the function $\mathcal{M}(\modparvec)$ is non-linear and does in general not have an analytical form because the nuclear model code may use numerical methods to solve differential equations or implements a stochastic simulation.
Consequently, there is no simple update formula to locate the parameter vector $\vec{p}_1$ associated with the maximum of the conditional pdf $\pi(\modparvec \,|\, \obsvec)$.

The idea of the modified Levenberg-Marquardt (LM) algorithm is to still rely on the formula for the conditional mean \cref{eq:cond_mvn_hidmean} and exploit a linear approximation of the nuclear model to locate a parameter vector closer to the maximum.
Introducing the quantities,
\begin{align}
    \mat{A} &= 
        \mat{J}^T \covexp^{-1} \mat{J} + \mat{P}_0^{-1} + \lambda \mathcal{I} \,,
    \\
     \vec{b} &= 
     \mat{J}^T \covexp^{-1} \left(
        \obsvec - \modrefvec
     \right) +
     \mat{P}_0^{-1}
     \left(
        \vec{p}_0 - \modparrefvec
     \right) \,,
\end{align}
the conditional mean vector is given by (if $\lambda=0$):
\begin{equation}
    \vec{p}_\textrm{prop} = \modparrefvec + \mat{A}^{-1} \vec{b} 
    \label{eq:LMproposal} \,.
\end{equation}
Consult \cref{apx:LM_details} for a derivation of this update formula.
We discuss the purpose of the term $\lambda\mathcal{I}$ in a moment.
It is not evident by visual inspection that this formula (with $\lambda=0$) is equivalent to \cref{eq:cond_mvn_hidmean} with the appropriate substitutions of \cref{eq:LMcondsubst} but they can be transformed into each other by making use of the Woodbury matrix identity~\citep{woodbury_inverting_1950}.
The Woodbury identity is briefly discussed in \cref{apx:Woodbury_identity}.

The LM algorithm is an iterative algorithm.
It is guaranteed to converge to a local optimum. If there is only one optimum, it is guaranteed to converge to the global optimum. The following elaboration will make apparent this feature.
In each iteration it relies on a linear approximation of the non-linear model constructed at the  obtained proposal vector $\vec{p}_\textrm{prop}$ of the previous iteration denoted by $\vec{p}_\textrm{prev}$.
Therefore the linear approximation employed in the current iteration is characterized by $\modparrefvec=\vec{p}_\textrm{prev}$, $\modrefvec=\mathcal{M}(\vec{p}_\textrm{prev})$ and $\mat{J}=\mat{J}(\vec{p}_\textrm{prev})$.

In the case of a mildly non-linear model, the proposal (with $\lambda=0$) in \cref{eq:LMproposal} is close to the true maximum of the pdf.
For stronger degrees of non-linearity, the current linear approximation can only be trusted in close vicinity of the current reference parameter vector $\modparrefvec$ and the proposed parameter vector $\vec{p}_\textrm{prop}$ may lie outside this region of trust.
This would also invalidate the trust in the pertinence of $\vec{p}_\textrm{prop}$.

The innovation of the LM algorithm is the introduction of the additional term $\lambda\mathcal{I}$ with $\lambda > 0$ and $\mathcal{I}$ a diagonal matrix.
If the proposal parameter vector is too far away from the reference parameter vector to rely on the linear approximation, the value of $\lambda$ is increased.
Consequently the matrix $\mat{A}$ becomes more diagonal and so its inverse.
As the vector $\vec{b}$ is the gradient of $\ln\pi(\modparvec \,|\, \obsvec)$ evaluated at $\modparrefvec$, i.e., $b_i = \partial (\ln\pi(\modparvec \,|\, \obsvec)) / \partial {p_i}$,
the parameter update degenerates gradually to the gradient ascent method with an increasing value of $\lambda$ and at the same time the step size becomes smaller.
The adjustment of $\lambda$ is therefore a mechanism to smoothly transition between an update using the formula for the conditional mean of the MVN distribution if the nuclear model is only mildly non-linear and the gradient ascent method for larger degrees of non-linearity.

The adjustment of $\lambda$ is effected based on the comparison between the expected increase of the value of $\ln\pi(\modparvec \,|\, \obsvec)$ according to the linear approximation and the actual increase by evaluating the nuclear physics model.
Let $f_\textrm{ex}(\vec{p}) = \ln\pi(\modparvec \,|\, \obsvec)$ be the value based on the exact nuclear physics model (applied in \cref{eq:mvn_likelihood_exactmodel}) and $f_\textrm{lin}(\modparvec) = \ln\tilde{\pi}(\modparvec \,|\, \obsvec)$ the value based on the linear approximation constructed at $\modparrefvec$.
The criterion for adjustment is the gain, see, e.g.,~\citep{madsen2004},
\begin{equation}
    \rho = \frac{
        f_\textrm{ex}(\vec{p}_\textrm{prop}) - f_\textrm{ex}(\vec{p}_\textrm{ref})
    }{
        f_\textrm{lin}(\vec{p}_\textrm{prop}) - f_\textrm{lin}(\vec{p}_\textrm{ref})
    } \,.
\end{equation}
The strategy proposed by Marquardt~\citep{marquardt_algorithm_1963} is to select after each iteration a new value of lambda according to the prescription
\begin{equation}
    \lambda' =
    \begin{cases}
    2 \lambda & \textrm{if } \rho < 0.25 \\
    \lambda & \textrm{if } 0.25 \leq \rho < 0.75 \\
    \lambda/3 & \textrm{if } 0.75 \leq \rho
    \end{cases} \,.
\end{equation}
Noteworthy, if a proposed parameter vector $\vec{p}_\textrm{prop}$ is associated with a lower value of the posterior pdf in \cref{eq:bayesupdate}, it is rejected and a new proposal computed on the basis of the updated value of $\lambda$.
Because each rejection leads simultaneously to a reduction of the step size and a new step direction closer aligned to the direction of steepest ascent, the LM algorithm is guaranteed to converge to an optimum.
We employed as stopping criterion the relative difference of the logarithmized posterior pdf associated with the parameter vector proposals of consecutive iterations.
In the current configuration of the pipeline, the optimization terminates if the relative difference falls below $10^{-5}$.

In the case of a large number of model parameters, it is computationally attractive to optimize only the parameters which are sensitive to the experimental data.
Afterwards, due to prior correlations between sensitive and insensitive parameters, also insensitive parameters must be updated.
This update procedure is described in \cref{apx:comp_posterior_expectation}.

\subsection{Taylor expansion of logarithmized pdf}
\label{subsec:taylor_logpdf}

Once the vector $\vec{p}_1$ associated with the maximum of the posterior distribution in \cref{eq:covobs_factorized} is located using the LM algorithm, one may be tempted to apply the conditioning formulas \cref{eq:cond_mvn_hidmean,eq:cond_mvn_hidcov} with the substitutions in \cref{eq:LMcondsubst} for one further improvement of the posterior center vector $\vec{p}_1$ and to obtain the posterior covariance matrix $\mat{P}_1$.
This strategy only works if the statistical model is consistent with the experimental data, i.e., the remaining differences between the posterior model prediction and the experimental data can be explained by the statistical and systematic uncertainties associated with the experiments.
Otherwise, in the case of the experimental data and/or the model being deficient, the results are  misleading.
For instance, the update of $\vec{p}_1$ could undergo a comparatively large adjustment at odds with the found solution by the LM algorithm, and the associated value of $\pi(\modparvec \,|\, \obsvec)$ would be drastically reduced.
The reason for this behavior is explained in~\cref{apx:necessity2ndorder}.

If experimental data or the nuclear model have deficiencies, a better solution to obtain the posterior covariance matrix $\mat{P}_1$ is to perform a second-order Taylor expansion at the posterior maximum.
Misspecified experimental data or models remain still problematic but the covariance matrix obtained in this way is at least consistent with the vector $\vec{p}_1$ found by the LM algorithm.
The form of a second-order Taylor approximation of the logarithmized posterior pdf is given by
\begin{multline}
    \ln \pi(\modparvec \,|\, \obsvec) = \pi_0 + 
    \mat{G}
    \left(
        \modparvec - \modparrefvec
    \right) \\
    + \frac{1}{2}
    \left(
        \modparvec - \modparrefvec
    \right)^T
    \mat{H}
        \left(
        \modparvec - \modparrefvec
    \right) \,.
    \label{eq:logposteriorpdf_2ndorderapprox1}
\end{multline}
Noteworthy, the gradient $\mat{G}$ is not associated with the nuclear model but with the posterior pdf, i.e., $G_i = \partial (\ln \pi(\modparvec \,|\, \obsvec)) / \partial p_i$.
Ideally, if the LM algorithm has converged and $\modparrefvec = \vec{p}_1$, i.e., the expansion is done exactly at the parameter vector corresponding to the posterior maximum, the gradient vanishes.
We want to extract the posterior covariance matrix $\mat{P}_1$ from this expression.
Comparing the expansion in \cref{eq:logposteriorpdf_2ndorderapprox1} with the logarithmized pdf of a multivariate normal distribution in \cref{eq:logmarlike}, we can identify the posterior covariance matrix $\mat{P}_1$ as
\begin{equation}
    \mat{P}_1 = -\mat{H}^{-1} \,.
\end{equation}
Therefore in order to determine the posterior covariance matrix, one needs to numerically determine the Hessian matrix. Its elements are given by the second-order derivatives with respect to the model parameters,
\begin{equation}
    H_{ij} = 
    \left.
    \frac{
        \partial^2 (\ln \pi(\modparvec \,|\, \obsvec)) 
    }{
        \partial p_i \partial p_j
    }
    \right|_{\vec{p}=\modparrefvec} \,.
\end{equation}
For the numerical evaluation of the full Hessian matrix, the number of required nuclear model invocation scales quadratically with the number of adjustable model parameters.
As the aim of the pipeline is large scale evaluations taking into account potentially hundreds of model parameters, the full determination of the Hessian matrix may be impractical or even infeasible in the case of a computationally expensive nuclear model.

Therefore we implemented an approximation of the full Hessian matrix, which exploits the mathematical structure of the pdf in \cref{eq:bayesupdate} when prior and likelihood are multivariate normal distributions.
In this case, the full Hessian matrix can be written as (see \cref{apx:comp_posterior_covmat})
\begin{equation}
    \mat{H} = -(\mat{U} + \mat{J}^T \mat{Q} \mat{J} + \mat{R}) \,,
\end{equation}
with $\mat{J}$ being the Jacobian matrix of the nuclear model appearing in \cref{eq:linapprox_model2}, $\mat{Q}$ being the inverse experimental covariance matrix and $\mat{R}$ being the inverse prior parameter covariance matrix.
The elements of the matrix $\mat{U}$ are determined by a product involving second-order derivatives of the nuclear model, elements of the experimental covariance matrix and differences between model predictions and experimental data.
The approximation consists in replacing the exact matrix $\mat{U}$ by an approximation that neglects second-order derivatives involving two different model parameters.
This approximation is explained in more detail in \cref{apx:comp_posterior_covmat}.

\section{IT BUILDING BLOCKS}
\label{sec:IT_building_blocks}

The implementation of an evaluation as a sequence of scripts formulated at a high-level of abstraction leads to concise and modular scripts.
Concise scripts are easier to read and facilitate the understanding of the important decisions taken in the evaluation.
For instance, the pipeline should not contain the code for fitting parameters with all the gory details of the Generalized Least Squares (GLS) or similar fitting method but instead invoke a function with an emblematic name to effect such fitting.
This idea does not mean that the code implementing lower-level functionality, such as the GLS approach, should be inaccessible.
It just means that such code should not clutter the scripts in the pipeline to draw away the attention from important aspects, such as the selection of experimental datasets for the evaluation.

Therefore this section is dedicated to---in lack of a better name---IT building blocks.
We discuss functionality that should in our opinion be available \textit{out-of-the-box} to evaluators who wish to design an evaluation pipeline.
We outline our approach to the compartmentalization of such functionality and the design considerations that guided us.
Specifically, we elaborate on the following:

\begin{itemize}
    \item 
\textbf{Handling of information in the Bayesian context:}
In the previous section we outlined the mathematical machinery relied upon in our prototype of an evaluation pipeline.
We believe that thinking about an evaluation in terms of, e.g., experimental measurements vectors $\obsvec$, experimental covariance matrices $\covexp$, and mapping matrices $\totmap$ may be not too intuitive.
Instead, the construction of such mathematical objects should be automatically derived from a more intuitive representation of information in terms of specific types of systematic components and their association with experiments.

\item
\textbf{Handling of experimental data:}
The combination of information of various experiments is at the heart of nuclear data evaluation.
Therefore it is an essential requirement to have flexible means to search and retrieve relevant experimental data, in particular data in the EXFOR library.
Besides an increased convenience in manual evaluation work, a better handling of experimental data facilitates the renormalization of experimental data according to the latest evaluations of monitor reactions and opens new possibilities for uncertainty quantification (UQ) and quality assurance based on advanced statistical methods and machine learning.
In the course of an evaluation, it may also be helpful to augment experimental data with additional information derived from the raw data, e.g., a covariance matrix derived from the individual systematic uncertainty components.
Adding such information should be facilitated as much as possible from the perspective of an evaluator.
Also facilitating linking together relevant data, such as evaluations of standard reactions with experimental data where they have been employed as monitor reactions, can have a positive impact on the quality and transparency of evaluations.

\item
\textbf{Mapping of model predictions to EXFOR:}
Not all measurements documented in the EXFOR library can be directly related to the output of nuclear model codes, such as TALYS employed in our prototype of the pipeline.
As an example, many experiments measure ratios of nuclear reactions whereas nuclear model codes provide the individual reactions as output. 
The translation of results from nuclear model codes to predictions comparable with experimental data is an essential requirement of any evaluation and should therefore be easy to achieve by an evaluator.

\item
\textbf{Parallelization of nuclear model invocations:}
With the availability of ever increasing computing power reflected in growing computational demands of evaluation approaches, cluster computing is not any longer only reserved to large research projects but can be a viable option even for a single evaluator.
Ideally, evaluators should not be required to care about \textit{how} computations are run in parallel but instead empowered to focus completely on \textit{what} they want to compute. 
\end{itemize}

The following subsections provide more details on design considerations to implement the functionality of each IT building block.
These building blocks have been implemented as \textit{R packages}, i.e., modules that can be used in the programming language R~\cite{Rlanguage}.
The restriction to one programming language is certainly a disadvantage but other languages, such as Python, provide equivalent functions and data types, which makes the discussion of design considerations still worthwhile.
A valuable improvement in the future would be an implementation of the IT building blocks so that they can be conveniently employed in any popular programming language.

\subsection{Handling of information in the Bayesian context}
\label{subsec:handling_info_bayesian_context}

In \cref{subsec:info_representation} we explained that it is computationally advantageous to avoid the explicit construction of covariance matrices and instead maintain a representation of the form
\begin{equation}
    \covobs = \totmap \covtotunobs \totmap^T + \covstaterr
    \label{eq:covobs_handling}
\end{equation}
with the covariance matrix $\covtotunobs$ associated with systematic uncertainties and model parameter uncertainties and the diagonal covariance matrix $\covstaterr$ associated with statistical uncertainties.

Besides the aspect of computational efficiency, there is another advantage of this representation in terms of information management.
It is always possible to construct the full covariance matrix on the basis of uncertainty components but without further knowledge impossible to decompose $\covobs$ back into its individual components.

In~\cref{eq:basic_stat_model} of \cref{subsec:info_representation}, we introduced the statistical model for nuclear data evaluation that includes model parameters, systematic errors, and statistical errors.
As this equation will be relevant in the following discussion, we reiterate it here for the convenience of the reader,
\begin{equation}
    \obsvec = \totrefvec + \totmap \totunobs + \staterrvec \,.
    \label{eq:obsvec_handling}
\end{equation}
The vector $\obsvec$ contains the experimental measurements and the vector $\totunobs$ contains model parameters and systematic error components.
The matrix $\totmap$ reflects how systematic error components and model parameters---summarized under the term \textit{systematic components}---are summed up and assigned to experimental data points.

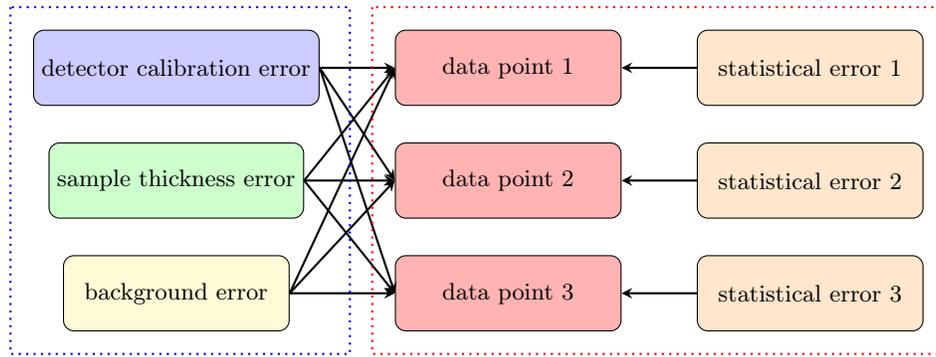
\begin{figure*}[t]
\begin{tikzpicture}[node distance=1.5cm]
\tikzstyle{startstop} = [rectangle, rounded corners, minimum width=3cm, minimum height=1cm,text centered, draw=black, fill=red!30]
\tikzstyle{io} = [trapezium, trapezium left angle=70, trapezium right angle=110, minimum width=3cm, minimum height=1cm, text centered, draw=black, fill=blue!30]
\tikzstyle{process} = [rectangle, minimum width=3cm, minimum height=1cm, text centered, draw=black, fill=orange!30]
\tikzstyle{decision} = [diamond, minimum width=3cm, minimum height=1cm, text centered, draw=black, fill=green!30]
\tikzstyle{arrow} = [thick,->,>=stealth]
    \node (syserr1) [startstop, fill=white!80!blue] {detector calibration error};
    \node (syserr2) [startstop, fill=white!80!green, below of=syserr1] {sample thickness error};
    \node (syserr3) [startstop, fill=white!80!yellow, below of=syserr2] {background error};
    \node (data1) [startstop, right=1cm of syserr1] {data point 1};
    \node (data2) [startstop, below of=data1] {data point 2};
    \node (data3) [startstop, below of=data2] {data point 3};
    \node (staterr1) [startstop, fill=white!80!orange, right=1cm of data1] {statistical error 1};
    \node (staterr2) [startstop, fill=white!80!orange, below of=staterr1] {statistical error 2};
    \node (staterr3) [startstop, fill=white!80!orange, below of=staterr2] {statistical error 3};
    \draw [arrow] (staterr1) -- (data1);
    \draw [arrow] (staterr2) -- (data2);
    \draw [arrow] (staterr3) -- (data3);
    \draw [arrow] (syserr1.east) -- (data1.west);
    \draw [arrow] (syserr1.east) -- (data2.west);
    \draw [arrow] (syserr1.east) -- (data3.west);
    \draw [arrow] (syserr2.east) -- (data1.west);
    \draw [arrow] (syserr2.east) -- (data2.west);
    \draw [arrow] (syserr2.east) -- (data3.west);
    \draw [arrow] (syserr3.east) -- (data1.west);
    \draw [arrow] (syserr3.east) -- (data2.west);
    \draw [arrow] (syserr3.east) -- (data3.west);
    \draw[red,thick,dotted] ($(data1.north west)+(-0.3,0.3)$) rectangle ($(staterr3.south east)+(0.3,-0.3)$);
    \draw[blue,thick,dotted] ($(syserr1.north west)+(-0.3,0.3)$) rectangle ($(syserr3.south east)+(0.8,-0.3)$);
    
\end{tikzpicture}
\caption{Bayesian network reflecting the typical relation between measured data points and associated systematic and statistical errors.
Systematic errors are assumed to be mutually independent but each systematic error can possibly affect all data points.
Statistical errors are also mutually independent and each of them is associated with exactly one data point.
}
\label{fig:bayesian_network}
\end{figure*}

The R package named \textit{nucdataBaynet}~\citep{gschnabel-nucdataBaynet}---an abbreviation of \textit{nuclear data in a Bayesian network}---provides the functions to do Bayesian inference exploiting the decomposition in \cref{eq:obsvec_handling} for efficient computation.
To discuss the data structures employed by this package, we consider an experiment with three measured data points being affected by the following systematic error components: (1) a detector calibration error, (2) sample thickness error, and (3) background noise.
There can be other systematic errors but it suffices to consider these three in our discussion.
These systematic errors can usually be assumed independent of each other.
For instance, the measurement device to determine the thickness of the specimen is not coupled by any physical process to the measurement procedure to determine the calibration of the detector.
The diagram in \cref{fig:bayesian_network} depicts the situation as a so-called \textit{Bayesian network}, e.g.~\citep{pearl_causality_2000}.
Each node is associated with a value.
An arrow from a node $\mathcal{A}$ to another node $\mathcal{B}$ indicates that a certain part of the value associated with $\mathcal{B}$ is given by a linear transformation of the value of $\mathcal{A}$.
In the graph, systematic errors contribute to the distortion of all three data points whereas each statistical error is connected to a single measured data point.
As an important remark, systematic components do not have to be connected to all data points, e.g., when considering data points stemming from several independent experiments.

The depicted structure of the relation between measured data points, systematic errors and statistical errors is very typical and we second the recommendation in~\citep{larson_concise_2008} that it may be regarded as the default structure for the representation of information of a nuclear experiment.

For this reason, we decided to base the \textit{nucdataBaynet} package implementing the statistical functionality outlined in the previous section on this specific representation.
In the remainder of this section, we sketch how this representation has been translated to data structures and how these data structures are used to establish the mapping of systematic components to measured data points.

\subsubsection{Experimental data and statistical uncertainties}
The experimental data points together with their statistical uncertainties are stored in a tabular data structure known as \textit{data frame} in Python and R.
For their handling, we use the R package \textit{data.table}~\citep{Rdatatable} which offers support for efficient searching, sorting and grouping of rows in data frames.
This data structure with some example contents is depicted in \cref{tbl:expDt311}.
\begin{table}[ht]
\centering
\begin{tabular}{rlrlrrr}
  \hline
IDX & EXPID & DIDX & REAC & L1 & DATA & UNC \\ 
  \hline
380 & 13764002 & 573 & (N,TOT) & 2.00 & 3255.80 & 127.05 \\ 
  381 & 13764002 & 574 & (N,TOT) & 1.99 & 3659.20 & 116.40 \\ 
  382 & 13764002 & 575 & (N,TOT) & 1.99 & 3246.60 & 128.05 \\ 
  383 & 13764002 & 576 & (N,TOT) & 1.99 & 3408.70 & 123.96 \\ 
  384 & 13764002 & 577 & (N,TOT) & 1.98 & 3359.80 & 124.44 \\ 
  385 & 13764002 & 578 & (N,TOT) & 1.98 & 2449.20 & 162.79 \\ 
   \hline
\end{tabular}
\caption{Structure of the data frame containing experimental data as used by package \textit{nucdataBaynet}.}
\label{tbl:expDt311}
\end{table}

The essential columns are \textit{IDX}, \textit{DATA} and \textit{UNC}.
The column \textit{IDX} indicates the position in $\obsvec$, the colum \textit{DATA}, the measured value at that position and \textit{UNC} the associated statistical uncertainty.
Depending on the type of systematic component, e.g., systematic error or model parameter, some of the other columns may be relevant to map the systematic components to the measured data points.

\subsubsection{Systematic components}
\label{subsubsec:syscomps}

Systematic components are also stored in a data frame.
An example is provided in \cref{tbl:sysDt312}.
These components are enclosed in a blue dotted rectangle in \cref{fig:bayesian_network}.
\begin{table}[ht]
\centering
\begin{tabular}{rlrllrr}
  \hline
IDX & EXPID & DIDX & ERRTYPE & DATA & UNC \\ 
  \hline
  1 & REACEXP-N,INL-01 &   1 & pw & 0.00 & 100.00 \\ 
    2 & REACEXP-N,INL-01 &   2 & pw & 0.00 & 100.00 \\ 
    3 & REACEXP-N,INL-01 &   3 & pw & 0.00 & 2.00 \\ 
    4 & EXPID-13764002 &   1 & sys-rel & 0.00 & 0.10 \\ 
    5 & EXPID-22316003 &   1 & sys-abs & 0.00 & 100.00 \\ 
    6 & EXPID-23134005 &   1 & sys-rel & 0.00 & 0.10 \\ 
    7 & EXPID-32201002 &   1 & sys-rel & 0.00 & 0.06 \\ 
    8 & EXPID-40532014 &   1 & sys-rel & 0.00 & 0.10 \\ 
   \hline
\end{tabular}
\caption{Structure of the data frame containing systematic components as used by package \textit{nucdataBaynet}.
Column \textit{GPTYPE} has been omitted for better display.}
\label{tbl:sysDt312}
\end{table}
The essential columns are \textit{IDX}, \textit{ERRTYPE}, \textit{DATA} and \textit{UNC}.
The column \textit{IDX} indicates the position of the systematic component in $\totunobs$, the column $\textit{DATA}$ stores its value, and \textit{UNC} the associated uncertainty.
The column \textit{ERRTYPE} indicates the type of systematic component, which can be, e.g., a relative normalization error (\textit{sys-rel}) or a piecewise linear function (\textit{pw}). 
Typically, systematic errors are a priori expected to be absent, hence the zeros in the \textit{DATA} column.
This is the default assumption in the GLS method as employed for nuclear data evaluation, even though systematic errors are usually not made explicit there.  
The non-zero uncertainties assigned in column \textit{UNC} allow for non-zero posterior expectations.

The important assumption taken is that all systematic components are mutually independent, which allows to store the uncertainties in a column of this data frame instead of managing a full covariance matrix.
This independence assumption must sometimes be abandoned, e.g., when dealing with energy dependent systematic uncertainties.
We are going to discuss this case later. 
For the moment we assume the independence assumption to hold.

\subsubsection{Mapping of systematic components to experimental data}
\label{subsubsec:mapping_sys_exp}
Having introduced the two data structures for experimental data and systematic components, we can discuss the mapping of systematic components to experimental data points, illustrated by the arrows in \cref{fig:bayesian_network}.
There can be many types of systematic components, such as relative systematic errors, absolute systematic errors, and parameters characterizing the shape of a piecewise linear function.
The type of systematic component is indicated in the column \textit{ERRTYPE}.
In the example in~\cref{tbl:sysDt312}, the identifier \textit{pw} denotes a piecewise linear function and the corresponding elements in the \textit{DATA} and \textit{UNC} column contain the prior expectation and uncertainty, respectively, of the values at the mesh points.
Analogously, the identifiers \textit{sys-rel} and \textit{sys-abs} refer to relative and absolute systematic errors, respectively.
It is crucial to be able to map systematic components to measured data points.
Considering \cref{eq:covobs_handling,eq:obsvec_handling}, this means to be able to construct the matrix $\totmap$ effecting the mapping.

During the conception of the pipeline, it was evident from the beginning that we cannot implement the mappings for all conceivable types of systematic components and therefore our objective was to design an extensible architecture to make the package as future-proof as possible.
The final implementation corresponds directly to the following picture:
Each type of systematic component is associated with a handler which knows how to deal with this particular type.
A manager module keeps track of the available handlers.
A request to the manager to construct the mapping matrix $\totmap$ for a given pair of a data frame with experimental data \textit{expDt} and another one with systematic components \textit{sysDt}, the manager splits the data frame \textit{sysDt} in chunks where each chunk corresponds to a specific type.
For each chunk, the manager asks in turn the available handlers if they know how to map this type of systematic component.
The first handler answering in the affirmative is tasked with the construction of the mapping matrix for that chunk.
At the very end, the manager combines all the results of the different handlers to a final result, which is the matrix $\totmap$ to map all systematic components at once to the experimental data.
In the diagram showing the Bayesian network, the matrix $\totmap$ defines the connections between the systematic components and the measured data points. 

In the language of informatics, each handler is a module providing the two functions:
\begin{itemize}
    \item \textit{getErrTypes()}: informs about the types of systematic components the handler module is able to map.
    \item \textit{map(\textit{expDt}, \textit{sysDt})}: returns the matrix implementing the mapping of systematic components in \textit{sysDt} to the experimental data in \textit{expDt}. The data frame \textit{expDt} contains all experimental data whereas \textit{sysDt} is a chunk of the full data frame containing only the \textit{ERRTYPE} the handler knows how to map.
\end{itemize}
These functions are the required interface to interact with the manager.
Handlers can and usually do possess more functions for their setup.
To make available a new type of systematic component, one has to implement a handler containing the two functions mentioned above.

Beside the essential columns \textit{IDX}, \textit{DATA} and \textit{UNC} in data frame \textit{expDt} and \textit{IDX}, \textit{ERRTYPE}, \textit{DATA} and \textit{UNC} in data frame \textit{sysDt}, individual handlers typically rely on some of the other columns to establish the mapping.
Columns not needed by a particular handler are ignored by that handler.
This flexibility is very helpful as new columns can be introduced to implement the functionality of a new handler without interfering with the functioning of already existing ones.

To make the functioning of a handler more comprehensible, we outline the handler implementing the mapping of normalization errors.
This handler receives from the manager the subset of rows in data frame \textit{sysDt} where the column \textit{ERRTYPE} contains the string \textit{sys-rel}.
We take as an example the row in~\cref{tbl:sysDt312} where the column \textit{IDX} is 4.
The string \textit{EXPID-13764002} in the \textit{EXPID} column of \textit{sysDt} informs the handler that this systematic error component is associated with rows in \textit{expDt} that contain the string \textit{13764002} in the column \textit{EXPID}.
The handler then identifies these rows in \textit{expDt} and retrieves the indices in \textit{IDX}.
Taking~\cref{tbl:expDt311} as example, the first index there $380$ in combination with the index $4$ in \textit{sysDt} leads to the specification $S_{380,4} = 3255.80$ where the value is the cross section to have the mapping of a \textit{relative} systematic error component.
The other elements of $\totmap$ associated with other rows of \textit{expDt} are assigned analogously.
This example also shows that the resulting matrix $\totmap$ is sparse.
The mapping of a normalization error requires only one non-zero element per row.
This sparsity of the mapping matrix is very typical for systematic error components and exploited in the \textit{nucdataBaynet} package by relying on the support of sparse matrices provided by the \textit{Matrix} package~\citep{RMatrix}.

Noteworthy, the functionality of the handler dealing with normalization errors is more general.
For instance, a normalization error can be assigned to experimental data points based on the string in the \textit{REAC} column.
Thereby each experimental data point of a specific reaction channel is associated with the same normalization error.
Such a normalization error can be introduced in addition to normalization errors at the level of datasets.
The next section on the handling of experimental data elaborates on the flexibility in data retrieval from the EXFOR library.
For instance, it is straight-forward to add to the data frame \textit{expDt} a column with the institution where the experiment was performed or a column indicating the detector type. 
Normalization errors could also be associated with data points based on those attributes, which enables the creation of elaborated error structures while maintaining computational tractability thanks to the decomposition of the covariance matrix explained before.

\subsubsection{Correlated systematic components}
The assumption of mutual independence of the systematic components does not always hold.
Some systematic errors of the experiment may be fully or partially correlated over energy.
Also Gaussian processes imposed on energy-dependent TALYS parameters, i.e., functions of energy, represent a prior that correlates the parameter values at nearby energies.
From a technical point of view, there is no difference between a Gaussian process on an energy-dependent parameter and a systematic experimental error correlated over energy.
Both cases are addressed by the introduction of off-diagonal elements in the covariance matrix $\covtotunobs$.
As the data structure to represent systematic components outlined in  \cref{subsubsec:syscomps} only allows the storage of uncertainties, i.e., assuming $\covtotunobs$ to be diagonal, another data structure is needed to account for the correlations defined by a Gaussian process.

To discuss this case, we show in \cref{fig:sysDt_with_GP} an example of a data frame for systematic components which also contains a Gaussian process 
specification for energy-dependent systematic components.
\begin{table}[ht]
\centering
\begin{tabular}{lllrrr}
  \hline
EXPID & ERRTYPE & GPTYPE & DATA & UNC & EN \\ 
  \hline
EXPID-32201002 & sys-rel & NA & 0.00 & 0.06 & NA \\ 
EXPID-40532014 & sys-rel & NA & 0.00 & 0.10 & NA \\ 
TALYS-d1adjust n & endep & sqrexp & 1.00 & 0.10 & 1.00 \\ 
TALYS-d1adjust n & endep & sqrexp & 1.00 & 0.10 & 2.00 \\ 
   \hline
\end{tabular}
\caption{Example of a data frame for systematic components \textit{sysDt} with a Gaussian process specification. Some columns not relevant in the current discussion are omitted in this table for better display.
The \textit{ERRTYPE} named \textit{talyspar\_endep} is abbreviated.
NA stands for \textit{not available} or \textit{not answered} and is a special value in R.}
\label{fig:sysDt_with_GP}
\end{table}
The assignment of the GP is indicated by the string \textit{sqrexp} in the column \textit{GPTYPE}.
Here the GP is imposed on the energy-dependent TALYS parameter associated with the input keyword \textit{d1adjust n} but GPs can also be assigned to systematic experimental errors.
As an aside, we implemented a specific handler for the mapping (or in this context more typical diction: propagation) of model parameters to measured data points.
The handler is responsible for rows with \textit{ERRTYPE} being either \textit{talyspar\_endep} or \textit{talyspar}.

The important column for correlated systematic components is \textit{GPTYPE}.
This column indicates the type of Gaussian process imposed on the systematic components in order to introduce correlations.
As for the mapping of systematic components to measured data points, specialized GP handlers are responsible for the different Gaussian process types.
The implementation of a GP handler for a specific \textit{GPTYPE} can rely on any column in the \textit{sysDt} data frame to construct the parts of the covariance matrix $\covtotunobs$ associated with this GP type.
The GP handler for the \textit{GPTYPE} identified by \textit{sqrexp}, in particular, requires the column \textit{EN}, which indicates the energy.
This information is used in the construction of the covariance matrix on the basis of the covariance function defined in \cref{eq:covfun_sqrexp}.

A GP handler needs to implement the following two functions:
\begin{itemize}
    \item \textit{getGPTypes()}: returns a list of strings of GP types supported by the handler.
    \item \textit{cov(sysDt, gpDt)}: returns the covariance elements of systematic components \textit{sysDt} of systematic components associated with a Gaussian process type supported by the handler. 
\end{itemize}
Besides these two functions, GP handlers usually have additional functions for their setup.
A manager module aware of the available handlers splits the data frame of systematic components in chunks according to the \textit{GPTYPE} and delegates the individual chunks to the appropriate GP handlers to compute the parts of the covariance matrix.
This manager module is the same that handles the mapping from systematic components to measured data points discussed in \cref{subsubsec:mapping_sys_exp}.
In a request to the manager module to return the complete covariance matrix $\covtotunobs$, the manager merges the covariance elements returned by the GP handlers with the diagonal elements corresponding to the uncertainties in the \textit{UNC} column in the data frame \textit{sysDt}, see \cref{fig:sysDt_with_GP}.
In the current implementation, values in the \textit{UNC} column in \textit{sysDt} are ignored if a GP is assigned to the systematic component.

Covariance functions usually depend on hyperparameters, such as the amplitude $\delta$ and the length scale $\lambda$ in the case of the squared exponential covariance function.
These specifications are stored in a data frame, which we call in the following \textit{gpDt}.
An example of such a data frame is given in \cref{tbl:gpDt}.
\begin{table}[ht]
\centering
\begin{tabular}{rlllr}
  \hline
IDX & EXPID & GPTYPE & PARNAME & PARVAL \\ 
  \hline
  1 & TALYS-d1adjust n & sqrexp & sigma & 0.10 \\ 
    2 & TALYS-d1adjust n & sqrexp & len & 2.00 \\ 
    3 & TALYS-d1adjust n & sqrexp & nugget & 0.00 \\ 
   \hline
\end{tabular}
\caption{Example of a data frame with the specification of hyperparameters of the Gaussian processes.}
\label{tbl:gpDt}
\end{table}
This data frame is required by the GP handlers to compute the covariance elements in $\covtotunobs$.
For this reason, it must also be passed as argument to the \textit{cov} function listed above.

As a final remark regarding the \textit{nucdataBaynet} package, the structure imposed on the experimental data by storing them in the outlined data structures is exploited in the computation of the inverses and determinants appearing the GLS formulas \cref{eq:cond_mvn_hidcov} and marginal likelihood formula \cref{eq:logmarlike} by relying on the matrix identities in \cref{eq:invcovobs_woodbury,eq:logdetcovobs_detlemma}.
The same identities are also employed in the analytical computation of the derivatives of the logarithmized marginal likelihood with respect to hyperparameters in Gaussian processes or any other type of uncertainty associated with a systematic component.
The derivatives of inverse matrices and logarithmized determinants appearing in this representation of the logarithmized marginal likelihood can be evaluated analytically, see \cref{apx:derivative_logdet,apx:derivative_invmat}. 
These derivatives are taken into account in the adjustment of hyperparameters and uncertainties by locating the maximum of the marginal likelihood using the L-BFGS algorithm~\citep{byrd_limited_1995}.

\subsection{Retrieval and handling of experimental data}
\label{subsec:handling_experimental_data}

The management of experimental data is an essential requirement for the creation of evaluation pipelines with a focus on automation and reproducibility.
Both the retrieval of data using flexible capabilities of search and the handling of retrieved data are important.

The \textit{EXFOR library}~\citep{otuka_towards_2014} maintained and continuously extended by the International Network of Nuclear Reaction Data Centres (NRDC) contains a large collection of nuclear cross section data and related quantities.
It relies on the EXFOR format~\citep{schwerer_exfor_2015} to store and organize the information associated with experiments.

The \textit{EXFOR format} is a text-format, hence readable by humans and hierarchically structured with the top two levels being so-called \textit{entries} and \textit{subentries}.
It exhibits a rigid structure and at times complex syntactical rules, such as those for the designation of the reaction system.
A detailed description of this format is available in~\citep{schwerer_exfor_2015}.
Unfortunately, it is sometimes perceived as impractical to extract relevant data.
A possible reason may be that popular programming languages, such as Java, Python, Matlab and R, do not natively support the extraction of information stored in the EXFOR format.

Several development activities were concerned with easier access to the EXFOR library and data in the EXFOR format.
The EXFOR web retrieval system~\citep{zerkin_experimental_2018} enables the user to search data based on a comprehensive catalogue of criteria and retrieve the result in a variety of formats including JSON and XML, which are better suited for numerical processing than the original EXFOR format.
Quick plotting in the web browser is also supported.
Another initiative to facilitate working with the data in EXFOR was the development of the \textit{X4toC4} code~\citep{cullen_program_2001}, which allows the translation of the EXFOR format to a tabular format named \textit{C4}.
The \textit{x4i} code by D.A. Brown and a pure Python 3 implementation, \textit{x4i3}~\citep{fedynitch_afedynitchx4i3_2020}, provide an interface to the EXFOR library.

Our philosophy on the management of experimental data in the EXFOR library was to convert the data from the EXFOR format to the JSON format first.
The JSON format enables the storage of hierarchically organized data and supports strings and numbers.
Thanks to these features, the logical structure of the EXFOR format and all data stored therein can be perfectly preserved.
Therefore, considering only the format itself, there is no advantage of the JSON over the EXFOR format to store the data of nuclear experiments as they are equivalent in terms of logical structure and information.
However, the JSON format is a standard format commonly employed for the exchange of information between web servers and web browsers.
Consequently, it is supported by a wide variety of programming languages---in contrast to the EXFOR format.

We used the R package \textit{exforParser}~\citep{gschnabel-exforParser} to translate the EXFOR to the JSON format.
As this parser does not modify the logical organization of the data, it can probably cope robustly with future updates and extensions of the EXFOR format, keeping the need to update the parser itself to a minimum.

Besides the convenient extraction and manipulation of experimental data facilitated by the JSON format, it is also important to have flexible capabilities of search to locate relevant data.
At the time of writing, more than 23k experiments (=number of entries) and more than 150k reactions (=number of subentries) are recorded in the EXFOR library.

Our approach to address this need was to create a document-oriented database containing the JSON objects with EXFOR data.
The fundamental concept of a document-oriented database is a document, a bundle of information belonging to one conceived entity.
As an example, a document could store information about a person, such as their hair color, eye color and age.
For comparison, a SQL database is based on the notion of tables where information about an entity may be distributed over several tables and one table usually contains partial information of several entities.

One of the popular document-oriented databases is \textit{MongoDB}, which enables the direct storage of \textit{JSON} objects.
It provides a flexible API (=application interface) to search, retrieve and modify data, which can be used from a large variety of programming languages and from the command line.

A script to convert the experimental data in the EXFOR format to the JSON format and feed it into a MongoDB database has been made available in the repository \textit{createExforDb} on GitHub~\citep{gschnabel-createExforDb}.
A Dockerfile automating the full installation of the MongoDB EXFOR database including a manual is also available online~\citep{gschnabel-compEXFOR-docker}.
More details about the translation of the EXFOR library to a MongoDB database are available in~\citep{schnabel_computational_2020}.

\subsubsection{Searching experimental data}

After this general discussion on how we prepared the experimental data in the EXFOR library, we provide a brief walk-through of searching and retrieving data to exemplify the convenient handling of experimental data.

An example search query using the MongoDB query language is shown in \cref{lst:mongodb_searchquery}.
\begin{lstlisting}[language=json,firstnumber=1,float,
caption=Example of a search query in the MongoDB query language. This query was used in the pipeline to retrieve neutron-induced cross sections of $^{56}$Fe., captionpos=b, label=lst:mongodb_searchquery]
{ "$and": [
  { "BIB.REACTION": { 
    "$regex":
        "\\(26-FE-56\\(N,[^)]+\\)[^,]*,,SIG\\)",
    "$options": ""
  }},
  { "DATA.TABLE.DATA": { "$exists": true }},
  { "DATA.TABLE.EN": { "$exists": true }}
]}
\end{lstlisting}
%
Search queries are expressed as JSON objects.
Search criteria are defined for specific fields, such as the \textit{REACTION} field in the example.
In the hierarchical representation of EXFOR, some fields are subfields of other fields, e.g., the \textit{REACTION} field is a subfield of the \textit{BIB} (=bibliography) section.
Fields deeper in the hierarchy are addressed by adding their parent field as a prefix, such as \textit{BIB.REACTION} in the example.

A large variety of elementary search filters are available and two are used in the example.
The filter in line four is a regular expression, e.g.,~\citep{noauthor_regular_nodate}.
A regular expression defines a search \textit{pattern} that is matched against text.
Its syntax allows the specification of sophisticated patterns.
For instance, the part \colorbox{gray!10}{N,[\^)]+} matches a sequence starting with the string \colorbox{gray!10}{N,} and then matches a number of one or several arbitrary characters that are not a closing bracket.
This example provides just a small glimpse at the vast possibilities to define search patterns using regular expressions.
Of course, an end-user may not be bothered with this syntax and common queries can and we think should be packaged in functions or scripts, also executable on the command line.

The second criterion employed two times in line seven and eight in the example allows to match only documents that contain a specific field.
In the case of the example, the document must contain the fields \textit{DATA.TABLE.EN} and \textit{DATA.TABLE.DATA}, i.e., the data table must contain a column with incident energies and the measured quantity.
Several other elementary search filters exist.
For example, fields containing numbers can be matched based on comparison operators, such as \textit{larger-than} and \textit{smaller-than}, and fields containing arrays can be matched based on whether they contain a specific range of values.
This search filter can be useful if one is interested in experimental data measured in a specific range of incident energies.

Such elementary search filters can be combined using the logical operators \textit{and} and \textit{or}.
The listing shows the application of the \textit{and} operator specifying that all elementary search filters must match at the same time.
More complex combinations of logical comparison operators are possible.

The outlined features of the query language hint at the flexibility to formulate search queries.
As MongoDB and its query language can cope with EXFOR data in its original nested structure and the heterogeneous appearance of data in different subentries, e.g., fields missing in one subentry and present in another one, we think it is a good and future-proof way to handle information in the EXFOR library.
The conversion of the EXFOR library into a MongoDB database can be done in dozens of minutes on a contemporary laptop, hence such a database can be easily kept in sync with the continuously evolving EXFOR library.
As an important remark, other implementations of document-oriented databases than MongoDB exist which provide equivalent functionality.
The EXFOR library has also been translated to a CouchDB database~\citep{gschnabel-exfor-couchdb-docker}, which comes with a more permissive license.

\subsubsection{Retrieving experimental data}

The MongoDB API is accessible from a variety of programming languages and also from the command line.
As the pipeline and all supporting packages are written in the programming language R, we sketch with short code snippets how experimental data can be retrieved in this language.
Access to a MongoDB database in R is possible via the \textit{mongolite} package~\citep{Rmongolite} and a small package \textit{MongoEXFOR}~\citep{gschnabel-MongoEXFOR} was developed providing---in our opinion---slightly more convenient access to EXFOR data.
The latter package is a thin wrapper around the former package.
The following code examples demonstrate the access to information in EXFOR via the interface of the \textit{MongoEXFOR} package.

There are two options for the retrieval of experimental data given in EXFOR subentries:
\begin{enumerate}
\item
    A result can be returned as a collection of \textit{nested lists}, each of these containing the information of one EXFOR subentry.
    A nested list in R is roughly equivalent to a JSON object.
    A code skeleton to obtain a result in this format is given by
    \begin{lstlisting}[language=R,firstnumber=1,linewidth=8cm,
caption=Code skeleton to retrieve data from the MongoDB EXFOR database, captionpos=n,numbers=none]
it = exforIterator(queryStr)
while (!is.null((curSub <- it$getNext()))) {
  # code to prepare the data
  # in the nested list in curSub
}
\end{lstlisting}
\item
    A result can be returned as a \textit{data frame}, which is an Excel-sheet like table, summarizing the information of several subentries.
    A code skeleton to obtain the data frame is given by
    \begin{lstlisting}[language=R,numbers=none,linewidth=8cm,
    caption=Code skeleton to retrieve data from the MongoDB EXFOR database, captionpos=n]
result = db$find(queryStr, {
     # code to preprocess the data
     # in each retrieved entry
    })
\end{lstlisting}
\end{enumerate}
Both options require the argument \textit{queryStr} in the function call, a string that contains the query specification passed to the MongoDB API.
An example of a query string was given in \cref{lst:mongodb_searchquery}.

The advantage of the first option being that the full information of a specific subentry is obtained which can then be processed before it enters the evaluation.
We used this option in the rule-based correction of uncertainty information in the evaluation pipeline.
In this task setting, the presence and absence of certain columns in the data table of an EXFOR subentry determines the sequence of actions taken.
For instance, the existence of a total uncertainty component is assessed via the R expression
\begin{lstlisting}[language=R,captionpos=n,numbers=none]
    is.null(curSub$DATA$TABLE$`ERR-T`)
\end{lstlisting}
and if present, the consistency with statistical and systematic uncertainty components assessed.
This expression also demonstrates the convenience of access to any information in an EXFOR subentry by specifying the chain of field names separated by a dollar sign to traverse the hierarchical structure of the subentry.
The backticks enclosing the last field name are necessary to avoid the interpretation as a subtraction operation.

The second option is to summarize the information of several subentries in a \textit{data frame}, a data structure akin to a table in Excel.
We use the example code snippet in \cref{lst:example_retrieval} to discuss the features of this retrieval option.
\begin{lstlisting}[language=R,caption=Retrieval of EXFOR data from several subentries in the form of a data frame.,captionpos=b,
label=lst:example_retrieval,float]
result = db$find(queryStr,{
  exforid = ID
  reac = BIB$REACTION
  detector_raw = nullToNA(BIB$DETECTOR)
  detector = str_match(detector_raw,
                       '\\(([^)]+)\\)')[2]
  energies = DATA$TABLE$EN
  measurement = DATA$TABLE$DATA
  list(ID=exforid, REAC=reac, DET=detector, 
       EN=energies, XS=measurement)
})
\end{lstlisting}

The \textit{find} function evaluates the code between lines two and nine enclosed by curly braces in the context of each matching subentry in the MongoDB EXFOR database.
Differently stated, variables such as \textit{BIB\$REACTION} in line three contain the \textit{REACTION} string of the subentry currently processed.
Any valid R code can be executed in the context of the subentry, hence anything from basic matrix operations to advanced statistical algorithms is possible.
As an aside, the ability to execute arbitrary R code in the context of a subentry also enables the implementation of a correction system.
R code can be stored along the data in the subentry and when the data is accessed, the R code can be executed first to correct and transform the experimental data before retrieval.
In this way, corrections recommended by experts or those used in previous evaluations can be automatically applied before the data is further processed for an evaluation.

The code in line four and five is an example of slightly more elaborate preprocessing.
First, in line four it is attempted to retrieve the character string in the \textit{DETECTOR} field.
Due to the variable being wrapped by the function \textit{nullToNA}, a missing \textit{DETECTOR} field leads to the assignment of the special value NA (=not available).
Functionality as provided by the function \textit{nullToNA} is very important to deal robustly with missing values because many fields in EXFOR subentries are not guaranteed to be present but we may still want to take their information into account if they are. 

To explain line five, we have to take a brief excursion to the conversion of the EXFOR library.
As the field content of subentries has been preserved from the translation of the EXFOR library to a MongoDB database, so-called EXFOR codes and free text descriptions may occur together.
EXFOR codes are abbreviations put between brackets referencing, e.g., detector types and institutions.
For instance, the string \textit{(SCIN)} in the field \textit{DETECTOR} indicates the utilization of a scintiliation detector in an experiment.
Fields often contain free text that provides additional information to humans, which is often the explanation of the EXFOR code.
For instance, the \textit{DETECTOR} field of an EXFOR subentry may contain \textit{(NAICR) NA-I SPECTROMETER}.

Returning to the discussion of the code snippet, the function \textit{str\_match} from the \textit{stringr} package~\citep{Rstringr} is used in line five to extract only the EXFOR code from the \textit{DETECTOR} field.
Irrespective of the fact that the regular expression passed as second argument looks daunting to the uninitiated, this line of code demonstrates that a concise instruction is enough to retrieve only the EXFOR code from a field---an important requirement to convert EXFOR data into formats understood by processing codes.
As an aside, the effortless manipulation of strings and data in high-level languages such as Python and R was one of the reasons why the logical structure and field contents of the EXFOR library were preserved in the translation to a MongoDB database.
Derived formats and databases suited for numerical processing can be easily produced using the powerful and flexible facilities of these languages.

Finally, we discuss the meaning of line eight and nine in the code snippet.
After the retrieval of data and their assignment to variables, the content of the resulting data frame needs to be specified.
The last statement \textit{list(...)} in the code block enclosed by curly braces defines the columns, their name, and content of the resulting data frame.
An excerpt from the resulting data frame using the code snippet is illustrated in \cref{tbl:example_retrieval_result_excerpt}.
\begin{table}[ht]
\centering
\begin{tabular}{lllrr}
  \hline
ID & REAC & DET & EN & XS \\ 
  \hline
10022010 & (26-FE-56(N,P) \ldots) & NAICR & 14.60 & 113.00 \\ 
  10031005 & (26-FE-56(N,P) \ldots) & NAICR & 14.80 & 109.00 \\ 
  10037004 & (26-FE-56(N,EL) \ldots) & SCIN & 5.05 & 2129.00 \\ 
  10037004 & (26-FE-56(N,EL) \ldots) & SCIN & 5.58 & 1916.00 \\ 
  10037005 & (26-FE-56(N,TOT) \ldots) & SCIN & 5.05 & 3607.00 \\ 
   \hline
\end{tabular}
\caption{Excerpt from the resulting data frame obtained by the code snippet in \cref{lst:example_retrieval}.
The full EXFOR reaction specification is truncated in column \textit{REAC} for better display.}
\label{tbl:example_retrieval_result_excerpt}
\end{table}
%

The intention of this walk-through was to give the reader an idea of the retrieval options and a glimpse on the ease and flexibility to retrieve data.
The functionality of this IT building block puts an evaluator in the position to retrieve relevant data by writing a few lines of code, which may be considered an essential requirement for the implementation of an evaluation pipeline.
Furthermore, the convenient extraction of features associated with experimental data, such as the detector type in the example, in combination with the possibility of large scale Bayesian computation provided by the \textit{nucdataBaynet} package discussed in \cref{subsec:handling_info_bayesian_context} enables the consideration of correlations between experiments due to detectors, monitors, institutions, etc. on an unprecedented scale in the field of nuclear data evaluation.
It also opens the door to conveniently retrieve and prepare the data in a format required by advanced statistical algorithms and machine learning methods.

\subsection{Mapping of model predictions to EXFOR}
\label{subsec:mapping_model_predictions}

Nuclear model codes, such as TALYS~\citep{koning_modern_2012} and EMPIRE~\citep{herman_empire_2007,herman_empire-32_2013}, are capable to predict a large variety of observables.
These codes write their predictions in the form of numeric tables to files.
Sometimes the results of model codes cannot be directly compared to the experimental data in the EXFOR library.
This is the case for measurements of sums and ratios of reaction cross sections.
However, the availability of predictions for all measured quantities is essential in evaluations where a nuclear model is employed as fitting function.
Furthermore, because we want to enable evaluations with a large amount of experimental data with the vision to take into account all available and trustworthy experimental data of the EXFOR library, the mapping of model predictions to experimental data needs to be solved in a general and extensible manner.

In this section, we sketch the design and functionality of the R package \textit{talysExforMapping}~\citep{gschnabel-talysExforMapping} for this purpose.
This package addresses the need to translate predictions of the nuclear model code TALYS to experimental data in EXFOR.
Even though the package is tailored to TALYS, its general philosophy and basic structure can be adopted for other codes as well, such as EMPIRE.

It was clear from the beginning that the large variety of observables stored in the EXFOR library in combination with the complex EXFOR format and different options within the format to represent the same information makes it unfeasible to come up with a comprehensive solution to the mapping of predictions to EXFOR data in one shot.
Consequently, the implementation of mapping functionality as an IT block was guided by the following desiderata:
\begin{enumerate}
    \item Support for the mapping of new observable types can be incrementally added whenever the need arises.
    \item As it is not clear in advance what information may be relevant to determine the appropriate mapping for an observable, functions to implement a mapping function should be provided with all information in an EXFOR subentry.
\end{enumerate}

Desideratum (1) was addressed by employing the concept of handler modules, as it was already done in \cref{subsec:handling_info_bayesian_context} for the management of information in the Bayesian context.
Each handler is responsible for the mapping of predictions to a specific type of observable.
Support for the mapping to a new observable type is then established by the implementation of a new handler.
Desideratum (2) was addressed by passing the complete information of an EXFOR subentry to the functions of a handler module.

A handler should be able to tell which predictions it needs from a model code so that they can be mapped to the types of observables in EXFOR subentries.
Observables of interest in the subentries are referenced in a data frame called \textit{expDt} in the following.
The structure of this data frame with some example content is illustrated in \cref{tbl:expDt}.
\begin{table}[ht]
\centering
\begin{tabular}{rlrlrr}
  \hline
IDX & EXPID & DIDX & REAC & L1 & DATA \\ 
  \hline
  1 & 10529004 & 207 & (26-FE-56(N,INL) \ldots) & 1.00 & 1043.80 \\ 
    2 & 10529004 & 208 & (26-FE-56(N,INL) \ldots) & 1.00 & 967.60 \\ 
    3 & 10529004 & 209 & (26-FE-56(N,INL) \ldots) & 1.00 & 731.20 \\ 
    4 & 10529004 & 210 & (26-FE-56(N,INL) \ldots) & 1.01 & 568.50 \\
   \hline
\end{tabular}
\caption{Structure of data frame \textit{expDt} with example content.
Columns \textit{L2} and \textit{L3} are omitted for better display.}
\label{tbl:expDt}
\end{table}
%
The essential colums are \textit{IDX}, \textit{EXPID} and \textit{DIDX}.
The purpose of column \textit{IDX} is to impose an order among the observables.
The column \textit{EXPID} indicates the EXFOR ID of the subentry where the measured value of the observable is recorded.
The column \textit{DIDX} indicates the row in the data table of the subentry containing the measured value.
The information in the other columns is redundant because it can always be retrieved from the referenced subentry.
The meaning of \textit{L1}, \textit{L2} and \textit{L3} is dependent on the EXFOR reaction string in \textit{REAC}.
For instance, the code \textit{SIG} at the end of the reaction string denotes cross sections and then the column \textit{L1} contains the associated incident energy and the other columns \textit{L2} and \textit{L3} are ignored.

The predictions required to map to the observables in \textit{expDt} are referenced in a data frame \textit{needsDt}.
Such a data frame with example content is shown in \cref{tbl:needsDt}.
\begin{table}[ht]
\centering
\begin{tabular}{cccclr}
  \hline
IDX & PROJ & ELEM & MASS & REAC & L1 \\ 
  \hline
  1 & N & FE &  56 & CS/EL & 1.50 \\ 
    2 & N & FE &  56 & CS/EL & 2.00 \\ 
    3 & N & FE &  56 & CS/REAC/100000/TOT & 1.00 \\ 
    4 & N & FE &  56 & CS/REAC/100000/TOT & 1.00 \\ 
    5 & N & FE &  56 & CS/REAC/100000/TOT & 1.00 \\ 
   \hline
\end{tabular}
\caption{Structure of data frame with example content to define required TALYS predictions.
The column names \textit{PROJ} and \textit{ELEM} are abbreviations of the full column names \textit{PROJECTILE} and \textit{ELEMENT} for better display.
Columns \textit{L2} and \textit{L3} are omitted for the same reason.}
\label{tbl:needsDt}
\end{table}
The column \textit{REAC} combines the definition of the reaction and measured quantity.
The prefix \textit{CS/} indicates a cross section and other prefixes exist which are associated with angular distributions, energy spectra and double differential cross sections.
Afterwards follows the specification of the reaction.
The naming convention mirrors closely the names of the TALYS output files.
The meaning of the columns \textit{L1}, \textit{L2}, \textit{L3} depends on the quantity.
In the case of a cross section, \textit{L1} indicates the incident energy and \textit{L2} and \textit{L3} are ignored.

As a side note, a situation with different energy meshes in \textit{expDt} and \textit{needsDt} as in this example is usual in nuclear data evaluation.
A basic requirement for handlers is therefore the ability to map the predictions of the computational mesh to the energies of the experiments.
The handlers currently implemented in \textit{talysExforMapping} rely on linear interpolation.

After the introduction of the data frame \textit{expDt} referencing observables of interest in a list of EXFOR subentries \textit{subents} and the data frame \textit{needsDt} referencing the required model predictions, we discuss how the handlers are employed.
There is one manager module that keeps track of the handlers and delegates the work to the appropriate handlers.
It provides the following interface:
\begin{itemize}
    \item \textit{canMap(subents)}: 
    Informs for each subentry in the list \textit{subents} whether a handler is available that knows how to translate predictions of TALYS to the observables in that subentry.
    \item \textit{needs(expDt, subents)}:
    Returns a data frame in the form of \cref{tbl:needsDt} with the predictions required from TALYS to enable the mapping to the observables stated in data frame \textit{expDt}.
    \item \textit{map(expDt, needsDt, subents)}:
    Translates the predictions of TALYS given in the data frame \textit{needsDt} to the observables referenced in \textit{expDt}.
    The data frame \textit{needsDt} is of the form as exemplified in \cref{tbl:needsDt} and must possess the additional column \textit{V1} with the associated predictions.
\end{itemize}
The individual handlers provide the same functions as the manager module but with different parameters as they work on the level of a single subentry.
The manager module also provides other functions for registering handlers and querying information about them.
From the point of view of mathematics, it is often useful to have the translation of the predictions in \textit{needsDt} to the observables in \textit{expDt} as a mapping matrix $\totmap$.
This functionality is also offered by the manager module.

In summary, due to the extensible architecture of the package \textit{talysExforMapping} and the fact that it deals with the unaltered structure and information of the EXFOR subentry, it can probably organically co-evolve with the original EXFOR library.

\subsection{Parallelization of nuclear model invocations}
\label{subsec:parallelization_nuclear_model}
Modern nuclear data evaluation relying on a nuclear model code is computationally expensive.
Evaluation methods based on Monte Carlo sampling, such as UMC-G~\citep{smith_unified_2008,capote_investigation_2008}, UMC-B~\citep{capote_new_2012},
BFMC~\citep{bauge_evaluation_2007,bauge_evaluation_2011} and
BMC~\citep{koning_bayesian_2015} require a large number of model code invocations with varied model parameter sets to achieve convergence.
Evaluation methods based on optimization, such as~\citep{helgesson_fitting_2017} employing the prior-aware LM algorithm and~\citep{kawano_covariance_1997} using the Kalman filter, require several invocations of the nuclear model to compute a gradient or Jacobian matrix.
A conceptual overview of some of these methods is presented in~\citep{capote_nuclear_2010,herman_covariance_data_2011}.
A quantitative comparison of some of the methods, also including a study of the impact of model defects, has been performed in~\citep{helgesson_comparison_2018}.

In the prototype of the pipeline, the joint optimization of 147 TALYS parameters requires 148 model calculations in each iteration and each model invocation takes tens of minutes.
These parameters are related to the optical model parameters for neutrons, protons, deuterons, tritons, He-3 and alpha particles and level density parameters of the target and daughter nuclides.
However, from the perspective of the pipeline, the underlying physics does not matter as long as the parameters are continuous quantities.
Any nuclear model code exposing parameters via input files or an API could be used.
Considering the computational demand, the parallel execution of nuclear model calculations on a multi-processor machine or computer cluster becomes a necessity.

The implementation of the functionality of this IT building block was guided by the following desiderata:
\begin{itemize}
    \item The implementation of parallelization functionality should be such that it is easy to set it up on a wide variety of different multi-core and cluster environments.
    \item The development of functionality to be parallelized should be possible on the local machine, i.e., functions and packages can be run both locally and remotely on the multi-core or cluster environment in the same way, which is good for testing and quality assurance.
    \item There should be as few restrictions as possible on what can be parallelized, i.e., everything from basic text processing to the invocation of a nuclear models code should be doable.
\end{itemize}
Several solutions exist addressing parallelization of calculations but none of those we are aware of fully addressed all points of our list.
Scheduling engines, such as the Sun Grid Engine (SGE)~\citep{Gentzsch2001SunGE}, manage the queuing of calculation requests and distribution of calculations to compute nodes.
Based on our experience, not all computer clusters at scientific institutions have a scheduling engine installed.
Of course, if it is available it can be relied upon.
Software frameworks such as MapReduce~\citep{dean_mapreduce_2008} and Spark~\citep{zaharia_apache_2016} enable the processing of structured data, text and images but we felt they are not the right tool to invoke nuclear model codes that produce several thousand output files in each run which can be immediately discarded after execution.
Also the programming language R provides support for the parallel execution of R functions on chunks of input data by means of the \textit{parallel} package part of the R language~\citep{Rlanguage}.
It uses sockets for communication between compute nodes.
The functionality of this package addresses many points of our list.
Unfortunately, the use of sockets is not possible in some scenarios due to firewall rules.

Considering our use case, which is running several instances of the same model code in parallel, the design of our solution was based on the following assumptions:
\begin{itemize}
    \item Individual calculation requests are bulky taking a few minutes or longer so that latency due to network communication or other overhead is negligible.
    \item Individual calculations are independent of each other so they do not need to exchange information between each other. 
    \item All computational units use Linux, provide SSH access and have access to a shared file system.
\end{itemize}
The last assumption is a reflection of our experience with scientific cluster environments.

Several R packages were implemented building on each other to enable the parallelization of computations.
The packages \textit{interactiveSSH}~\citep{gschnabel-interactiveSSH} and \textit{rsyncFacility}~\citep{gschnabel-rsyncFacility} enable the execution of commands in a bash prompt on a remote machine and the transfer of files via SSH.
Building on top of these packages, the package \textit{remoteFunctionSSH}~\citep{gschnabel-remoteFunctionSSH} enables the remote execution of functions written in R in a transparent way.
The package \textit{clusterSSH}~\citep{gschnabel-clusterSSH} builds on top of \textit{remoteFunctionSSH} to enable the parallel execution of R functions on chunks of data on a multi-core machine or computer cluster.
Finally, the package \textit{clusterTALYS}~\citep{gschnabel-clusterTALYS} enables the parallelization of TALYS calculations and retrieval of the output produced.
These packages rely only on the remote execution of commands via SSH and the transfer of files to exchange data and code between the local machine and the remote machine, hence they enable parallelization as long as assumption three above holds.
As a side note, in the meantime an R package named \textit{ssh}~\citep{Rssh} has become available, which appears to be a more robust implementation of the functionality provided by \textit{interactiveSSH}.

We are not going to discuss the technicalities of these packages here.
Instead, we first highlight the functionality of the package \textit{remoteFunctionSSH} as it was a catalyst to implement the required functionality for the parallelization of TALYS model calculations.
Afterwards, we provide a concise example to run TALYS calculations in parallel using the \textit{clusterTALYS} package.

An example of the usage of the \textit{remoteFunctionSSH} package is shown in \cref{lst:remFunExample}.
\begin{lstlisting}[language=R,caption=Definition of a function and its execution on a remote machine,captionpos=b,float,label=lst:remFunExample]
remHnd = initSSH(...)
fun = function(x,y) { x + y}
remfun = remHnd$createRemoteFunction(fun)
remote_result = remfun(2,4)
local_result = fun(2,4)
\end{lstlisting}
The setup of the remote connection is done in line one.
The parameters of the function \textit{initSSH} omitted for conciseness in the code snippet are the host address, user credentials and a directory on the local machine and on the remote machine, which are used for data transfer.
Line two defines a function named \textit{fun} in the local R session.
This function is transferred to the remote machine and associated with the local variable \textit{remfun}.
The invocation of the function \textit{remfun} in line four executes the function \textit{fun} remotely and returns the result as usual in the local R session.
For comparison, the same function is also executed on the local machine in line five.

The important feature is that from the viewpoint of the user, there is no difference between invoking \textit{fun} and its remote counterpart \textit{remfun} because both take exactly the same parameters and return the same result.
The ability to design and test functions in the local R session and then run them on a remote machine in the same way as if they were local functions is tremendously helpful for the rapid development of functionality.
This example featured a simple sum of two numbers but any valid R code can be executed remotely, implementing anything from simple arithmetical operations to the invocation of a scheduling engine, such as SGE, or the invocation of a nuclear model code.

Finally we highlight the package \textit{clusterTALYS} which enables the parallel execution of TALYS on a computer cluster.
A basic example of the usage of this package is shown in \cref{lst:clusterTALYS}.
\begin{lstlisting}[language=R,caption=Setup and execution of several TALYS calculations in parallel including the retrieval of predictions.,captionpos=b,float,label=lst:clusterTALYS]
talysClust = initClusterTALYS(...)
paramset = list(projectile = "n",
                  element = "Fe",
                  mass = 56,
                  energy = c(1,2,3)
                )
outSpec = data.table(REAC = "CS/TOT",
                      L1 = c(1,2,3,4),
                      L2 = 0, L3 = 0
                    )
paramsetList = replicate(3, paramset,
                         simplify=FALSE)
runObj = talysClust$run(paramsetList, outSpec,
                        pollTime=30)
result = talysClust$result(runObj) 
\end{lstlisting}
An object is initialized in line one which bundles the parallelization functionality.
The parameters of function \textit{initClusterTalys} are omitted for conciseness.
In line two a list with TALYS input keywords and their values is defined.
Noteworthy, the energy keyword takes an array of numbers, which are written to a separate file as required by TALYS during the setup of a TALYS calculation.
Similar enhancements were introduced for energy dependent model parameters.
Specifications, such as \textit{"rvadjust(10) n" = 1.05} with the energy in bracket are translated to additional files with energy/value tables as required by TALYS.
Line seven defines a data frame with the predictions of interest, which is of the form as exemplified in \cref{tbl:needsDt} without the columns \textit{PROJ}, \textit{ELEM} and \textit{MASS}. 
The parameter set in \textit{paramSet} is stored three times in a list \textit{paramsetList}.
TALYS is then launched in line 13 for each parameter set in \textit{paramsetList}.
The result retrieved in line 15 is a list of data frames where each data frame corresponds to a calculation with a specific parameter set in \textit{paramsetList}.

This example demonstrated the seamless integration of TALYS into the programming language and statistical environment of R, regarding both parallel execution of the model code and retrieval of generated predictions.
The convenient interaction with R is very beneficial for the creation of an evaluation pipeline, as it facilitates the formulation of code logic at a high-level of abstraction.
Furthermore, advanced methods of statistics, machine learning and optimization available in R thanks to efforts of a large number of contributors can be easily applied on TALYS.
Because R code can be run not only on the command line but also in an interactive computer programming environment (REPL), such as RStudio~\citep{Rstudio}, on-demand and hands-on data analysis are also facilitated.

\section{THE NUCLEAR DATA EVALUATION PIPELINE}
\label{sec:evaluation_pipeline}
The mathematical and IT building blocks discussed in the two previous sections are a means to an end, which is to provide best estimates and uncertainty information of nuclear observables on the basis of information from nuclear models and experiments.
Generally, the evaluation process can be divided into the following stages:
\begin{enumerate}
    \item \textbf{Collection of experimental data.} 
    Data suitable to be used in the evaluation have to be identified.
    Ideally, they are already added to the EXFOR library where they can be retrieved via a targeted search.
    For more recent measurements, they have to be extracted from the original publication or by direct communication with the authors of the experiment.
    \item \textbf{Correction of experimental data.}
    Experimental data need to be renormalized according to the latest evaluations of monitor reactions.
    Outliers and inconsistencies in the experimental data have to be identified and appropriate measures taken.
    These measures may be the exclusion of datasets, an in-depth analysis of original experimental setups to figure out the problem with the data or an automatic correction relying on rules of thumb or algorithms of statistics and machine learning.
    Noteworthy, these measures are not mutually exclusive.
    For instance, data can be manually corrected by human experts as good as possible and remaining inconsistencies addressed by an automated statistical treatment.
    \item \textbf{Fitting a model to the corrected experimental data.}
    The model may be a nuclear physics model or a simple mathematical function, such as a piecewise linear function.
    To have the advantages of both, which are good extrapolation in the case of the nuclear physics model and great flexibility to mimic the trends in the data for the mathematical model, both models may be combined to a meta-model, which is exactly the idea of model defects, e.g., \citep{leeb_consistent_2008,neudecker_impact_2013,schnabel_differential_2016,helgesson_fitting_2017}.
    Predictions of this meta-model are the sum of the nuclear physics model and the flexible mathematical model.
    Another popular approach is to replace the exact nuclear physics model by a MVN distribution, as done in UMC-G~\citep{smith_unified_2008,capote_investigation_2008}.
    A mean vector and covariance matrix are estimated from a sample of model predictions with varied parameter sets.
    Nowadays, there seems to be a convergence to Bayesian methods for model fitting.
    They enable the inclusion of prior knowledge and resulting estimates come always with associated uncertainties.
    \item \textbf{Generation of data files.}
    The final results of the evaluation need to be cast into formats understandable by verification, processing and simulation codes.
    The format adopted worldwide by major institutions linked to nuclear physics is the ENDF-6 format~\citep{trkov_endf-6_2018}.
    An emerging alternative to the ENDF-6 format is the GNDS format~\citep{noauthor_specifications_2020}.
\end{enumerate}

Some evaluations were informed by feedback from integral experiments.
The above mentioned sequence of stages remains valid if one broadens the scope of the term experimental data to include integral experiments as well.
From the point of Bayesian statistics, the distinction between so-called differential data and integral experimental data is irrelevant.
Any information can be included in the Bayesian fitting procedure as long as the link to the other pieces of information, e.g., other experimental data or model parameters, is mathematically well defined.

Having said that, at present integral and differential data are considered differently for technical reasons.
In the case of integral simulations, complicated processing beyond the simple ACE format is often needed, such as self-shielding calculations, macroscopic group cross sections tabulated as a function of specific parameters and buckling factors.
Such processing often relies on approximations for computational tractability or for compatibility with the limitations of specific file formats.
Ignoring these effects in a Bayesian evaluation using integral data bears the risk of adjusting differential data too much in an attempt to compensate for the unaccounted errors due to approximations during processing.
First attempts to account for such effects are presented in~\citep{sjostrand_monte_2019,siefman_data_2020} using the MLO approach.
The long runtime of neutron transport codes for certain benchmark experiments is another obstacle to the inclusion of integral data in Bayesian methods for nuclear data evaluation.

In the following we present our specific implementation of a nuclear data evaluation pipeline passing through the four stages.
Our implementation has to be regarded as a prototype without the pretense that selected experimental data, employed corrections or statistical algorithms are the ultimate answer to everything.

For instance, a shortcoming in the current version of the pipeline is that experimental data retrieved from the EXFOR library are not renormalized according to the latest evaluations of the employed monitor reaction.
More generally, a variety of errors may affect measurement results, discussed in detail in~\citep{smith_experimental_2012}.
Ideally, an in-depth analysis of the experiments included in an evaluation is performed to determine the error sources and their potential impact.
However, this has not been done for this example evaluation.
Another issue is that a crucial assumption of the statistical model (in the nuclear physics sense) implemented in TALYS does not hold for $^{56}$Fe.
Because the proton number in $^{56}$Fe is close to a magic number, the level density is very low and effects of its structure can be observed up to about 6 MeV.
The statistical model assumes these effects to be completely averaged out which is only the case if the average level width is much larger than the average level spacing.
No-model fits, e.g., based on Gaussian processes~\citep{iwamoto_generation_2020}, may provide better fits to the experimental data in this energy range.

Yet, we do believe that the coherent assembly of innovative approaches, such as the correction of experimental data based on both rules and an statistical algorithm, and the specification of Gaussian process priors on energy-dependent model parameters can serve as inspiration for future nuclear data evaluation efforts.
As the pipeline is modularly designed and can be run at the push of a button, modifications in any step can be gradually implemented and their impact systematically evaluated.
For instance, another model for fitting than the nuclear model code TALYS could be used in the pipeline.
In the evaluation of neutron standards~\citep{carlson_international_2009} one prefers a flexible mathematical fitting function, such as a piecewise linear model, to avoid potential distortions of the result due to the rigidity of a nuclear model.

\begin{table*}[t]
 \def\colwidth{10.5cm}
 \centering
    \begin{tabular}{|c|l|}
        \hline
        \textbf{Step} & \textbf{Description} \\
        \hline
        1 & \multicolumn{1}{p{\colwidth}|}{Retrieval of relevant experimental data} \\
        2 & \multicolumn{1}{p{\colwidth}|}{Generation of predictions based on a reference calculation} \\
        3 & \multicolumn{1}{p{\colwidth}|}{Rule-based correction of experimental uncertainties} \\
        4 & \multicolumn{1}{p{\colwidth}|}{Correction of systematic experimental uncertainties using MLO} \\
        5 & \multicolumn{1}{p{\colwidth}|}{Evaluation of the Jacobian associated with the reference calculation} \\
        6 & \multicolumn{1}{p{\colwidth}|}{Setup of Gaussian processes for energy-dependent model parameters} \\
        7 & \multicolumn{1}{p{\colwidth}|}{Optimization of TALYS parameters using the LM algorithm} \\
        8 & \multicolumn{1}{p{\colwidth}|}{Calculation of a MVN approximation of the posterior pdf} \\
        9 & \multicolumn{1}{p{\colwidth}|}{Generation of representative random files} \\
        \hline
    \end{tabular}
    \caption{Sequence of steps in the current implementation of the nuclear data evaluation pipeline}
    \label{tbl:pipeline_steps}
\end{table*}

 In the following sections we go through each step, see~\cref{tbl:pipeline_steps}, and discuss how the functionality of the mathematical building blocks provided by the \textit{nucdataBaynet} package and the IT building blocks are applied in the pipeline.
 Each step is associated with a script in~\citep{gschnabel-eval-fe56} whose filename contains the step index as a prefix followed by a description roughly matching the one in the table.
 The correspondence to the four general stages is as follows:
 Step one corresponds to stage one.
 Steps two to six are associated with stage two.
 Please note that for this attribution, we also considered the prior on model parameters as a type of experimental data. 
 Steps seven and eight are implementing the stage three.
 Finally, step nine corresponds to stage four.

 
 The pipeline is open-source and has been published on GitHub~\citep{gschnabel-eval-fe56} along with a manual with guidance on installation, usage, and the structure of the pipeline.

 

\subsection{Retrieval of experimental data}
\label{subsec:pipeline_step1}

In the first step, relevant experimental data were retrieved from the MongoDB EXFOR database, see~\cref{subsec:handling_experimental_data} for the discussion of this IT building block.
To that end, we formulated a search query implementing the following criteria:
\begin{enumerate}
\item The \textit{REACTION} field must match the regular expression: \\ \colorbox{gray!10}{\\(26-FE-56\\(N,[\^)]+\\)[\^,]*,,SIG\\)}, i.e., neutron induced cross sections of $^{56}$. 
\item Both the \textit{EN} and \textit{DATA} column must be present in the data table of the EXFOR subentry.
\end{enumerate}
We are aware that the regular expression is certainly not the most user-friendly way to define reactions to be searched.
Furthermore, reaction strings exist that denote relevant reaction data for $^{56}$Fe but are not matched by this regular expression.
The comprehensive and completely automated retrieval of relevant data is difficult due to several ways to denote the same observable in an EXFOR reaction string.
This issue is planned to be addressed in WPEC SG50 dealing with the creation of a machine-readable, curated experimental database. 

The obtained subentries compatible with the search criteria above are then only considered in the evaluation if:
\begin{enumerate}
    \item They have valid uncertainty information, i.e., they contain any of the fields \textit{DATA-ERR}, \textit{ERR-S}, and \textit{ERR-T}.
    \item It is known how to produce a TALYS prediction that can be related to the data in the subentry.
\end{enumerate}
Of the subentries fulfilling these criteria, only data points between 2 and 30\,MeV are selected for the evaluation.

The validity of uncertainty information is checked by a function in a package named \textit{exforUncertainty}~\citep{gschnabel-exforUncertainty}.
More functionality of this package is discussed in step three dealing with the rule-based correction of uncertainties.
The determination whether TALYS predictions can be related to the measurements in the subentries is done via the package \textit{talysExforMapping}, see~\cref{subsec:mapping_model_predictions} for an explanation of this IT building block.

\subsection{Generation of predictions based on a reference calculation}
\label{subsec:reference_calculation_step2}

A calculation of TALYS with default parameters is performed to obtain reference predictions of the observables present in the experimental data.
These predictions are the basis to translate relative experimental uncertainties to absolute ones to avoid the underestimation of evaluated quantities---a manifestation of the phenomenon known as Peelle's pertinent puzzle (PPP)~\citep{peelle1987peelle} in the field of nuclear data.
Excerpts of the informally distributed memorandum by Peelle can be found in~\citep{chiba_suggested_1991}. 

TALYS parameters are adjusted by making use of the \textit{adjust} keywords in the TALYS input file, e.g., \textit{rvadjust n 1.05}.
An effective parameter $\tilde{p}$ is therefore the multiplication factor applied to the default value $p_\textrm{ref}$ used by TALYS:
\begin{equation}
    p = \tilde{p} \times p_\textrm{def}
\end{equation}

Furthermore, each parameter $\tilde{p}$ was further constrained to the range $[0.5, 1.5]$ by applying the transformation
\begin{equation}
\tilde{p} = x_0 + \delta \times 
\frac{1 - \exp(-k(t-x_0))}
{1 + \exp(-k(t-x_0))}
\label{eq:talysparamtrafo}
\end{equation}
and setting $x_0=1, \delta=0.5, k=4$.
Making use of the transformed parameters $t$, unconstrained optimization algorithms can be applied and the restriction on the range still enforced.
For the sake of conciseness, any mention of model parameters in the following discussion including other subsections always refers to the transformed ones.
As a side note, a variety of other transformations is conceivable that may be better in certain situations.
A convenient feature of the employed transformation is that a transformed parameter value equals approximately the original parameter value if $\tilde{p}$ is not too far away from one.

    
    Instead of a reference calculation to avoid PPP as done in the pipeline,
    there are other options to remove or at least alleviate the undesired bias.
    For instance, a smooth curve can be fit to the experimental data and used as the reference in the conversion.
    Statistical fluctuations are 'washed out` and consequently the bias reduced.
    This option may be better if the shape of the reference prediction was not well aligned with the experimental data.
    
    Another approach is to apply an iterative fitting strategy where the conversion of absolute to relative uncertainties in the current iteration is based on the intermediate fit of the previous iteration.
    This approach was suggested by Chiba and Smith~\citep{chiba_suggested_1991} and followed for the evaluation of neutron standards where two iterations of the Generalized Least Squares method were undertaken.
    Relative uncertainties were converted to absolute ones based on the result of the first iteration.
    Also in~\citep{helgesson_fitting_2017} the importance of adjusting the covariance matrix due to the presence of relative uncertainties during the iterative fitting procedure was emphasized.
    This latter paper employs the prior-aware Levenberg-Marquardt algorithm outlined in \cref{subsec:LM_with_prior} which relies on an additional \textit{damping term} compared to the approach of Chiba and Smith.
    A simulation study presented in~\citep[chap. 6.3]{schnabel_large_2015} gives indication that a damping term is indeed necessary to ensure convergence.
    

\subsection{Rule-based correction of experimental uncertainties}
\label{subsec:pipeline_step3}
The information on uncertainties of experiments in EXFOR is often incomplete.
The idea put forward in~\citep{helgesson_combining_2017} is to introduce or modify uncertainty components based on pragmatic and reasonable rules.
In that work, for instance, datasets were discarded if the specified uncertainties were deemed too low; and an additional uncertainty component was introduced if only either a statistical or systematic uncertainty component provided.
The rule-based approach enables the construction of complete uncertainty information.

The rules implemented in the pipeline differ from those of~\citep{helgesson_combining_2017} and should be considered as placeholders to be substituted by better ones in the future.
They have been bundled in the R package \textit{exforUncertainty}~\citep{gschnabel-exforUncertainty}.
We briefly outline the approach adopted for the rule-based construction of uncertainty information here.
As a side note, the pipeline stores in datatables the datasets retrieved and the datasets used in the evaluation.
The generation of a report about discarded experiments during pipeline execution is therefore possible, even though not implemented in the current version of the pipeline prototype.

First, an attempt is made to extract systematic and statistical uncertainties.
Statistical uncertainties are retrieved from either the \textit{ERR-S} or the \textit{DATA-ERR} column in the subentry.
A situation of both fields being present is not accepted and the subentry discarded from the evaluation.
If the statistical error is given in percent in the subentry, it is converted to an absolute uncertainty using the predictions of the reference calculation obtained in the previous step of the pipeline.

Systematic uncertainties are retrieved from columns named \textit{ERR-0}, \textit{ERR-1}, etc.
Uncertainties in one column are assumed to be fully correlated, and uncertainties of different columns uncorrelated.
The information whether a systematic uncertainty component is absolute or relative is preserved.
Moreover, if an uncertainty component declared as absolute is a constant proportion of the experimental cross section in the \textit{DATA} column, it is converted to a relative one.

Once the available information on statistical and systematic uncertainties is extracted from a subentry, potentially missing information is reconstructed according to the following rules:

\begin{itemize}
\item
  If only the total uncertainty component in column \textit{ERR-T} is available, both the statistical and systematic uncertainty component are assumed to be 10\%.
  In effect, the information in \textit{ERR-T} is completely ignored.
\item
  If either the statistical or systematic uncertainty component is missing and the total uncertainty is available in the column \textit{ERR-T}, the missing uncertainty component will be reconstructed.
  Let us denote by $\delta_\textrm{miss}$ the missing uncertainty component, by $\delta_\textrm{avail}$ the available one, and by $\delta_\textrm{tot}$ the total uncertainty component.
  The missing uncertainty component is then reconstructed as
  \begin{equation}
      \delta_\textit{miss} = 
      \sqrt{\delta_\textit{tot}^2 - \delta_\textrm{avail}^2} \,.
  \end{equation}
  A negative difference under the root indicates an inconsistency in the uncertainty information of the subentry and is penalized by assuming 10\% for the missing uncertainty component.
\item
  If statistical, systematic and total uncertainty component are provided simultaneously but
  the quantity
  \begin{equation}
      \Delta^2 := \delta_\textit{tot}^2 - \delta_\textit{sys}^2 - \delta_\textit{stat}^2
  \end{equation}
  does not vanish, the absolute value of $\Delta$ is added as both a statistical and a  systematic uncertainty component.
\end{itemize}

The amount of uncertainty introduced in response to missing or inconsistent uncertainties is ad-hoc because something had to be done.
A better approach would be to perform a statistical analysis of uncertainties provided in similar experiments to determine a reasonable value for the missing uncertainty.
The automated filling in of missing data by relying on available data of similar cases is known as \textit{imputation} in statistics. 
Yet another approach would be to base the introduction of additional uncertainty on templates that contain ranges for typical uncertainty components, e.g., detector calibration uncertainty.
Such templates have been recently created for certain experiments~\citep{neudecker_template_2018,neudecker_applyingtemplates_2020,neudecker_templates_2021}.
We also think that the validation of uncertainties using the \textit{physial uncertainty bounds} method presented in~\citep{neudecker_validating_2020} contains relevant ideas in this context.

\subsection{Correction of experimental uncertainties based on statistics}
\label{subsec:pipeline_step4_MLO_correction_experiment}
Nuclear experiments are complex endeavors and it may happen that not all sources of uncertainties are always recognized.
A discussion of this possibility in different experimental situations and effects in evaluations can be found in~\citep{capote_unrecognized_2020}.
The presence of unrecognized (or at least unreported) uncertainties can be detected if several datasets from different experiments are inconsistent with each other.
The difference between the measured values of different datasets is then too large to be explainable by the reported uncertainties.

The most desirable resolution of such a situation would be to figure out the forgotten uncertainty and take it into account in the evaluation.
This manual approach can be time-consuming and the timely success or even being successful at all is not guaranteed.
Still, in evaluations automated to a large extent such as those of the TENDL library~\citep{koning_tendl_2019}, something must be done about inconsistent data.
We feel that due to the complexity of proper uncertainty quantification, the inclusion of an automatic treatment of inconsistent data is also valuable in manual evaluations of specific isotopes.
Even if one does not accept the automatic correction of uncertainties in such evaluations, an algorithm to automatically correct uncertainties can be used as a safeguard informing the evaluator of missing uncertainties and their potential magnitude.

The solution implemented in the pipeline is to introduce additional systematic uncertainties and adjust their value based on the maximization of the marginal likelihood (MLO) in \cref{eq:logmarlike}.
Another possibility could be the exclusion of inconsistent experimental datasets using an outlier detection algorithm.
We think algorithms for outlier detection and uncertainty correction can be beneficially applied side-by-side in future nuclear data evaluation pipelines.


The experimental covariance matrix $\covexp$ is the result of the rule-based approach implemented as the previous step of the pipeline.
As an important reminder, we never explicitly build up this matrix but always use the representation in \cref{subsec:info_representation}, i.e.,
\begin{equation}
    \covexp = \syserrmap \mat{U}_\textrm{sys} \syserrmap^T + \covstaterr \,. 
\end{equation}
The covariance matrix $\mat{U}_\textrm{sys}$ is associated with systematic errors and the covariance matrix $\covstaterr$ with statistical errors.

To automatically determine the additional systematic uncertainties in the case of inconsistent data, we augment the experimental covariance matrix in the following way:
\begin{equation}
    \covexp^\textrm{aug}(\vec{\lambda}) = \covexp + \mat{S}_\textrm{extra} \mat{U}_\textrm{extra} \mat{S}_\textrm{extra}^T \,. 
    \label{eq:augexpcov}
\end{equation}
The matrix $\mat{U}_\textrm{extra}$ is diagonal and each element $\lambda_i^2$ represents the square of an additional normalization uncertainty associated with a specific dataset.
The augmented experimental covariance matrix $\covexp^\textrm{aug}$ is therefore a function of $\vec{\lambda}$, which are regarded as fitting parameters in MLO.
The element $e_{ij}$ in the i-th row and j-th column of $\mat{S}_\textrm{extra}$ is non-zero if the i-th element of $\obsvec$ belongs to the experimental dataset associated with the extra normalization uncertainty $\lambda_j$.
In this case, the element $e_{ij}$ is given by the reference cross section obtained in the second step associated with the i-th element of the experimental measurement vector $\obsvec$.
This construction means that each $\lambda_i$ is a relative and fully correlated normalization error for a specific dataset.

The experimental covariance matrix expresses how far measured values are expected to be away from the unknown true values.
In an evaluation and for the application of the MLO approach, we always need a model for what the truth may look like.
For the determination of additional experimental systematic uncertainties, we do not want to rely on a nuclear physics model.
If such a model were deficient, the adjusted values of the additional uncertainties would not reflect unrecognized errors in the experiment but rather deficiencies of the nuclear model.
Therefore, instead of using a physics model, we use a piecewise linear model as defined in \cref{eq:pwlinint}.
As long as the mesh is dense enough, it can very well approximate any continuous function.

We recall the important statement that a covariance matrix of the values at the mesh points $\mat{P}$ in combination with linear interpolation induces a Gaussian process, i.e., the covariances between values at arbitrary energies can be evaluated according to \cref{eq:covpwlinint}.
The mapping of the values at the energy mesh of the piecewise linear function to the experimental energies is established by the matrix $\modparmap$ whose construction was described in \cref{eq:Sgplinint}. 
The full observational covariance matrix $\covobs$ can be written as
\begin{equation}
    \covobs = \covexp^\textrm{aug} + \modparmap \mat{P} \modparmap^T \,.
    \label{eq:covobs_step4}
\end{equation}
Again, as explained in \cref{subsec:info_representation}, this matrix is never explicitly stored but always dealt with in the representation
\begin{equation}
    \covobs = \totmap \covtotunobs \totmap^T + \covstaterr
    \label{eq:covobs_repr_step_four}
\end{equation}
and being of the following form in the current step:
\begin{equation}
    \covtotunobs = \begin{pmatrix}
    \mat{U}_\textrm{sys} & \mat{0} & \mat{0} \\
    \mat{0} & \mat{U}_\textrm{extra} & \mat{0} \\
    \mat{0} & \mat{0} & \mat{P}
    \end{pmatrix}
    \;\;\textrm{and}\;\;
    \totmap = \begin{pmatrix}
    \syserrmap \\
    \mat{S}_\textrm{extra} \\
    \modparmap
    \end{pmatrix}^T \,.
    \label{eq:covmat_repr_step_four}
\end{equation}

There is a strong prior expectation that cross section curves are smooth above a certain incident energy.
In such regions, we believe that non-expert humans are mostly guided by the shape of an envisaged cross section curve underlying the experimental data to make a judgement of whether datasets are consistent or not.
In other words, we believe that neither the absolute values of this envisaged cross section curve nor the steepness of its slope matter.
What matters is its smoothness, i.e., the rate of change of the slope.
If too rapid variations in the cross section curve are required to be within experimental uncertainties close to the points of the various datasets, the datasets would be deemed inconsistent.

One way forward to formalize this notion about reasonable shapes is to softly constrain the second derivative of the piecewise linear function to fit the cross section curve.
We briefly sketch the essential ideas here.
\Cref{apx:prior_2nd_derivative} contains a more rigorous discussion in terms of equations.
In order to construct the corresponding prior covariance matrix $\mat{P}$ we consider the formula for numeric differentiation,
\begin{equation}
    \frac{\textrm{d}f(E_i)}{\textrm{d}E} \approx 
    \frac{f(E_{i+1}) - f(E_i)}{E_{i+1}-E_i}
    \label{eq:numericdiff} \,,
\end{equation}
where the energies $E_i$ and $E_{i+1}$ are neighbors on the reduced energy grid as introduced in the context of \cref{eq:basic_linearinterpol} for the sparsification of the Gaussian process.
The numerical differentiation rule defined in \cref{eq:numericdiff} is a linear operator and can thus be rewritten as a matrix multiplication.
Let $\vec{f}$ denote the vector of function values at the energies of the mesh and $\vec{\Delta}$ the vector of first derivatives at the energies $E_1, E_2, \dots, E_{M-1}$ of the mesh.
We can then construct a matrix $\mat{S}_\textrm{diff}$ such that $\vec{\Delta} \approx \mat{S}_\textrm{diff} \vec{f}$.
Importantly, according to the prescription in \cref{eq:numericdiff}, we cannot evaluate the first derivative of the function at the highest energy of the mesh $E_M$.

The vector of second derivatives $\vec{\Delta}^2$ can be calculated by another application of $\mat{S}_\textrm{diff}$.
With a slight abuse of notation because the dimension of $\mat{S}_\textrm{diff}$ is reduced in the second application, this can be written as $\vec{\Delta}^2 = \mat{S}_\textrm{diff} \mat{S}_\textrm{diff} \vec{f}$.
Due to $\mat{S}_\textrm{diff}$ having one more column than row, this matrix cannot be inverted.
This is a reflection of the fact that integration---the inverse operation to differentiation---is only unique up to an integration constant.
To make $\mat{S}_\textrm{diff}$ invertible, we augment it by inserting one row of the form $(1,0,\dots,0)$ at the top so that $\mat{S}_\textrm{diff}^\textrm{aug} \mat{S}_\textrm{diff}^\textrm{aug}\vec{f}$ yields an augmented vector $\vec{\Delta}_\textrm{aug}^2 = (f(E_1), \Delta_1, \vec{\Delta}^2)^T$.
The first element is the function value $f(E_1)$, the second element $\Delta_1$ the first derivative at $f(E_1)$, and each subsequent element at position $i$ represents the second derivative at $f(E_{i-2})$.
Because $\mat{S}_\textrm{diff}^\textrm{aug}$ is a square matrix and invertible, the original parameters can be expressed by the transformed ones, i.e., $\vec{f} = \mat{R} \vec{\Delta}^2$ with $\mat{R}=(\mat{S}_\textrm{diff}^\textrm{aug}\mat{S}_\textrm{diff}^\textrm{aug})^{-1}$.
The sandwich formula yields the covariance matrix $\mat{P}$ associated with the original parameters as a function of the covariance matrix $\tilde{\mat{P}}$ of the transformed parameters, i.e., $\mat{P} = \mat{R} \tilde{\mat{P}} \mat{R}^T$. 

For simplicity and computational efficiency, we assume $\tilde{\mat{P}}$ to be diagonal.
We set the prior variance for $f(E_1)$ and its first derivative to large values to express our ignorance about the cross section value and the slope.
The determination of the variances associated with second derivatives were guided by the cross section at the peak of the curve.
These assignments were, however, somewhat ad-hoc and a more principled approach for their determination merits investigation in the future.

Due to the fact that the prior covariance matrix of the piecewise linear model is diagonal in the space of second derivatives and there is a bijective and linear mapping between the function values at the locations of the mesh and the values of the second derivaties at the locations of the mesh, it is computationally more efficient to use the second derivatives as model parameters.
Therefore, in \cref{eq:covmat_repr_step_four}, the block $\mat{P}$ in the covariance matrix $\covtotunobs$ needs to be substituted by $\tilde{\mat{P}}$ and the block $\modparmap$ in the Jacobian $\modparmap$ needs to be replaced by $
    \modparmap
    \left( 
        \mat{S}_\textrm{diff}^\textrm{aug}
        \mat{S}_\textrm{diff}^\textrm{aug}
    \right)^{-1}
$.

At this stage we have discussed the construction of all parts of the observational covariance matrix $\covobs$ specified in \cref{eq:covobs_step4}.
With the additional specifications $\muobs=\vec{0}$ and $\vec{x}_\textrm{obs}=\obsvec$, we can evaluate the marginal likelihood given in \cref{eq:logmarlike} and adjust the additional normalization uncertainties in $\vec{\lambda}$ (which determine $\mat{U}_\textrm{extra})$ to maximize it.
Importantly, the systematic errors of datasets in different reaction channels were assumed to be independent.
This assumption enabled the application of MLO to each reaction channel in isolation.

The important result of this step is the term in $\covobs$ that corresponds to the augmented experimental covariance matrix $\covexp^\textrm{aug}$, see \cref{eq:augexpcov}, based on the adjusted additional systematic uncertainties $\vec{\lambda}$\,.

Finally, we want to hint at possible improvements of this approach for uncertainty correction.
Instead of manually designing a Gaussian process using an intuition of how humans would identify problematic data, one could also infer the properties of the Gaussian process in a data-driven way from a collection of reaction systems.
This idea was explored in~\citep{schnabel_first_2018} with the aim to determine a model defect prior for reaction channels, which can then be used in an evaluation procedure.
The Gaussian approach construction described there could also be a way forward to extend the idea of MLO uncertainty correction to energy ranges where resonant-like structures start to emerge. 

Another idea hinting at a potential improvement was presented in the context of validation~\citep{forrest_statistical_2007,forrest_detailed_2008}.
There, various quantities derived of cross section curves, such as the peak cross section, were regarded as functions of quantities associated with properties of the nuclei, such as the asymmetry.
These functions were fitted globally to a collection of experimental data and the empirical distribution of deviations from these derived quantities constructed.
Problematic reaction systems were then identified as outliers with respect to this empirical distribution.
We think that such an approach could be turned around and not only used for validation but also for uncertainty correction in the future.

\subsection{Evaluation of the Jacobian associated with the reference calculation}
\label{subsec:eval_jacob_step5}
At this stage of the pipeline, we have a fully specified experimental covariance matrix $\covexp^\textrm{aug}$ required for the statistical inference carried out in subsequent steps.
This statistical inference comprises the determination of hyperparameters associated with the Gaussian processes for energy-dependent model parameters in the next step and the adjustment of TALYS model parameters thereafter.
Both steps mentioned rely on the Jacobian matrix $\mat{J}$ of the nuclear model whose elements are given by $J_{ij}=\partial \sigma_i / \partial p_j$ with $\sigma_i$ being the prediction associated with the i-th element of the experimental measurement vector $\obsvec$ and $p_j$ being the j-th TALYS model parameter.
More details on how the Jacobian matrix is going to be used in these steps are explained then.

The numerical evaluation of elements in the Jacobian matrix is based on the formula
\begin{equation}
    \frac{\partial \sigma_i}{\partial p_j} =
    \frac{\sigma_i(\modparrefvec + h p_i \vec{\Delta}_i) - \sigma_i(\modparvec)}
    {h p_i} \,.
    \label{eq:numdiff_findiff}
\end{equation}
The only non-zero element of the vector $\vec{\Delta}_i$ at index $i$ is one.
Using this specification, the step size $h$ of the finite difference approximation denotes the magnitude of the perturbation as proportion of the reference parameter value.
Too large values of $h$ cause the numerical derivative to be a worse approximation to the exact derivative.
Too small values of $h$ on the other hand make the result of the numerical evaluation more prone to round-off errors as the ratio of two very small numbers is taken.
The value that has been employed in the pipeline is $h=0.01$, i.e., a one percent perturbation of each TALYS parameter.
Depending on the parameter, a one percent perturbation may not be the optimal choice.
Lower values can in principle produce more precise gradient approximations unless associated differences of function values are dominated by numerical noise.
This tradeoff has to be taken into account and an optimal choice needs to be informed by the specifics of the simulation. 
For TALYS and the parameters considered, we found a one percent perturbation to work reasonably well.

Please note a more accurate formula for the numeric differentiation would be
\begin{equation}
    \frac{\partial \sigma_i}{\partial p_j} =
    \frac{\sigma_i(\modparrefvec + h \vec{\Delta}_i) - \sigma_i(\modparvec - h \vec{\Delta}_i)}
    {2 h p_i}
\end{equation}
because the lowest-order term of the error cancels out.
However, this formula requires approximately twice as many evaluations of the physics model as the formula in \cref{eq:numdiff_findiff}.
Because about thousand model parameters were considered adjustable in our example evaluation of neutron-induced cross sections of $^{56}$Fe , we relied on \cref{eq:numdiff_findiff} to cut down on computational resources needed.
Despite this reduction of the computational cost, we still had to rely on a scientific cluster with about 150 processing cores to evaluate the Jacobian matrix, which took a few hours.
The parallelization of model calculations was done with the help of the \textit{clusterTALYS} package discussed as an IT building block in~\cref{subsec:parallelization_nuclear_model}.

\subsection{Setup of Gaussian processes for energy-dependent model parameters}
\label{subsec:setup_GP_step6}

Nuclear models are often if not most of the times deficient.
This statement probably sounds rather rude to the ears of theoretical physicists who work hard to develop models with more accurate descriptions of physics processes.
We acknowledge these efforts and deeply respect these people pushing the limits, yet we are uncompromising concerning our statement.

A model is potentially deficient if we have reason to believe that its predictions cannot perfectly reproduce the truth.
Therefore even a very good model with an excellent capability of extrapolation and predictions within a few percent of trustworthy data is potentially deficient if the differences between predictions and data cannot be explained away by experimental uncertainties.


Work to incorporate model defects into nuclear data evaluations was presented in~\citep{pigni_uncertainty_2003,leeb_covariances_2005,leeb_consistent_2008,neudecker_impact_2013} and later the connection to Gaussian processes identified and formalized~\citep{schnabel_large_2015,schnabel_differential_2016}.
Some work dealing with the notion of model defects in a more general setting outside nuclear data predates the efforts in the field of nuclear data, such as~\citep{blight_bayesian_1975,kennedy_bayesian_2001}.


We implemented an idea proposed and explored in~\citep{helgesson_treating_2018} to use the capability of TALYS to allow variations of parameters as function of incident energy to simulate the treatment of model defects.
Effectively, this measure injects more flexibility into the nuclear physics model and it becomes capable of capturing a broader range of shapes.
The advantage of this approach being that physics constraints are automatically taken care of by the nuclear model code TALYS.

The idea is the following.
Nuclear model parameters, such as the real radius $r_v$ of the optical model, by default assumed constant for a specific isotope, are regarded as functions of incident energy, e.g., $r_v(E)$.
These functions are assumed to be piecewise linear functions, i.e., the function is characterized by the function values on a mesh of energies.
Intermediate values are obtained by linear interpolation.
Therefore, each energy-dependent model parameter in the model parameter vector $\modparvec$ is represented as a collection of model parameters, i.e., the values of the parameter at a specific incident energy.
In other words, we expand a parameter $p_k$ in $\vec{p} = (\dots, p_k, \dots)^T)$ assumed to be energy-dependent as $\vec{p} = (\dots, p_k(E_1), p_k(E_2), \dots, p_k(E_M), \dots)$ with $E_1, E_2, \dots, E_M$ being the energies of the predefined mesh.
As a reminder, the model parameters $p_k$ here are the transformed ones defined in \cref{eq:talysparamtrafo}. 

The covariance of parameter values at different mesh points are defined by a squared exponential covariance function,
\begin{multline}
    \kappa_k(E_i,E_j) = \Cov{p_k(E_i)}{p_k(E_j)} = \\
    \delta_k^2 \exp\left(
        -\frac{1}{2\lambda_k^2} (E_i - E_j)^2
    \right) + \eta_{ij} \,.
    \label{eq:covfun_step_six}
\end{multline}
The amplitude $\delta_k$ represents the a priori standard deviation of the parameter values and is considered to be independent of incident energy.
The length-scale $\lambda_k$ is a measure of similarity of a parameter at nearby incident energies.
The so-called nugget parameter $\eta_{ij}$ is assigned a small numerical value, e.g., $10^{-4}$, if $E_i$ and $E_j$ are equal and zero otherwise.
It is introduced for numerical stability and its impact on the result is usually negligible.
These parameters of the covariance function are called hyperparameters.
In the evaluation of $^{56}$Fe in the pipeline, we allow the amplitude and length-scale  to be different for each energy-dependent model parameter.

At this point it becomes clear that energy-dependent TALYS parameters are not conceptually different from energy-independent ones from an evaluation point of view.
They are represented by values in the parameter vector $\modparvec$ and the covariance functions are employed to compute the associated blocks in the prior covariance matrix $\mat{P}$.

The determination of the hyperparameters $\delta_k$ and $\lambda_k$ is done via marginal likelihood optimization---the same approach that has already been employed in the determination of additional systematic uncertainties of experiments in step four discussed in \cref{subsec:pipeline_step4_MLO_correction_experiment}.
Therefore, we deal again with the observational covariance matrix in the representation of \cref{eq:covobs_repr_step_four}.
The important difference is that $\mat{P}$ is now the prior covariance matrix associated with nuclear model parameters and not the function values of a piecewise linear function at the mesh points as it was in step four.
Analogously, each element of the Jacobian matrix $\modparmap$ represents now the derivative of a prediction $\sigma_i$ with respect to a model parameter $p_j$, i.e., $(\modparmap)_{ij}=\partial \sigma_i / \partial p_j$.
This Jacobian matrix has already been computed in step five.

In more detail the MLO approach is carried out in the following way:
We adjust the hyperparameters in the covariance functions in \cref{eq:covfun_step_six} which alter the prior covariance matrix $\mat{P}$ in \cref{eq:covmat_repr_step_four} and consequently the value of the logarithmized marginal likelihood in \cref{eq:logmarlike}.
To determine the values of the hyperparameters to maximize the marginal likelihood, we use the L-BFGS algorithm~\citep{byrd_limited_1995}.
The package \textit{nucdataBaynet} supports these operations.

\subsection{Optimization of TALYS parameters using the LM algorithm}
\label{subsec:optim_talys_step7}

At this stage of the pipeline, all covariance matrices required for an evaluation are determined.
The experimental covariance matrix was constructed using a rule-based approach in step three and additional relative systematic uncertainties introduced in step four to account for unrecognized or unreported systematic uncertainties. The final experimental covariance matrix was denoted as $\covexp^\textrm{aug}$.
In the previous step---step six---we determined the hyperparameters of the Gaussian processes imposed on energy-dependent model parameters, which were necessary to compute the associated elements of the prior covariance matrix $\mat{P}$.

The posterior distribution
\begin{equation}
    \pi(\modparvec \,|\, \obsvec) \propto
    \ell(\obsvec \,|\, \mathcal{M}(\modparvec)) 
    \pi(\modparvec)
\end{equation}
with
\begin{align}
    \ell(\obsvec \,|\, \mathcal{M}(\modparvec)) &=
    \mathcal{N}(\obsvec \,|\, \mathcal{M}(\modparvec); \covexp^\textrm{aug}) \\
    \pi(\modparvec) &= \mathcal{N}(\modparvec \,|\, \vec{p}_0; \mat{P})
\end{align}
is therefore fully determined.
The predictions of the nuclear model TALYS are denoted by $\mathcal{M}(\modparvec)$ and this model link is non-linear.

In order to locate the parameter vector $\vec{p}_\textrm{max}$ associated with the maximum of $\pi(\modparvec \,|\, \obsvec)$, we employed the modified Levenberg-Marquardt algorithm outlined in \cref{subsec:LM_with_prior}.

Importantly, parameters for adjustment were selected based on their sensitivity to experimental data.
Each column in the sensitivity matrix $\totmap$ reflects the impact of an adjustment of a specific parameter on the predicted observables.
A parameter was only taken into account for adjustment if the maximum of the values in the corresponding column $\totmap$ squared was larger than one.
This criterion reduced the number of parameters from roughly thousand to about 150 parameters.

As a final remark, an alternative formulation of the parameter exclusion criterion in terms of relative changes of cross sections induced by relative changes of model parameters may be easier to work with and to develop an intuition about it.

\subsection{Calculation of a MVN approximation of the posterior pdf}
\label{subsec:posterior_approx_step8}

After the localisation of the model parameter vector that corresponds to the maximum of the posterior distribution in the previous step, associated uncertainty information needs to be determined as well.
Due to the model link between parameters and observables being non-linear, neither the posterior distribution of parameters nor model predictions has to be multivariate normal.
To deal with a general distribution, Monte Carlo methods, such as importance sampling~\citep{mcbook_owen_2013} and Markov Chain Monte Carlo (MCMC)~\citep{brooks_handbook_2011}, can be used to obtain representative samples of it.
Alternatively, the shape of the posterior distribution can be approximated by an analytic expression.
We opted for the latter option and approximated the posterior distribution by a multivariate normal distribution as described in~\cref{subsec:taylor_logpdf}.
Even though this decision was born out of the necessity of computational tractability, we think it may be defensible on the following grounds:
\begin{itemize}
    \item Whenever experimental data strongly constrain the parameters, the model link is in good approximation linear within the ranges of parameter values allowed by the posterior uncertainties.
    Due to the assumption of a multivariate normal prior and likelihood, the posterior is then also in good approximation multivariate normal.
    \item The posterior distribution of model parameters weakly constrained by the experimental data is close to the prior distribution, which we specified as multivariate normal.
    \item Stronger deviations of the approximative posterior parameter distribution from the exact distribution have to be expected for intermediate cases.
    If the objective is not unbiased parameter estimation but uncertainty quantification on the observable side, which is typically the case in nuclear data evaluation, the distortion of the distribution on the observable side is not critical as long as posterior uncertainties are somewhat compatible with those associated with the exact distribution.
\end{itemize}
Regarding the second point, we note that inappropriate priors are always problematic in Bayesian statistics.
Here we are only concerned about how close the posterior approximation may be to the true posterior, regardless of whether the probablistic assumptions in the statistical model are a good reflection of reality.

The third point may be considered an optimistic intuition. Certainly mathematical toy scenarios can be constructed which invalidate this assumption.
It would be interesting and pertinent to study possible distortions in evaluations due to such an approximation in typical evaluation scenarios.
However, such studies are outside the scope of this paper.

As described in~\cref{subsec:taylor_logpdf}, the computation of the full Hessian matrix required to obtain the posterior parameter covariance matrix can be unpractical.
Its computation for about 150 model parameters requires more than ten thousand model invocations, each lasting tens of minutes.
To improve the time complexity from $N^2$ with $N$ being the number of model parameters, we exploited the approximation briefly described in \cref{subsec:taylor_logpdf} and in more detail in \cref{apx:comp_posterior_expectation} and \cref{apx:comp_posterior_covmat}.

Consequently, the second order Taylor approximation described in~\cref{subsec:taylor_logpdf} leads to the  following approximation of the posterior distribution,
\begin{equation}
    \pi(\modparvec \,|\, \obsvec) = \mathcal{N}(\modparvec \,|\, \vec{p}_\textrm{max}, \tilde{\mat{P}}_1) \,,
    \label{eq:postpdf_approx_48}
\end{equation}
with $\vec{p}_\textrm{max}$ being the parameter vector associated with the maximum of the posterior distribution and $\tilde{\mat{P}}_1$ an approximate posterior covariance matrix constructed according to the prescription detailed in \cref{apx:comp_posterior_covmat}.

\subsection{Generation of representative random files}

Two options are popular to perform uncertainty quantification in nuclear applications.
One option is to propagate the covariance matrices associated with observables in evaluated nuclear data files (ENDF files) to the quantities of interest in the specific application.
The other option followed, e.g., in the Total Monte Carlo approach~\citep{koning_towards_2008} is to rely on a representative sample of ENDF files where each file contains the values of observables obtained by a draw from the posterior distribution.
Each ENDF file is then employed in the simulation of the nuclear application and the resulting empirical distribution of the quantities of interest analyzed, e.g., by plotting histograms or by computing summary statistics such as the mean vector and standard deviation.

The pipeline produces both a collection of random files and a best ENDF file with covariance matrices.
For this purpose, a sample of parameter vectors is drawn from the approximative posterior distribution in \cref{eq:postpdf_approx_48} obtained in the previous step.
We stress here that the probability distribution constructed in step eight to draw samples and to extract a covariance matrix is different from the one used in the TMC approach applied for TENDL.

TALYS calculations are then performed for each of these parameter sets.
The TEFAL code converts the output of each calculation to a corresponding ENDF file.
A modified version of the TASMAN code loops over the outputs of the individual calculations and generates a `best' ENDF file that also includes covariance matrices.
Our modification of TASMAN~\citep{gschnabel-tasmanPatch} enabled the use of precalculated TALYS outputs instead of letting TASMAN generate a sample of parameter sets and invoke TALYS.
TALYS, TEFAL and TASMAN are part of the T6 evaluation system~\citep{koning_modern_2012}.

One may question the necessity of sampling for the construction of an ENDF file.
For the multivariate normal approximation of the posterior distribution, we argued that this approximation is justified if posterior uncertainties constrain the model parameters to a domain where the nuclear model is approximately linear.
In this case linear error propagation would be sufficient.
However, even if the model link between observables of the experimental data and the selected parameters used in the LM fitting is linear, the model link between predictions of reaction channels without experimental data and model parameters may not be linear.

Irrespective of this theoretical justification, the primary motivation to rely on this procedure for the generation of an ENDF file with covariance information was convenience.
In the view of a constrained time budget, we wanted to avoid the need to implement functions to augment an ENDF file with covariance information.
The TASMAN code can take care of this but derives the covariance matrix from a sample of model calculations.
We are optimistic that the emerging GNDS format~\citep{noauthor_specifications_2020} is going to improve the management of evaluated nuclear data.

\section{DISCUSSION OF PRACTICALITIES AND RESULTS}

Even though the focus of the paper is on the conceptual level, we want to present results to convey an idea of the working of the algorithms and features of the final evaluation.
We also discuss practicalities to demonstrate the feasibility of the employed algorithms.

For the evaluation of neutron-induced reactions of $^{56}$Fe, the search of experimental data in the EXFOR library according to the search specification in step one,  see~\cref{subsec:pipeline_step1}, yielded data for nine reaction channels, which are (n,2n), (n,a), (n,d), (n,el), (n,inl), (n,n+p), (n,p), (n,t) and (n,tot), amounting to a total number of 4333 data points.
Most data points (3895) are associated with (n,tot) and most of the remaining data points (324) are associated with (n,p). 

\begin{figure}[b]
    \centering
    \includegraphics{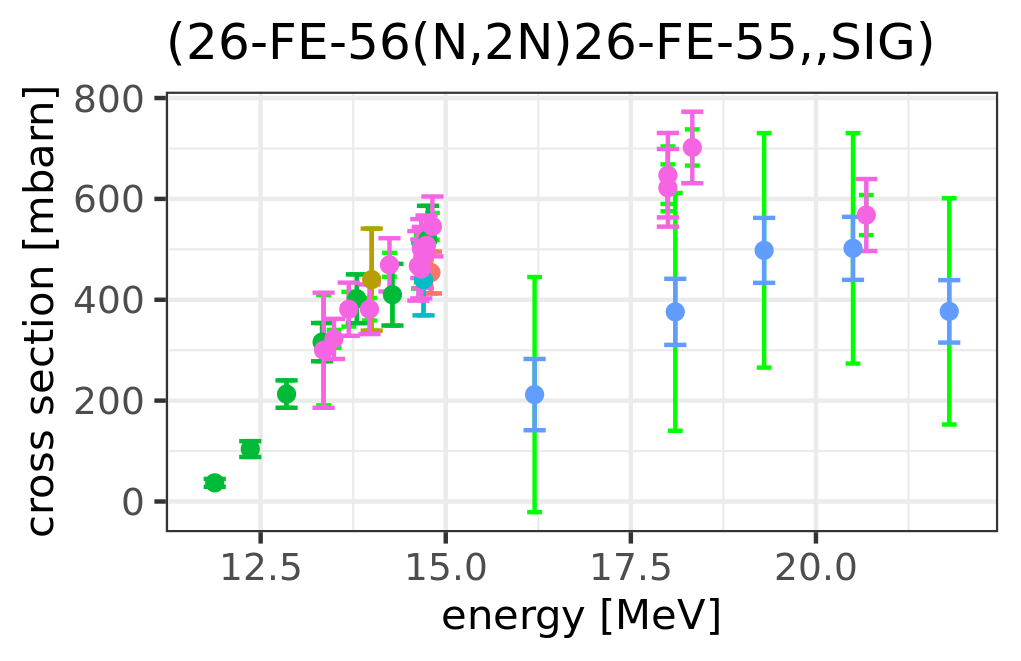}
    \caption{Example of the automatic correction of experimental systematic uncertainties based on the MLO approach in step four of the pipeline.
    The error bars attached to the data points are total uncertainties, i.e., including statistical and systematic uncertainties, after the rule-based correction.
    The extension of the error bars of the blue dataset indicated by the green extension line represents the increase of the total uncertainty determined by the MLO approach due to the inconsistency between datasets.
    }
    \label{fig:example_MLO_correction}
\end{figure}

The effect of the automatic addition of normalization uncertainties to experimental datasets via the MLO approach performed in step four (\cref{subsec:pipeline_step4_MLO_correction_experiment}) is shown at the example of the (n,2n) reaction channel in~\cref{fig:example_MLO_correction}.
To obtain this result, we fixed the dataset with EXFOR identfication number 23171003 to have a statistical uncertainty of 5\% and a normalization uncertainty of 0.5\%.
With this imposed assumption, the MLO approach introduces an additional normalization uncertainty to one dataset that also visually appears as an outlier.

\begin{figure}[t]
    \centering
    \includegraphics{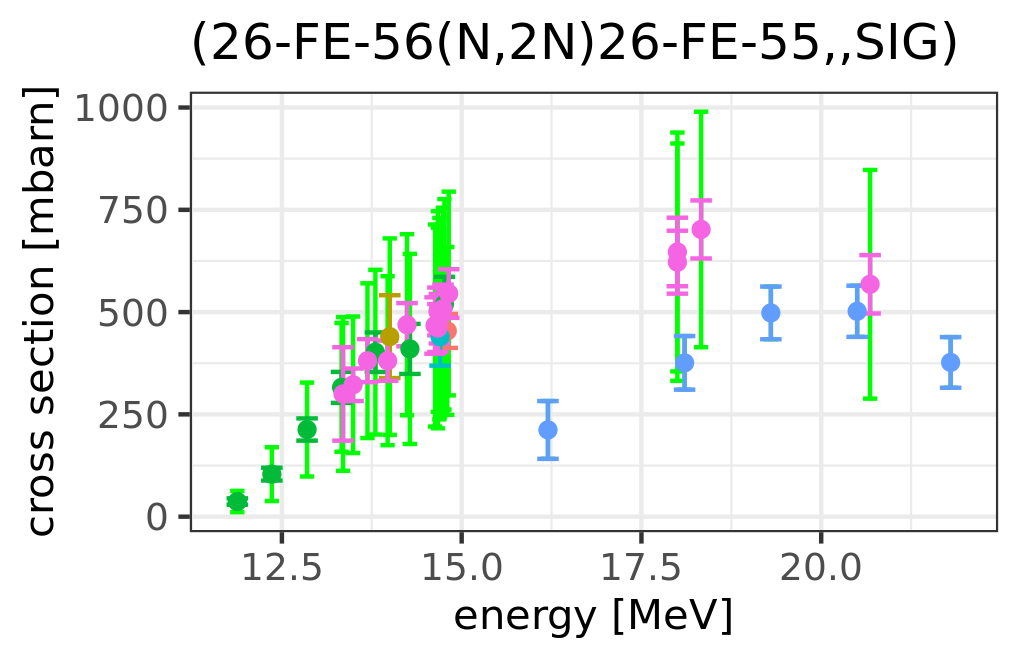}
    \caption{Example of the automatic correction of experimental systematic uncertainties based on the MLO approach if not fixing the uncertainty of dataset with EXFOR identification number 23171003.
    The extension of the error bars by the green lines represents the increase of the total uncertainty due to an additional normalization uncertainty determined by the MLO approach.
    }
    \label{fig:example_MLO_correction_hypothetical}
\end{figure}

Without this manual fix of the uncertainty of one dataset, the MLO approach would add the additional normalization uncertainties to the datasets we intuitively regarded as consistent, see \cref{fig:example_MLO_correction_hypothetical}.
Moreover, in the region around 15\,MeV very good and almost mono-energetic neutron sources are available so that typical uncertainties of activation measurements are between two and five percent.
Such a large increase of uncertainties is therefore not plausible from an experimental point of view.

This behavior of the MLO approach certainly needs further exploration and a more thorough study in order to go towards more automated evaluations and less human involvement.
Regarding the current implementation of the MLO approach, the more uncertainty information is already precisely provided by a human expert, the higher the chance that the MLO approach determines an assignment of extra uncertainties compatible with our expectations.
Therefore, the codification of choices imposed by a human and the use of statistical algorithms to correct data are not mutually exclusive.
In such cases, an important requirement of a statistical algorithm is that extra uncertainties should only be introduced if the data are indeed inconsistent, i.e., the human expert has missed sources of uncertainty.

Irrespective of the difference in the results with and without fixing the uncertainty of a specific dataset, the MLO approach seems in both cases to be able to correct uncertainties to establish consistency among datasets.
This statement holds in general if the parametrization of the covariance matrix is flexible enough.
Specifically in the current implementation of the pipeline, the adjustable parameters associated with the covariance matrix allow for the introduction of additional normalization uncertainties.
If the inconsistency in the data is not a mere normalization issue, the solution found by the MLO approach will therefore not be ideal.
Noteworthy, as demonstrated in~\citep{schnabel_fitting_2018}, the MLO approach is not limited to the adjustment of normalization uncertainties but any other type of uncertainty, e.g., systematic uncertainties partially correlated over energy, can be determined as well.
Despite this possibility, in the case of a severely deficient dataset, it is probably better to remove it, potentially automatically using an outlier detection algorithm.

After the determination of extra experimental normalization uncertainties, the Gaussian process hyperparameters of energy-dependent model parameters were determined via the MLO approach in step six, see \cref{subsec:setup_GP_step6}.
The optical model parameters $v_1, d_1, w_1, v_{so1}, w_{so1}, r_\mathcal{C}$ for neutron, proton, deuteron, tritium, Helium-3, and alpha are regarded as energy dependent.
For further explanation of these parameters, consult the manual accompanying the nuclear reaction code TALYS.
The energy mesh of each of these 36 energy-dependent parameters consisted of 16 energies uniformly spaced between 0 and 30\,MeV amounting to a total number of 576 effective parameters associated with energy-dependent parameters, which are considered for adjustment.
Each of the 36 energy-dependent parameters is associated with two Gaussian process hyperparameters: the amplitude and the length-scale.
These 72 hyperparameters are considered as prior knowledge and have to be determined before the adjustment of the 576 effective parameters.

In order to adjust these hyperparameters by maximizing the logarithmized marginal likelihood in \cref{eq:logmarlike}, the Jacobian matrix (=sensitivity matrix) between model parameters and observables measured in the experiments is required.
This Jacobian matrix was computed in step five, see~\cref{subsec:eval_jacob_step5}.
Step five is a computationally expensive step because 926 model calculations were carried out in this tentative evaluation, each lasting about 20 minutes, to obtain the Jacobian matrix of dimension $4333\times 925$ due to 4333 experimental data points and 925 model parameters.
On the computer cluster at the division of applied nuclear physics at Uppsala university, we performed about 150 calculations in parallel, hence the execution time was around two hours.

Once the Jacobian had been computed, the adjustment of the 72 Gaussian process hyperparameters associated with energy-dependent parameters to maximize the logarithmized marginal likelihood in \cref{eq:logmarlike} took a bit more than a minute (without parallelization) using the L-BFGS-B algorithm~\citep{byrd_limited_1995}.
We forced all length-scales $\lambda_k$ to be in the range between 2 and 50\,MeV and all amplitudes $\delta_k$ to be in the range between 0.1 and 0.5.
In other words, the uncertainty associated with energy-dependent parameters was bound to be between 10\% and 50\%.
We imposed the minimum on the length-scales to make the optimization process robust against potential kinks in the experimental data.
The minimum imposed on the amplitudes prevented the complete removal of uncertainties associated with energy-dependent parameters insensitive to the experimental data.
This minimum mirrors the a priori uncertainty of 10\% imposed on energy-independent TALYS model parameters.
We want to stress here that these adjusted uncertainties have to be regarded as prior uncertainties even though they were informed by the experimental data.
The determination of hyperparameters by using the experimental data to obtain a fully specified prior is at the heart of \textit{empirical Bayesian methods}~\citep{robbins1956,maritz_empirical_2018} where it is interpreted as an approximation to a full Bayesian treatment.

The short runtime of a bit more than a minute to determine the 72 Gaussian process hyperparameters by optimizing the logarithmized marginal likelihood despite the occuring inversion of a covariance matrix of dimension $4333 \times 4333$ in each iteration is possible thanks to the decomposition of the covariance matrix presented in \cref{subsec:info_representation} and the exploitation of the matrix identities outlined in \cref{apx:linalg_identities}.
Especially, the availability of an analytical gradient of the logarithmized marginal likelihood thanks to the identities in \cref{apx:derivative_invmat,apx:derivative_logdet} significantly speeds up the computation compared to a numeric evaluation of the gradient.
All of this functionality has been made available in the R package \textit{nucdataBaynet}~\citep{gschnabel-nucdataBaynet} whose mathematical underpinnings were discussed in \cref{sec:mathematical_building_blocks} and technical implementation aspects in \cref{subsec:handling_info_bayesian_context}.

\begin{figure}[t]
    \centering
    \includegraphics{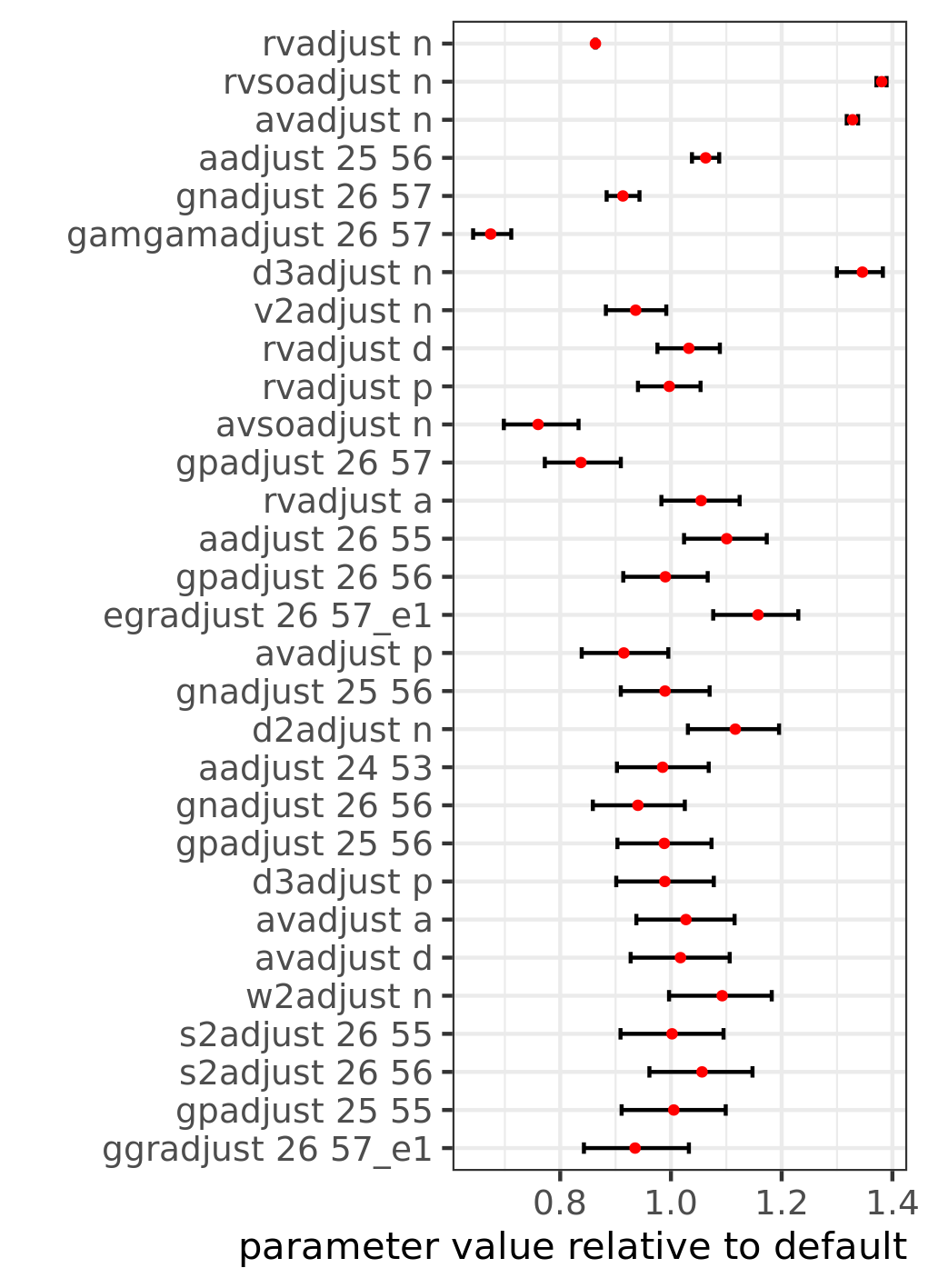}
    \caption{Posterior expectations and uncertainties of the energy-\textit{independent} model parameters most constrained by the experimental data.
    Parameters are sorted according to posterior uncertainty.}
    \label{fig:posterior_pars}
\end{figure}
\begin{figure}[t]
    \centering
    \includegraphics{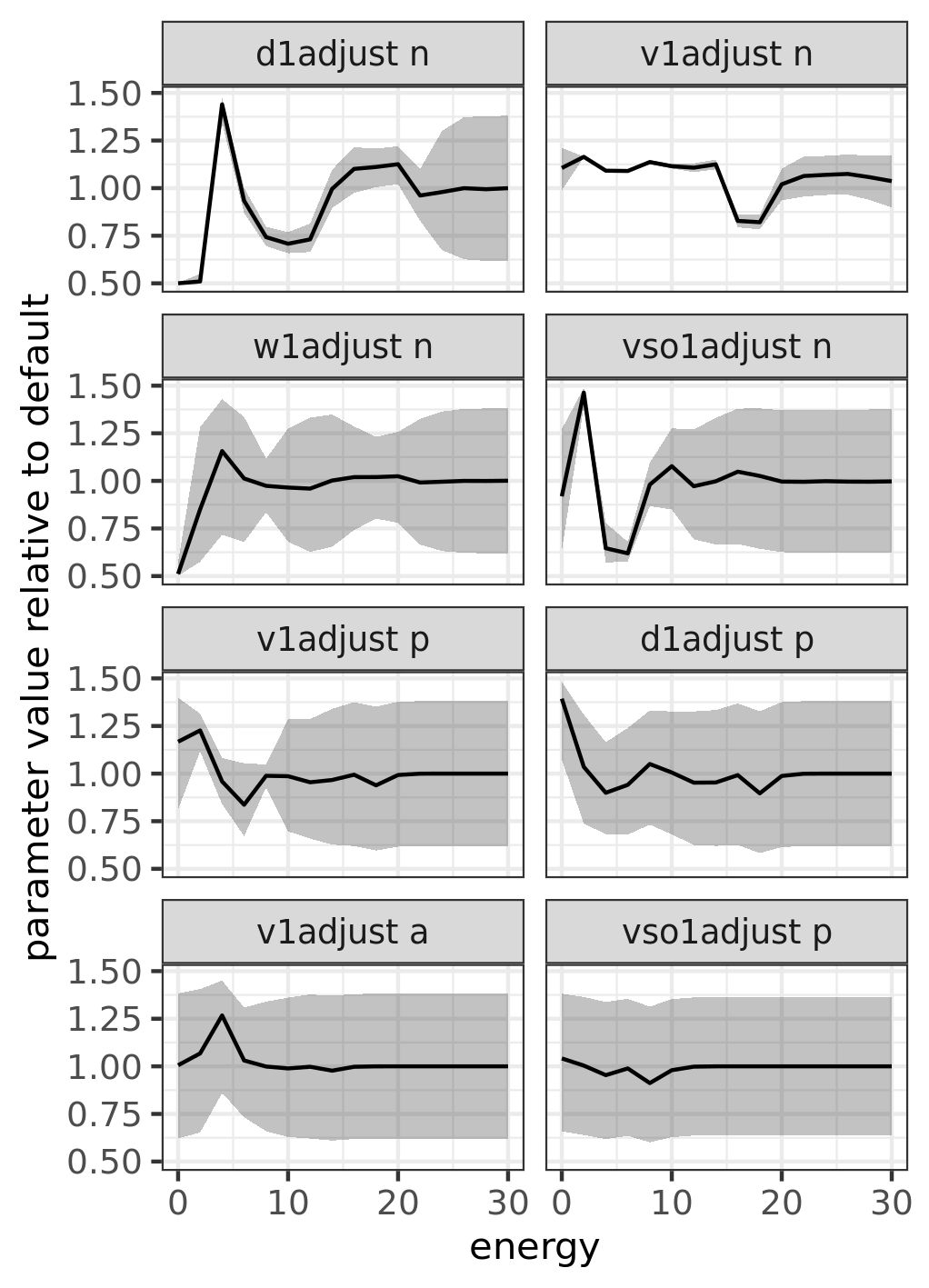}
    \caption{Posterior expectations and uncertainties of the energy-\textit{dependent} model parameters most constrained by the experimental data.}
    \label{fig:posterior_endep_pars}
\end{figure}

Before the invocation of the LM algorithm to adjust TALYS model parameters based on the prior knowledge about experimental data and parameters, the Jacobian matrix obtained in step five, see \cref{subsec:eval_jacob_step5}, was used to exclude insensitive parameters from adjustment.
A parameter was excluded if the maximal value in the corresponding column in the Jacobian matrix was smaller than one. This criterion reduced the number of parameters from 925 to 147 parameters.
The LM algorithm required about 15 iterations to obtain optimized values for the 147 model parameters.
The Jacobian associated with these parameters needs to be recomputed in every iteration, which amounts to 148 model invocations per iteration.
These model invocations were performed in parallel on the computer cluster.
Due to a time requirement of about twenty minutes per model calculation, the LM algorithm terminated after about five hours.

The posterior approximation evaluated in step eight, see \cref{subsec:posterior_approx_step8}, relies on the evaluation of the diagonal elements of the Hessian matrix associated with the logarithmized marginal likelihood.
The numeric evaluation of the second-derivatives of the marginal likelihood with respect to the 147 model parameters required $3\times 147$ model invocations.
Once these second-derivatives are available, the computation time to obtain the approximate posterior distribution for all 925 TALYS model parameters according to the approach in \cref{apx:comp_posterior_covmat} and \cref{apx:comp_posterior_expectation} is negligible.

\begin{figure*}[t]
    \centering
    \includegraphics{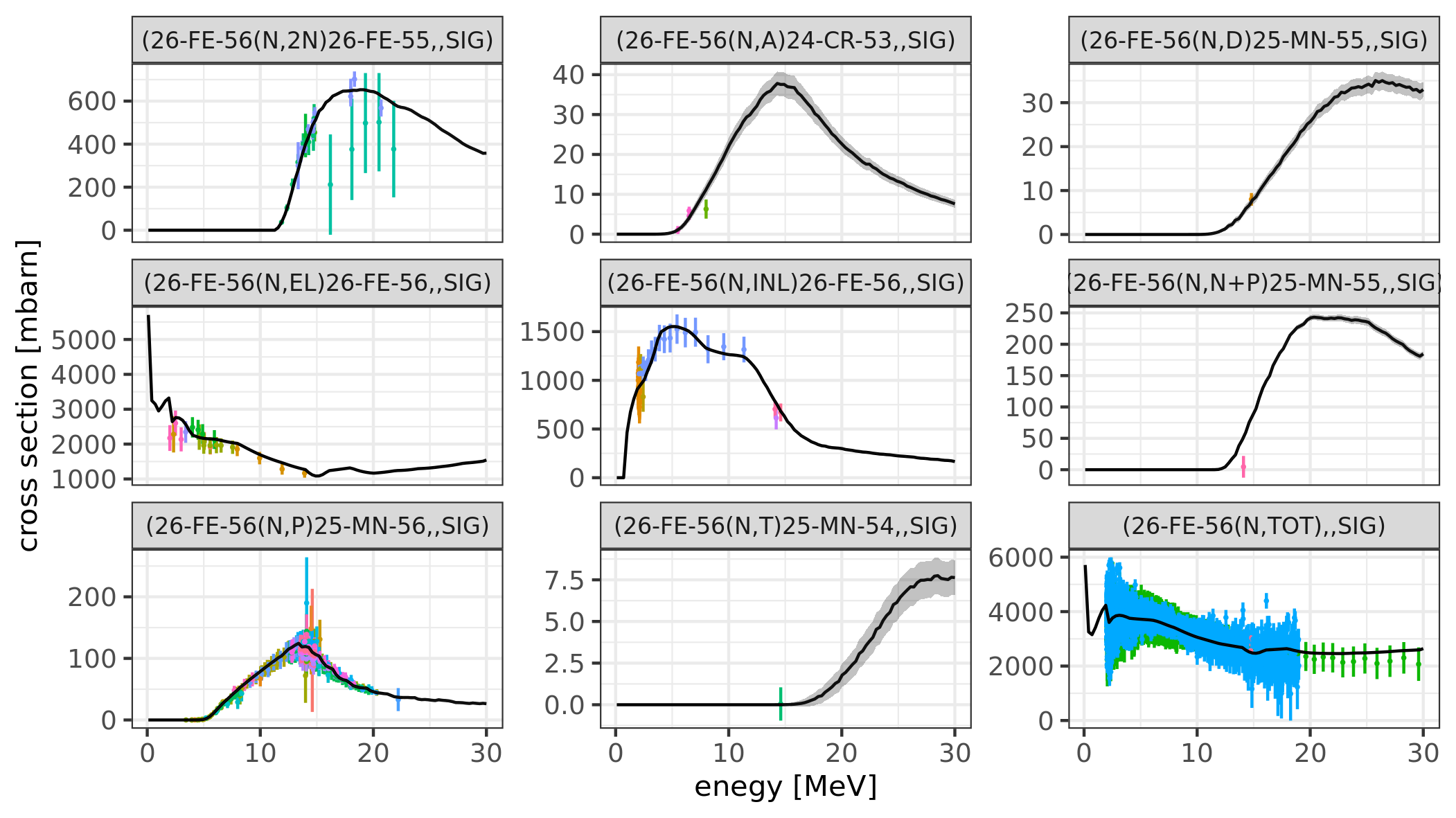}
    \caption{Posterior expectations and uncertainties of the evaluated cross section curves corresponding to the posterior of the model parameters.}
    \label{fig:posterior_xs}
\end{figure*}

The resulting posterior expectations and uncertainties for energy-independent model parameters most constrained by the experimental data are shown in \cref{fig:posterior_pars}.
We see that the posterior uncertainty of many parameters, e.g., \textit{gpadjust 25 55}, is not significantly reduced compared to the 10\% prior uncertainty.
Please note that parameter values are correlated in the posterior distribution and these correlations need to be taken into account for the sampling of parameter combinations.

The posterior expectations and uncertainties of the energy-dependent model parameters most constrained by the data are shown in \cref{fig:posterior_endep_pars}.
Among the 36 energy-dependent parameters, only a small proportion is significantly informed by the experimental data.
The visible spike in \textit{d1adjust n} at about 5\,MeV is a reflection of the parameter transformation explained in \cref{subsec:reference_calculation_step2} to constrain the parameter values to the interval $[0.5, 1.5]$.

Regarding the posterior expectations and uncertainties of the model parameters most informed by the experimental data, the inclusion of 147 model parameters in the LM optimization appears to be very cautious and in future large scale evaluations the criterion to exclude insensitive model parameters can be designed to be more aggressive.

The posterior for the observables is depicted in \cref{fig:posterior_xs}.
The fits appear to be reasonable and follow well the trends in the experimental data.
Many of the posterior uncertainties appear to be very small, maybe too small to be accepted in a real evaluation.
This may be an indication that Gaussian processes on energy-dependent parameters are not always enough to effectively address the deficiency of nuclear models, and the introduction of model defects on the observable side needs to be considered as well.

From a physics point of view, it is questionable whether the measurements in the EXFOR library tagged as (n,n+p) and (n,d) really measure different reaction processes.
As the deuteron is a weakly bound particle, the Q-values for the (n,n+p) and (n,d) reactions are not very different and consequently these reactions cannot be discerned using the activation method.
The extensible design of the IT building block discussed in \cref{subsec:mapping_model_predictions} enables the introduction of more sophisticated handlers that take into account the measurement method and the reaction string to determine the appropriate mapping from model predictions to the experimental data.
In the example of the (n,n+p) and (n,d) reaction, the appropriate mapping would then be to use the corresponding residual production cross section instead of the exclusive reaction cross sections if the experiment relied on activation.

The experimental data included in the example evaluation spanned the energy range from 2 to 30\,MeV, a number of 4333 data points.
To get a better grasp of what is possible in future evaluations, we also launched the pipeline with an extended collection of experimental data by including also data in the range from 0.1 to 2\,MeV increasing the number of data points to 21868.
The experimental covariance matrix (never explicitly computed) contains two dense blocks, each larger than $8000\times 8000$, due to the normalization uncertainty of the associated datasets. 
The EXFOR identifications for these datasets are 13511003 and 22316003.

Clearly, the nuclear models in TALYS are not expected to describe the resonances appearing in this energy range.
The sole purpose of this exercise was to determine whether the algorithms in the pipeline can cope with this amount of data.
All steps were performed without problems and the total runtime of the pipeline not significantly increased.
This can be explained by the fact that the runtime is almost exclusively determined by the runtime of the nuclear models code TALYS.
Each invocation of a TALYS calculation lasted more than ten minutes whereas the statistical algorithms despite the increased number of data points took one order of magnitude less time thanks to the speed ups explained in \cref{apx:linalg_identities}.

\section{SUMMARY AND CONCLUSIONS}


We presented a prototype of a nuclear data evaluation pipeline which covers all aspects of the nuclear data evaluation process, starting from the retrieval of experimental data and their preparation over fitting a nuclear-model and the generation of ENDF files.
The pipeline is an example of a fully reproducible evaluation.
It has been made available online~\citep{gschnabel-eval-fe56}.
A Dockerfile to facilitate the installation of all required components as a single bundle is also available online~\citep{gschnabel-eval-fe56-docker}.

Several innovations in evaluation methodology of the last years found their way into the pipeline:
\begin{itemize}
    \item Rule-based construction and correction of experimental uncertainties
    \item The introduction of extra systematic uncertainties of experimental data based on marginal likelihood optimization
    \item Gaussian process priors on energy-dependent model parameters
    \item The exact treatment of the non-linear model by using a modified Levenberg-Marquardt algorithm, which takes the experimental covariance matrix and prior parameter covariance matrix into account
    \item An approximation of the posterior distribution that also incorporates information on second-order derivatives of the nuclear model
\end{itemize}

The latter four items benefited from a unified representation of statistical errors, systematic errors and model parameters in the form of a simple Bayesian network, which enables to take into account a large collection of experimental datasets without the need of data reduction as a preparatory step.
The ability to deal with a large amount of correlated experimental data implies the promise to deal with exact experimental covariance matrices on the scale of EXFOR.
Both Monte Carlo and deterministic evaluation methods, such as GLS, are equally amenable to the advantages of this representation of experimental data.

The pipeline was also discussed in terms of building blocks of information technology.
We emphasized the need to make the following functionality available to creators of nuclear data evaluation pipelines:
\begin{itemize}
    \item Convenient programmatic access to experimental data
    \item The convenient execution of model calculations in parallel in multi-processor environments or on scientific clusters and the retrieval of results
    \item The convenient generation of model predictions of the observables whose measured values are recorded in the experimental datasets 
\end{itemize}
We outlined design considerations of our implementation of these IT building blocks and provided short usage examples to make the potential advantages for data treatment and the creation of pipelines more tangible.
In particular, the conversion of the EXFOR library to a JSON database in the form of a MongoDB database greatly facilitated the programmatic access to any information in the EXFOR library.
We anticipate that the ease of access to information in the EXFOR library will not only speed up the creation of evaluation pipelines but make nuclear data more accessible to advanced statistical treatment and machine learning methods, such as explored in~\citep{hirdt_data_2013,schnabel_fitting_2018,schnabel_estimating_2018,dwivedi2019trees,whewell_ml_2020}.

As another example, one important focus of the IT building block implementing the parallelization of model calculations was the functioning on a variety of multiprocessor or scientific cluster configurations, which is especially relevant in the scientific environment where the configuration of scientific clusters can be quite different across different institutions.

In the transformation or mapping of the output from nuclear model codes to the observables measured in the experiments, we emphasized the need to have an extensible code architecture that enables the incremental addition of support for new types of observables.
We argued that functions responsible for the mapping should have access to all information---including meta information---associated with experimental data because any information may be potentially relevant for a proper mapping.
Therefore, functions for mapping always received the full information of an EXFOR subentry in our implementation of this IT building block.

Several evaluation methods exist, either based on Monte Carlo sampling or deterministic update formulas, and advancements in the field of Bayesian statistics and machine learning will gradually also lead to improvements of methods for nuclear data evaluation.
We therefore stress the point that the value of the evaluation pipeline is not about the specific choices of methods or data but as a demonstration of an automated and reproducible evaluation involving advanced statistical methods and parallelization on a computer cluster.
During the discussion of the pipeline, we also pointed out possible improvements, such as:
\begin{itemize}
    \item The introduction of computational fields in the MongoDB JSON database to facilitate the search and access of information in the EXFOR library
    \item The automatic renormalization of experimental data in the EXFOR library according to the newest evaluations of monitor reactions  
    \item The enhancement of the rule-based approach for the construction of uncertainty information by relying on statistical data imputation and uncertainty templates
    \item The addition of outlier detection algorithms as a complement to automatic uncertainty correction via the MLO approach
    \item The introduction of model defects on the observable side to deal with deficient models if the increased flexibility achieved by energy-dependent model parameters is not enough
\end{itemize}

Besides these potential improvements, there is another one deserving special attention in the future.
Similar to EXFOR, existing {\em evaluated} data also represents historic knowledge which could be taken into account in the nuclear data pipeline for the creation of a new evaluation. Rather than insisting on re-evaluating everything from scratch, the release of new evaluations could be brought forward in time. Hence, ease of access to existing evaluated data, similar to EXFOR in JSON format, is required to make such evaluation processes efficient. The GNDS format is an important step in this direction.

There are certainly many other possible improvements.
The modular design of the presented evaluation pipeline which can be run at the push of a button enables rapid testing of new algorithms and modified assumptions.
The scripts constituting the pipeline~\citep{gschnabel-eval-fe56} as well as all supporting packages have been made available online.
A Dockerfile~\citep{gschnabel-eval-fe56-docker} also available online greatly facilitates the installation of all required components of the pipeline as a single bundle.

\section{Acknowledgments}
Part of this work has been carried out within the framework of the EUROfusion consortium and has received funding from the Euratom research and training programme 2014-2018 and 2019-2020 under grant agreement No 633053. The views and opinions expressed herein do not necessarily reflect those of the European Commission. Uppsala university acknowledges funding from SSM and SKC for contribution to this paper.

\bibliography{main}

\input{appendices}

\end{document}

%% file: package_list.tex
\usepackage{amssymb}
\usepackage{amsfonts}
\usepackage{amsmath}
\usepackage{makeidx}
\usepackage[bookmarksnumbered,colorlinks,bookmarks,breaklinks=true,linkbordercolor={1 1 1},linktocpage,citecolor=blue]{hyperref}
\usepackage{graphicx,color}
\usepackage{multirow}
\usepackage{dcolumn}
\usepackage{bm}
\usepackage{fancyhdr}
\usepackage{array}
\usepackage{threeparttable}
\usepackage{verbatim}
\usepackage[activate={true,nocompatibility},final,tracking=true,kerning=true,spacing=true,factor=1100,stretch=10,shrink=10]{microtype}
\usepackage[english]{babel}

\usepackage[utf8]{inputenc}
\usepackage{xcolor}
\usepackage{adjustbox}
\usepackage{dsfont}
\usepackage{float}
\PassOptionsToPackage{hyphens}{url}
\usepackage{cleveref}

\usepackage{float}
\usepackage{placeins}
\usepackage{tikz}
\usetikzlibrary{shapes.geometric, arrows,positioning,calc}

\usepackage{beramono} 
\usepackage[T1]{fontenc}

\usepackage{listings}
\usepackage{upquote}

%% file: math_definitions.tex
\newcommand{\Var}[1]{\operatorname{Var}\left[ #1 \right]}
\newcommand{\Cov}[2]{\operatorname{Cov}\left[ #1, #2 \right]}
\newcommand{\Exp}[1]{\mathds{E}\left[ #1 \right]}

\newcommand{\mat}[1]{\mathbf{#1}}

\def\muvec{\vec{\mu}}
\def\muobs{\vec{\mu}_\textrm{obs}}
\def\mutrue{\vec{\mu}_\textrm{true}}
\def\muhid{\vec{\mu}_\textrm{hid}}
\def\xobs{\vec{x}_\textrm{obs}}

\def\sigmamat{\mathbf{\Sigma}}
\def\covobs{\mathbf{\Sigma}_\textrm{obs}}
\def\covhid{\mathbf{\Sigma}_\textrm{hid}}
\def\covobshid{\mathbf{\Sigma}_\textrm{obs,hid}}

\def\obsvec{\vec{\sigma}_\textrm{exp}}
\def\covexp{\mathbf{\Sigma}_\textrm{exp}}

\def\modrefvec{\vec{\sigma}_\textrm{ref}}
\def\modparvec{\vec{p}}
\def\modparrefvec{\vec{p}_\textrm{ref}}
\def\syserrvec{\vec{\varepsilon}_\textrm{sys}}

\def\staterrvec{\vec{\varepsilon}_\textrm{stat}}

\def\syserrmap{\mathbf{S}_\textrm{sys}}
\def\modparmap{\mathbf{S}_\textrm{mod}}
\def\totmap{\mathbf{S}}

\def\totunobs{\vec{u}}
\def\covtotunobs{\mathbf{U}}
\def\totunobsref{\vec{u}_\textrm{ref}}

\def\covstaterr{\mathbf{D}}

\def\totrefvec{\vec{\sigma}_0}

%% file: preamble.tex
\def\etals{\textit{et al.}}

\colorlet{punct}{red!60!black}
\definecolor{background}{HTML}{EEEEEE}
\definecolor{delim}{RGB}{20,105,176}
\colorlet{numb}{magenta!60!black}

\lstdefinelanguage{json}{
    basicstyle=\footnotesize\ttfamily,
    numbers=left,
    numberstyle=\scriptsize,
    stepnumber=1,
    numbersep=8pt,
    showstringspaces=false,
    breaklines=true,
    frame=lines,
    backgroundcolor=\color{background},
    literate=
     *{0}{{{\color{numb}0}}}{1}
      {1}{{{\color{numb}1}}}{1}
      {2}{{{\color{numb}2}}}{1}
      {3}{{{\color{numb}3}}}{1}
      {4}{{{\color{numb}4}}}{1}
      {5}{{{\color{numb}5}}}{1}
      {6}{{{\color{numb}6}}}{1}
      {7}{{{\color{numb}7}}}{1}
      {8}{{{\color{numb}8}}}{1}
      {9}{{{\color{numb}9}}}{1}
      {:}{{{\color{punct}{:}}}}{1}
      {,}{{{\color{punct}{,}}}}{1}
      {\{}{{{\color{delim}{\{}}}}{1}
      {\}}{{{\color{delim}{\}}}}}{1}
      {[}{{{\color{delim}{[}}}}{1}
      {]}{{{\color{delim}{]}}}}{1},
}

\lstdefinelanguage{R}{
  basicstyle=\footnotesize\ttfamily, 
  numbers=left,
  numberstyle=\scriptsize,
  stepnumber=1,
  numbersep=5pt,
  backgroundcolor=\color{background},
  showspaces=false,
  showstringspaces=false,
  showtabs=false,
  frame=single,                   
  rulecolor=\color{black},        
  tabsize=2,
  captionpos=b,
  breaklines=true,
  breakatwhitespace=false,
  title=\lstname,                 
  keywordstyle=\color{blue},
  commentstyle=\color{green},
  stringstyle=\color{violet},
  escapeinside={\%*}{*)}
} 

\crefname{lstlisting}{listing}{listings}
\Crefname{lstlisting}{Listing}{Listings}

%% file: appendices.tex
\clearpage
\appendix

\input{appendix_statistics}
\input{appendix_LA_identities}
\input{appendix_updating}

\input{appendix_download}

%% file: appendix_statistics.tex
\section{GENERAL STATISTICS}

\subsection{Properties of variance and covariance operator}
\label{apx:covariance_properties}

Especially when working with linear combinations of random vectors governed by multivariate normal distribution, using the properties of the variance and covariance operator is very convenient to derive distribution parameters.

Given random vectors $\vec{x}$ and $\vec{y}$, we denote by $\Cov{\vec{x}}{\vec{y}}$ the matrix containing the covariances between elements of $\vec{x}$ and $\vec{y}$, i.e., $(\Cov{\vec{x}}{\vec{y}})_{ij} = \Cov{x_i}{y_j}$.

With $\alpha$ being a real scalar and $\mat{S}$ a matrix, the following properties hold:
\begin{align}
    \Cov{\vec{x}}{\vec{y}} &=
    \left( \Cov{\vec{y}}{\vec{x}} \right)^T \\
    \Cov{\alpha\,\vec{x}}{\vec{y}} &= \alpha \Cov{\vec{x}}{\vec{y}} \\
    \Cov{\mat{S}\vec{x}}{\vec{y}} &= \mat{S} \Cov{\vec{x}}{\vec{y}} \\
    \Cov{\vec{x} + \vec{y}}{\vec{z}} &= \Cov{\vec{x}}{\vec{z}} + \Cov{\vec{y}}{\vec{z}}
\end{align}

The variance operator is given by $\Var{\vec{x}} := \Cov{\vec{x}}{\vec{x}}$.
Therefore it always yields a symmetric matrix and has in addition the properties
\begin{align}
    \Var{\alpha\,\vec{x}} &= \alpha^2 \Var{\vec{x}} \,, \\
    \Var{\mat{S}\vec{x}} &= \mat{S}\Var{\vec{x}}\mat{S}^T \,, \\
    \Var{\vec{x} + \vec{y}} &= \Var{\vec{x}} + \Var{\vec{y}} + 2 \Cov{\vec{x}}{\vec{y}} \,.
\end{align}
For independent random variables, the last term containing the covariances between $\vec{x}$ and $\vec{y}$ vanishes.
Therefore, the variance of a sum of independent random variables is given by
\begin{equation}
    \Var{\sum_{i=1}^{N} \vec{x}_i} = \sum_{i=1}^{N} \Var{\vec{x}_i} \,.
\end{equation}

%% file: appendix_LA_identities.tex
{

\section{LINEAR ALGEBRA IDENTITIES TO SPEED UP COMPUTATIONS}
\label{apx:linalg_identities}


\newcommand\diff{\,\mathrm{d}}
\newcommand{\tr}{\mathrm{Tr}}

\newcommand{\sigmaexp}{\vec{\sigma}_\textrm{exp}}
\newcommand{\sigmaexpPart}[1]{\vec{\sigma}_{\textrm{exp},#1}}

\newcommand{\covexpBlock}[1]{\mat{B}_{#1}}

\newcommand{\invcovexp}{\tilde{\mat{B}}}
\newcommand*{\redcovexp}{\mat{B}_{\vec{z}}}
\newcommand{\sigmafit}{\vec{\sigma}_\textrm{fit}}
\newcommand{\modpar}{\vec{y}}
\newcommand{\priormodpar}{\vec{y}_0}
\newcommand{\postmodpar}{\vec{y}_1}
\newcommand{\priorcovpar}{\mat{A}_0}
\newcommand{\invpriorcovpar}{\mat{\tilde{A}}_0}
\newcommand{\postcovpar}{\mat{A}_1}
\newcommand{\marcov}{\mat{M}}
\newcommand{\marcovBlock}[1]{\mat{M}_{#1}}
\newcommand{\invmarcov}{\mat{\tilde{M}}}
\newcommand{\invmarcovBlock}[1]{\mat{\tilde{M}}_{#1}}

\newcommand{\jacob}{\mat{S}}
\newcommand{\jacobBlock}[1]{\mat{S}_{#1}}
\newcommand{\probvec}{\vec{z}}
\newcommand{\probvecEl}[1]{z_{#1}}

\newcommand{\adjcovexp}{\mat{C}}
\newcommand{\adjcovexpBlock}[1]{\mat{C}_{#1}}
\newcommand{\invadjcovexp}{\mat{\tilde{C}}}
\newcommand{\invadjcovexpBlock}[1]{\mat{\tilde{C}}_{#1}}
\newcommand{\kappavec}{\vec{\kappa}}
\newcommand{\kappavecEl}[1]{\kappa_{#1}}
\newcommand{\lambdavec}{\vec{\lambda}}
\newcommand{\lambdavecEl}[1]{\lambda_{#1}}

\newcommand{\valueT}{\textrm{T}}
\newcommand{\valueF}{\textrm{F}}

\subsection{Derivative of an inverse matrix}
\label{apx:derivative_invmat}
Assume a matrix $\marcov(\kappavec)$ which is a function of potentially several parameters $\kappa_l$ summarized in $\kappavec$. We denote its inverse by $\invmarcov(\kappavec)$.
The relation between these two matrices in terms of their components is given by
\begin{equation}
	\sum_{j} \marcov_{ij}(\kappavec) \invmarcov_{jk}(\kappavec) = \delta_{ij} \,,
\end{equation}
with $\delta_{ij}$ being one if $i=j$ and zero otherwise.
Taking the partial derivative with respect to an element $\kappa_{l}$ of $\kappavec$ gives
\begin{equation}
	\sum_{j} \left(
	\frac{\partial{\marcov_{ij}(\kappavec)}}{\partial \kappa_{l}}
	\invmarcov_{jk}(\kappavec)
	+
	\marcov_{ij}(\kappavec)
	\frac{\partial{\invmarcov_{jk}(\kappavec)}}{\partial \kappa_{l}}
	\right)  = 0 \,.
\end{equation}
This relation can be expressed in terms of matrix products,
\begin{equation}	
	\marcov(\kappavec)
	\frac{\partial{\invmarcov(\kappavec)}}{\partial \kappa_{l}}
	=
	-\frac{\partial{\marcov(\kappavec)}}{\partial \kappa_{l}}
	\invmarcov(\kappavec)
 \,.
\end{equation}
Multiplying by $\invmarcov(\kappavec)$ from the left yields the final result
\begin{equation}
	\frac{\partial{\tilde{\mat{M}}(\kappavec)}}{\partial \kappa_{l}} =
	-\tilde{\mat{M}}(\kappavec)
	\frac{\partial{\mat{M}(\kappavec)}}{\partial \kappa_{l}}
	\tilde{\mat{M}}(\kappavec) \,.		
\end{equation}

\subsection{Derivative of a logarithmized determinant}
\label{apx:derivative_logdet}

In the following we assume that a matrix $\mat{M}$ is a function of potentially several parameters $\kappa_1, \kappa_2, \dots$ summarized in a vector $\vec{\kappa}$.
In the pipeline, $\vec{\kappa}$ can contain systematic and statistical uncertainties and hyperparameters of Gaussian processes.
The logarithmized probability density function of a multivariate normal distribution contains the logarithmized determinant of a covariance matrix.
The availability of the gradient in analytic form of this logarithmized determinant with respect to a set of parameters $\vec{\kappa}$ is extremely useful in gradient based optimization algorithms and is indeed computable.

Using the chain rule, the derivate of $\ln\det\mat{M}(\kappavec)$ can be written as
\begin{equation}
	\frac{\partial \ln\det\marcov(\kappavec)}{\partial \kappavecEl{l}} =
	\frac{1}{\det\marcov(\kappavec)}
	\frac{\partial \det\marcov(\kappavec)}{\partial \kappavecEl{l}} \,.
	\label{apx:eq:logdetderive}
\end{equation}
Jacobi's formula \cite[p.~305]{harville_matrix_1997} provides us with the derivative of the determinant,
\begin{equation}
	\frac{\partial \det\marcov(\kappavec)}{\partial \kappavecEl{l}} =
	\tr \left(
	\textrm{adj}\big(\marcov(\kappavec)\big) 
	\frac{\partial\marcov(\kappavec)}{\partial\kappavecEl{l}}
	\right) \,.
	\label{apx:eq:detderive}
\end{equation}
The adjugate matrix appearing in this expression is defined by \cite[p.~192]{harville_matrix_1997}
\begin{equation}
	\marcov \, \textrm{adj}(\marcov) = \det(\marcov) \, \mathds{1}
	\;\Rightarrow\;
	\textrm{adj}(\marcov) = \det(\marcov) \mat{\marcov}^{-1}
	\label{apx:eq:adjmat}
\end{equation}
Inserting \cref{apx:eq:adjmat} into \cref{apx:eq:detderive} and the resulting expression into \cref{apx:eq:logdetderive} yields the final result:
\begin{equation}
	\frac{\partial \ln\det\marcov(\kappavec)}{\partial \kappavecEl{l}}
	= \tr \left(
	\left(\marcov(\kappavec)\right)^{-1}
	\frac{\partial\marcov(\kappavec)}{\partial\kappavecEl{l}}
	\right) \,.
\end{equation}
Please note that only the diagonal elements of the matrix product have to be computed to evaluate the trace.
For a matrix of size $N\times N$, this reduces the time complexity from $N^3$ to $N^2$.

\subsection{Matrix determinant lemma}
\label{apx:matrix_determinant_lemma}
For the derivation of the matrix determinant lemma in the version used in this paper, note that
\begin{equation}
\begin{split}
	\det\left(\mat{A}+\mat{U}\mat{V}^T \right) &= 
	\det\left(
	\mat{A}
	\left(\mathds{1} + \mat{A}^{-1}\mat{U}\mat{V}^T \right)
	\right) \\
	&= \det(\mat{A}) \det \left( \mathds{1} + \mat{A}^{-1}\mat{U}\mat{V}^T \right) \,.
\end{split}
\end{equation}
The application of Sylvester's determinant identity \cite[p.~416]{harville_matrix_1997}, i.e. $\det(\mathds{1}+\mat{A}\mat{B})=\det(\mathds{1}+\mat{B}\mat{A})$, yields
\begin{equation}
	\det\left(\mat{A}+\mat{U}\mat{V}^T \right) = \det(\mat{A})
	\det\left( \mathds{1} + \mat{V}^T \mat{A}^{-1}\mat{U} \right) \,.
\end{equation}
Now replace $\mat{U}$ by the matrix product $\mat{U}\mat{W}$ and extract $\mat{W}$ to obtain 
\begin{multline}
	\det\left(\mat{A}+\mat{U}\mat{W}\mat{V}^T \right) = \\
	\det(\mat{A})
	\det\left( \left(
	\mat{W}^{-1} + \mat{V}^T \mat{A}^{-1}\mat{U} \right)
	\mat{W} \right) \,.
\end{multline}
Making use of $\det(\mat{A}\mat{B})=\det(\mat{A})\det(\mat{B})$, we get the final result
\begin{multline}
	\det\left(\mat{A}+\mat{U}\mat{W}\mat{V}^T \right) = \\
	\det(\mat{A}) \det(\mat{W})
	\det\left(
	\mat{W}^{-1} + \mat{V}^T \mat{A}^{-1}\mat{U} \right) \,.
\end{multline}

\subsection{Woodbury identity}
\label{apx:Woodbury_identity}
The Woodbury matrix identity \citep{woodbury_inverting_1950}, \citep[p.~424]{harville_matrix_1997} enables the inversion of a matrix of a specific structure in an alternative way:
 \begin{multline}
    \label{eq:woodbury_identity}
    (\mat{A} + \mat{U} \mat{C} \mat{V})^{-1} = \\
    \mat{A}^{-1} - \mat{A}^{-1} \mat{U}
    (\mat{C}^{-1} + \mat{V} \mat{A}^{-1} \mat{U})^{-1} \mat{V} \mat{A}^{-1} \,
    \end{multline}
with $\mat{A}, \mat{U}, \mat{C},\textrm{and }\mat{V}$ being matrices of appropriate dimension.

The validity of the formula can be verified by multiplying both sides by $(\mat{A} + \mat{U} \mat{C} \mat{V})$ from the left. As the left side equals the identity matrix, it suffices to evaluate the right hand side of \cref{eq:woodbury_identity},
 \begin{equation}\begin{split}
    &(\mat{A} + \mat{U} \mat{C} \mat{V}) \times \\
    &[ \mat{A}^{-1} - \mat{A}^{-1} \mat{U}(\mat{C}^{-1} + \mat{V} \mat{A}^{-1} \mat{U})^{-1} \mat{V} \mat{A}^{-1} ] \\
    = &\mathds{1} + \mat{U} \mat{C} \mat{V} \mat{A}^{-1} \\
    &-(\mat{U} + \mat{U} \mat{C} \mat{V} \mat{A}^{-1}\mat{U}) 
    (\mat{C}^{-1} + \mat{V} \mat{A}^{-1} \mat{U})^{-1} \mat{V} \mat{A}^{-1} \\
    = &\mathds{1} + \mat{U} \mat{C} \mat{V} \mat{A}^{-1} \\
    &-\mat{U}\mat{C} (\mat{C}^{-1}+\mat{V}\mat{A}^{-1}\mat{U})
    (\mat{C}^{-1} + \mat{V} \mat{A}^{-1} \mat{U})^{-1} \mat{V} \mat{A}^{-1} \\
    = &\mathds{1} + \mat{U} \mat{C} \mat{V} \mat{A}^{-1} - \mat{U} \mat{C} \mat{V} \mat{A}^{-1} = \mathds{1} \,.
    \end{split}\end{equation}
}

\subsection{Products involving the inverse of sparse covariance or other positive-definite matrices}
\label{apx:prod_inv_mat}

In \cref{subsec:info_representation} we had a matrix of the form $\mat{Z} = \covtotunobs^{-1} + \totmap^T \covstaterr^{-1} \totmap$ whose inverse needs to be multiplied with vectors or matrices.
To carry out such products, in the \textit{nucdataBaynet} package we compute the Cholesky decomposition, e.g.,~\citep{harville_matrix_1997}, of covariance matrices and other positive-definite matrices first,
\begin{equation}
    \mat{Z} = \mat{L} \mat{L}^T\,,
\end{equation}
with $\mat{L}$ being a lower triagonal matrix.

To evaluate a matrix product of the form $\mat{R}=\mat{X} \mat{Z}^{-1}\mat{X}^T$, note that
\begin{equation}
\begin{split}
    \mat{R} &= \mat{X}(\mat{L}\mat{L}^T)^{-1}\mat{X}^T \\
            &= \left[
                \mat{X} (\mat{L}^T)^{-1}
                \right]
                \left[
                \mat{L}^{-1} \mat{X}^T
                \right] = \mat{M} \mat{M}^T
\end{split}
\end{equation}
Therefore it suffices to evaluate $\mat{M} = \mat{X}(\mat{L}^T)^{-1}$ and then compute the final result by $\mat{R}=\mat{M}\mat{M}^T$.
The computation of $\mat{M}$ can be formulated as the problem to solve the set of linear equations
\begin{equation}
    \mat{L}^T\mat{M} = \mat{X} \,,
\end{equation}
which can be efficiently done by back-substitution due to $\mat{L}^T$ being upper triagonal.

Also the computation of matrix products of the form $\mat{S}=\mat{X} \mat{Z}^{-1}$ can be restated as the solution to a set of linear equations in two stages,
\begin{equation}
    \mat{L}^T \mat{M} = \mat{X} \;\;\textrm{then} \;\;
    \mat{L} \mat{S} = \mat{M} \,,
\end{equation}
where the first equation can be solved by back-substitution and the second one by forward-substitution.

In addition to being positive-definite, we can expect $\mat{Z}$ mentioned at the beginning of this section to be sparse because:
\begin{itemize}
    \item The covariance matrix $\mat{U}$ associated with systematic components is diagonal or at least block-diagonal and so is its inverse.
    \item The sparseness of the mapping matrix $\mat{S}$ in combination with the diagonal structure of $\mat{D}$ leads to $\mat{S}^T \mat{D}^{-1} \mat{S}$ being sparse, too.
\end{itemize}

The Cholesky factor of sparse matrices is in general not sparse, hence the benefit of sparsity of $\mat{Z}$ would be lost for solving the systems of linear equations outlined above.

Therefore a reordering of rows of columns of the original matrix can be performed before computing the Cholesky decomposition so that resulting matrices are sparse.
A popular algorithm to determine the permutations are variants of the \textit{minimum degree ordering algorithm}~\citep{george_evolution_1989,davis_direct_2006}.
For the operations with sparse matrices including matrix products, solving systems of linear equations with sparse matrices and sparse Cholesky decompositions, we rely on the functionality of the R package \textit{Matrix}~\citep{RMatrix}.

%% file: appendix_updating.tex
\section{BAYESIAN INFERENCE}

\input{appendix_LM}

\input{appendix_smoothness_prior}
\input{appendix_taylor_approx_necessity}
\input{appendix_posterior_conditioning}

%% file: appendix_LM.tex
\subsection{Details on modified Levenberg-Marquardt algorithm}
\label{apx:LM_details}

The LM algorithm \citep{levenberg_method_1944,marquardt_algorithm_1963} extends the linear-least squares method to be applicable to non-linear models.
We use a modification of the LM algorithm to account for the experimental covariance matrix and prior parameter covariance matrix.

The objective is to locate the maximum of the posterior pdf
\begin{equation}
    \pi(\modparvec \,|\, \obsvec) = \frac{1}{\pi(\obsvec)} \ell(\obsvec \,|\, \modparvec) \pi(\modparvec) \,.
\end{equation}
The likelihood and prior parameter distribution are given by
\begin{align}
    \ell(\obsvec \,|\, \modparvec) &= \mathcal{N}(\obsvec \,|\, \mathcal{M}(\vec{p}); \covexp) \,, \\
    \pi(\modparvec) &= \mathcal{N}(\modparvec \,|\, \vec{p}_0; \mat{P}) \,.
\end{align}
The model link $\mathcal{M}(\vec{p})$ is non-linear.
The LM algorithm is an iterative approach and it relies in each iteration on a linear approximation of the non-linear model,
\begin{equation}
    \mathcal{M}_\textrm{lin}(\vec{p}) = \modparrefvec +
    \mat{J} \left( \modparvec - \modparrefvec \right) \,.
\end{equation}
The introduction of $\delta\modparvec = \modparvec - \modparrefvec$, $\vec{d} = \obsvec - \modrefvec$ allows us to write
\begin{multline}
    (-2) \ln \pi(\modparvec \,|\, \obsvec) =
    \left(
        \vec{d} - \mat{J} \, \delta\modparvec
    \right)^T
    \covexp^{-1}
    \left(
        \vec{d} - \mat{J} \, \delta\modparvec
    \right) 
    \\
    +
    \left(
        \delta\modparvec + \modparrefvec - \vec{p}_0
    \right)^T
    \mat{P}^{-1}
    \left(
        \delta\modparvec + \modparrefvec - \vec{p}_0
    \right)
    + \mathcal{C}
\end{multline}
with the constant $\mathcal{C}$ absorbing everything which is independent of $\delta\modparvec$.
The constant is of no significance as we are going to take derivatives with respect to elements in $\delta\modparvec$.

Now we isolate terms containing $\delta\modparvec$ and regroup the expression according to their order,
\begin{multline}
    \frac{1}{2}
    \delta\modparvec^T
    \left(
        \mat{J}^T \covexp^{-1} \mat{J} +
        \mat{P}^{-1}
    \right)
    \delta\modparvec \\
    -
    \delta\vec{p}^T
    \left(
    \mat{J}^T \covexp^{-1} \vec{d}
    + \mat{P}^{-1} (\vec{p}_0 - \modparrefvec)
    \right) + \mathcal{T}
    \label{eq:LMobjective}
\end{multline}
where $\mathcal{T}$ denotes the transpose of the terms explicitly written out.
To determine the vector $\delta\modparvec$ that minimizes this expression, we calculate the gradient of \cref{eq:LMobjective} and require it to vanish:
\begin{multline}
    \left(
        \mat{J}^T \covexp^{-1} \mat{J} +
        \mat{P}^{-1}
    \right)
    \delta\modparvec \\
    -
    \left(
    \mat{J}^T \covexp^{-1} \vec{d}
    + \mat{P}^{-1} (\vec{p}_0 - \modparrefvec)
    \right) = \vec{0} \,.
    \label{eq:gradLMobjective}
\end{multline}
Rearranging yields
\begin{equation}
    \left(
        \mat{J}^T \covexp^{-1} \mat{J} +
        \mat{P}^{-1}
    \right)
    \delta\modparvec
    =
    \left(
    \mat{J}^T \covexp^{-1} \vec{d}
    + \mat{P}^{-1} (\vec{p}_0 - \modparrefvec)
    \right) \,.
\end{equation}
Please note that the right-hand side is proportional to the gradient of \cref{eq:LMobjective} evaluated at the reference parameter vector, i.e., $\delta\modparvec=0$, used as expansion point for the construction of the linear approximation.
Therefore it is also proportional to the gradient of the logarithmized posterior distribution $\ln \pi(\modparvec\,|\,\obsvec)$ evaluated at $\modparvec=\modparrefvec$.

We introduce the following abbreviations:
\begin{align}
    \mat{A} &= 
        \mat{J}^T \covexp^{-1} \mat{J} + \mat{P}^{-1} + \lambda \mathcal{I}
    \\
     \vec{b} &= 
     \mat{J}^T \covexp^{-1} \left(
        \obsvec - \modrefvec
     \right) +
     \mat{P}^{-1}
     \left(
        \vec{p}_0 - \modparrefvec
     \right) \,.
\end{align}
The matrix $\mat{\mathcal{I}}$ is the identity matrix.
The introduction of the term $\lambda\mathcal{I}$ allows to make $\mat{A}$ more diagonal by increasing the value of $\lambda$.
The purpose of this mechanism will become apparent in a moment.

The update equation for $\delta\modparvec$ can be written as 
\begin{align}
    \mat{A} \delta\modparvec &= \vec{b} 
    \\
    \vec{p}_\textrm{prop} &= \modparrefvec + \delta\modparvec
\end{align}
With increasing $\lambda$, the matrix $\mat{A}$ becomes more similar to the identity matrix.
Consequently, the update degenerates gradually to the gradient descent method.
For $\lambda=0$, the update is equivalent to the GLS method.

%% file: appendix_smoothness_prior.tex
\subsection{Prior on second-derivative as smoothness prior}
\label{apx:prior_2nd_derivative}
Assume that we have tuples of cross values $y_i$ and associated energies $x_i$, i.e. $\{(x_i, y_i)\}_{i=1..N}$, in some reaction channel.
Let the energies $x_i$ be sorted in ascending order, i.e. $x_i < x_{i+1}$.
We can use a polygon chain to obtain cross sections for any intermediate energies~$x$,
\begin{equation}
\label{eq:polychain}
f(x) = \sum_{i=1}^{N}
\mathcal{I}_i(x) \big(
u_i(x) y_i +
v_{i+1}(x) y_{i+1} \big)
\end{equation}
with $\mathcal{I}_i(x) = 1 \textrm{ if } x_i < x < x_{i+1}, \textrm{and } 0 \textrm{ otherwise}$.
We have introduced the following abbreviatios which will be useful later on:
\begin{equation}
u_i(x) = \frac{x_{i+1} - x}{x_{i+1} - x_i} \;\;\textrm{and}\;\;
v_i(x) = \frac{x - x_{i-1}}{x_{i} - x_{i-1}}
\end{equation}
Given that the grid spanned by the $x_i$'s is sufficiently dense, any continuous function can be approximated with sufficient precision.

The prediction for cross sections $\tilde{y} = (\tilde{y}_1, \cdots, \tilde{y}_M)$ at locations $\tilde{x} = (\tilde{x}_1, \cdots, \tilde{x}_M)^T$ using \cref{eq:polychain} can be also written in matrix notation.
To that end, we introduce the sensitivity matrix
\begin{equation}
S_{ji} :=
\begin{cases}
  u_i(\tilde{x}_j) & \textrm{if } x_i < \tilde{x}_j < x_{i+1} \\
  v_i(\tilde{x}_j) & \textrm{if } x_{i-1} < \tilde{x}_j < x_{i} 
\end{cases}
\,.
\end{equation}
\Cref{eq:polychain} now becomes
\begin{equation}
  \tilde{y} = S y \,.
\end{equation}
In principle, we could introduce a prior on the values in the vector $y$.
However, sometimes we would like to express our knowledge in terms of allowed differences or even changes of differences.
This can be achieved by a reparametrization:
\begin{equation}
  \label{eq:firstordertrafo}
  y_i = \begin{cases}
    y_1 & \textrm{if } i = 1\\
    y_1 + \sum_{k=1}^{i-1} \Delta y_k & \textrm{if } i \ge 2
  \end{cases}
\end{equation}
We can summarize the new parameters in a vector $u = (y_1, \Delta y_1, \cdots, \Delta y_{N-1})^T$ and define the following matrix
\begin{equation}
  T_{ij} = \begin{cases}
  1 & \textrm{if } i \ge j \\
  0 & \textrm{otherwise}
  \end{cases}
\end{equation}
in order to identically rewrite \cref{eq:firstordertrafo} in matrix notation as
\begin{equation}
  y = T u \,.
\end{equation}
To get a parametrization in terms of second derivatives, we apply once again a similiar transformation,
\begin{equation}
  \Delta y_i = \begin{cases}
    \Delta y_1 & \textrm{if } i = 1 \\
    \Delta y_1 + \sum_{k=1}^{i-1} \Delta^2 y_k & \textrm{if } i \ge 2 \\
  \end{cases}
\end{equation}
We can summarize the new parameters in a vector $v = (y_1, \Delta y_1, \Delta^2 y_1, \cdots, \Delta^2 y_{N-2})^T$ and transform them to $u$ defined above using the following matrix
\begin{equation}
  R_{ij} = \begin{cases}
    1 & \textrm{if } i = j = 1 \\
    1 & \textrm{if } i \ge 2 \textrm{ and } i \ge j \\
    0 & \textrm{otherwise}
  \end{cases}  
\end{equation}
in order to obtain
\begin{equation}
  u = R v \,.
\end{equation}
The concatenation of all the matrices yields finally the prediction of the polygon chain in terms of the parameters corresponding to the (finite version) of the second derivatives.
\begin{equation}
  \label{eq:finalmapping}
  \tilde{y} = Z v \textrm{ with } Z = S T R 
\end{equation}
The prior can be formulated for the elements in $v$ and it can be assumed that no correlations between elements in $v$ exist,
\begin{equation}
    \pi(v) = \mathcal{N}(v \,|\, \vec{0}, \mat{D}) \,.
\end{equation}
The vector of zeros indicates that the best guess for the curvature of the cross section curve is zero, i.e., changes of the slope are penalized.

Even though the prior covariance matrix $\mat{D}$ for the values of the second-order derivative is diagonal, correlations are induced between the function values on the mesh due to the particular form of the mapping matrix $Z$.

%% file: appendix_taylor_approx_necessity.tex
\subsection{Necessity of second-order Taylor approximation of posterior pdf}
\label{apx:necessity2ndorder}
After the localisation of the posterior maximum by the LM algorithm, one may be tempted to apply the GLS formulas to compute the posterior covariance matrix as it only requires the Jacobian matrix of the nuclear model, i.e., first-order derivates of the nuclear model.
Here we explain why this approach is potentially flawed and also second-order derivatives of the model need to be taken into account.

The logarithmized posterior pdf is given by
\begin{equation}
\begin{split}
    \ln \pi(\vec{p}\,|\,\vec{\sigma}) =
    &-\frac{1}{2} (\vec{\sigma} - \mathcal{M}(\vec{p}))^T
    \mat{Q}
    (\vec{\sigma} - \mathcal{M}(\vec{p})) \\
    &-\frac{1}{2} (\vec{p} - \vec{s})^T \mat{R} (\vec{p} - \vec{s})
    + \mathcal{C} \,.
\end{split}
\end{equation}
The matrix $\mat{Q}$ is the inverse of the experimental covariance matrix $\covexp$ , the vector $\vec{\sigma}$ contains the measurements, the matrix $\mat{R}$ is the inverse of the prior parameter covariance matrix, the vector $\vec{s}$ denotes the prior expectation of $\vec{p}$ and $\mathcal{M}$ yields the vector with model predictions based on parameter set $\vec{p}$. The argument of $\mathcal{M}$ will be dropped from now on for notational convenience.
The term $\mathcal{C}$ summarizes the terms independent of $\vec{p}$ and ensures the proper normalization of the pdf.

We rewrite the right hand side in element-wise notation:
\begin{multline}
    -\frac{1}{2} \sum_{i=1}^{N}\sum_{j=1}^{N} (\sigma_i - \mathcal{M}_i) Q_{ij} (\sigma_j - \mathcal{M}_j) \\
    -\frac{1}{2} \sum_{m=1}^{M} \sum_{n=1}^{M} (p_m - s_m) R_{mn} (p_n - s_n) + \mathcal{C}.
\end{multline}
The upper limit $N$ of the first two summations is the number of experimental data points and the upper limit $M$ of the last two summations is the number of model parameters. 

Taking the first derivative with respect to a model parameter $p_{k}$ yields
\begin{multline}
G_k := \sum_{i=1}^{N}\sum_{j=1}^{N} (\partial_k \mathcal{M}_i) Q_{ij} (\sigma_j - \mathcal{M}_j) \\
- \sum_{n=1}^{M} R_{kn} (p_n - s_n) \,,
\label{eq:gradient_logpost_einstein}
\end{multline}
where we used the short-hand notation $\partial_k := \partial/\partial p_k$.
We denote this expression as $G_k$ because it is the k-th element of the gradient of the logarithmized posterior pdf with respect to the model parameters.

We take once again the derivative with respect to a model parameter $p_t$
\begin{multline}
H_{kt} :=
\sum_{i=1}^{N}\sum_{j=1}^{N} (\partial_k\partial_t \mathcal{M}_i) Q_{ij} (\sigma_j - \mathcal{M}_j) \\
-\sum_{i=1}^{N}\sum_{j=1}^{N} (\partial_k \mathcal{M}_i) Q_{ij} (\partial_t \mathcal{M}_j)
-R_{kt} \,.
\label{eq:hessian_with_2nd_model_deriv}
\end{multline}
This expression states the elements $H_{kt}$ of the Hessian matrix of the logarithmized posterior pdf.
We rewrite~\cref{eq:hessian_with_2nd_model_deriv} in matrix notation,
\begin{equation}
    \mat{H} = -(\mat{U} + \mat{J}^T \mat{Q} \mat{J} + \mat{R}) \,,
    \label{eq:hessian_with_2nd_model_deriv_matrix}
\end{equation}
with the elements of the matrix $\mat{U}$ being of the form 
\begin{equation}
    U_{kt} = -\sum_{i=1}^{N}\sum_{j=1}^{N}(\partial_k\partial_t \mathcal{M}_i) Q_{ij} (\sigma_j - \mathcal{M}_j) 
\end{equation}
and the elements of the Jacobian matrix $\mat{J}$ being $J_{ik} := \partial_k \mathcal{M}_i$.

For a linear model, the matrix $\mat{U}$ in \cref{eq:hessian_with_2nd_model_deriv_matrix} vanishes and recalling that for a linear model the relation between the Hessian and the covariance matrix is given by $\mat{P}_1=-\mat{H}^{-1}$, we recover the GLS solution for the posterior covariance matrix, $\mat{P}_1 = (\mat{J}^T \mat{Q} \mat{J} + \mat{R})^{-1}$.

For non-linear models, the neglect of the first term containing the second-order derivative of the model can be problematic for two reasons:
\begin{itemize}
    \item The posterior is not contracted enough so that the linear approximation of the model is not sufficient in the domain of the parameter space admissible by the posterior distribution.
    \item The model is deficient in a statistical sense, i.e., the residuals $(\sigma_j - \mathcal{M}_j)$ are large and/or not about as often positive as negative, i.e., the model systematically over- or underestimates the data.
    Even if the non-linearity of the model is mild, the first term may get inflated due to the large residuals so that its contribution to the Hessian cannot be neglected anymore.
\end{itemize}
By implication, the generalized least squares method has a higher chance of producing a valid posterior covariance matrix for a non-linear model if the model is not deficient.
The modelling of model defects by Gaussian processes could improve the situation for a nuclear model with deficiencies as the elements $Q_{ij}$ in the inverse experimental covariance matrix are decreased, hence also reducing the magnitude of the first term in \cref{eq:hessian_with_2nd_model_deriv}.

%% file: appendix_posterior_conditioning.tex
\subsection{Computation of the posterior covariance matrix}
\label{apx:comp_posterior_covmat}
In order to reduce the computational requirement of the LM algorithm, one may skip the computation of the elements in the Jacobian matrix associated with model parameters insensitive to the experimental data and set these elements to zero.

After the optimization by the LM algorithm, the posterior covariance matrix needs to be computed for all parameters, including those insensitive to the experimental data.
The reason being that correlations between parameters imposed by the prior lead to a propagation of information from the sensitive parameters to the insensitive ones.
For instance, Gaussian processes introduce correlations between energy-dependent optical model parameters, e.g., between the values of $r_v$ at 5\,MeV and 10\,MeV.
If $r_v$ at 10\,MeV gets excluded from the LM optimization, we have a situation where a sensitive parameter is already a priori correlated to an insensitive one.

In this section we explain the calculation of the full posterior covariance.
For the derivation we will take into account the specific structure of the Hessian matrix due to the presence of insensitive parameters and introduce an approximation to make the computation tractable for computationally expensive models, such as TALYS.

To determine the posterior covariance matrix of all parameters, we consider a second-order Taylor approximation of the logarithmized posterior pdf constructed at the reference parameter vector $\modparrefvec$:
\begin{multline}
    \ln \pi(\modparvec \,|\, \obsvec) = \\
    \ln \pi_0 + 
    \mat{G}
    \left(
        \modparvec - \modparrefvec
    \right)
    + \frac{1}{2}
    \left(
        \modparvec - \modparrefvec
    \right)^T
    \mat{H}
        \left(
        \modparvec - \modparrefvec
    \right) \,.
    \label{eq:logposteriorpdf_2ndorderapprox}
\end{multline}
The normalization constant is given by $\ln\pi_0 = \ln \pi(\modparrefvec \,|\, \obsvec)$ and $\mat{G}$ is the gradient and $\mat{H}$ the Hessian matrix of the logarithmized posterior pdf evaluated at $\modparrefvec$.
The form of the gradient was given in~\cref{eq:gradient_logpost_einstein} and that of the Hessian in~\cref{eq:hessian_with_2nd_model_deriv_matrix}.
Given the nuclear model can be well approximated by a second-order Taylor expansion in the parameter domain admissible by the posterior distribution, the posterior covariance matrix can be approximated as $\mat{P}_1 = -\mat{H}^{-1}$.

We split the parameter vector in blocks $\vec{p} = (\vec{p}_1^T, \vec{p}_2^T)^T$ with $\vec{p}_1$ containing the sensitive parameters and $\vec{p}_2$ containing the insensitive parameters.
The Jacobian matrix is divided into blocks accordingly, i.e., $\mat{J} = (\mat{J}_1, \mat{0})$.
Insensitive parameters do not have any impact on the observables under consideration, which leads to the block of zeros in the Jacobian.

By excluding parameters from optimization by the LM algorithm on the basis of the sensitivity matrix, we assume that the sensitivity to a specific parameter not only vanishes locally at the current expansion point but everywhere, i.e., $\partial/\partial p_k \mathcal{M}_i(\vec{p}) = 0$ for all predictions $\mathcal{M}_i$ and any parameter vector $\vec{p}$ in the admissible parameter domain.
The validity of this assumption can be assessed after the LM optimization by checking whether supposedly insensitive parameters are also insensitive according to the sensitivity matrix evaluated at the optimized parameter set.
Given this assumption holds, the matrix $\mat{U}$ in~\cref{eq:hessian_with_2nd_model_deriv_matrix} can be written in the partitioned form
\begin{equation}
    \mat{U} = \begin{pmatrix}
    \mat{U}_{11} & \mat{0} \\
    \mat{0} & \mat{0} 
    \end{pmatrix} 
\end{equation}
due to the vanishing partial derivatives $\partial_k\partial_t \mathcal{M}_i$ if $p_k$ or $p_t$ is an insensitive parameter.
The matrix $\mat{J}^T\mat{Q}\mat{J}$ is of the same structure due to the structure of $\mat{J}$.
Consequently, the Hessian matrix reads
\begin{equation}
    \mat{H} = -\begin{pmatrix}
    \mat{U}_{11} + \mat{R}_{11} + \mat{J}_1^T \mat{Q} \mat{J}_1 & \mat{R}_{12} \\
    \mat{R}_{21} & \mat{R}_{22}
    \end{pmatrix}
\end{equation}
and the full posterior covariance matrix can be computed as $\mat{P}_1 = -\mat{H}^{-1}$.
Noteworthy, the blocks of the inverse prior parameter covariance matrix are different from the inverses of the blocks in the the prior parameter covariance matrix, i.e., $\mat{R}_{ij} \neq (\mat{P}_{ij})^{-1}$.

The full computation of the matrix $\mat{U}$ may be too costly for a large number of parameters in combination with an expensive nuclear model.
A seemingly possible solution is to calculate only the diagonal elements of $\mat{U}$ and assume all off-diagonal elements being zero.
In a simulation study, we found that this construction may lead to $\mat{H}$ not being positive-definite.
A solution that still takes into account second-order derivatives of the model is to approximate $\mat{U}$ by a diagonal matrix $\tilde{\mat{U}}$ whose elements are given by 
\begin{equation}
    \tilde{U}_{ii} =
    \begin{cases}
        U_{ii} & \textrm{if } U_{ii} > 0 \\
        0 & \textrm{otherwise}
    \end{cases}
\end{equation}
Considering that the resulting approximative posterior covariance matrix is of the form,
\begin{equation}
    \tilde{\mat{P}}_1 = \left(
        \tilde{\mat{U}} + \mat{J}^T \mat{Q} \mat{J} + \mat{R} 
    \right)^{-1}
    \label{eq:approx_post_cov_posdiagU_apx} \,,
\end{equation}
we can see that $\tilde{\mat{U}}$ enters in the same way as the inverse prior covariance matrix $\mat{R}$.
Therefore $\tilde{\mat{U}}$ can be interpreted as an additional prior information that enforces an upper bound on the diagonal elements of the posterior covariance matrix.
Even though this approximation may be regarded as less conservative than the GLS formula $(\mat{J}^T \mat{Q} \mat{J} + \mat{R})^{-1}$ for the posterior covariance matrix, we think that it is a better approximation to the full Hessian matrix because it takes more information of the full Hessian matrix into account.
Preliminary simulation studies seem to support this view.
Moreover, as can be understood by~\cref{eq:newtonstep_with_cov} of~\cref{apx:comp_posterior_expectation}, an overestimation of posterior  uncertainties as can be the case in the GLS approach leads to a too large adjustment of the parameters.

\subsection{Computation of the posterior expectation of insensitive paramaters}
\label{apx:comp_posterior_expectation}

In \cref{apx:comp_posterior_covmat} we explained the computation of the full posterior covariance matrix including parameters optimized by the LM algorithm and those considered insensitive to the data and therefore excluded from optimization.
However, prior correlations not only change the posterior covariances associated with insensitive parameters but also the posterior expectation values.

To find the update prescription for insensitive parameters, we first consider the update for all parameters.
To this end, we need to locate the maximum of the posterior pdf in \cref{eq:logposteriorpdf_2ndorderapprox}.
As a necessary condition, the gradient of this pdf for the parameter vector associated with the maximum must vanish,
\begin{equation}
    \mat{G} + \mat{H} \left(
        \modparvec - \modparrefvec
    \right) = \vec{0} \,.
\end{equation}
Thus the parameter vector maximizing \cref{eq:logposteriorpdf_2ndorderapprox} reads
\begin{equation}
    \modparvec = \modparrefvec - \mat{H}^{-1}\mat{G} \,.
    \label{eq:newtonstep}
\end{equation}
If the full Hessian is not available due to the infeasibility to compute all second-order derivatives of the nuclear model, we can adopt the approximation suggested in \cref{apx:comp_posterior_covmat}.
Recalling that $\mat{P}_1 = -\mat{H}^{-1}$ and employing the approximation of the posterior covariance matrix in \cref{eq:approx_post_cov_posdiagU_apx}, the update formula can be restated as
\begin{equation}
    \modparvec = \modparrefvec + \tilde{\mat{P}}_1 \mat{G}
     \label{eq:newtonstep_with_cov}
\end{equation}

We already computed $\mat{G}$ in \cref{eq:gradient_logpost_einstein} for a posterior pdf with a multivariate normal prior and likelihood.
We restate it here in matrix notation,
\begin{equation}
    \mat{G} = \mat{J}^T \mat{Q} (\sigma - \sigma_\textrm{ref}) - \mat{R} (\modparrefvec - \vec{s}) \,,
\end{equation}
with $\mat{J}$ being the Jacobian matrix of the nuclear model.
This gradient is evaluated at $\modparrefvec$, where the values of sensitive parameters are from the solution of the LM algorithm and the values of insensitive parameters are given by the prior expectations. 

All ingredients of the update formula in $\cref{eq:newtonstep_with_cov}$ are fully specified now.
The final parameter vector $\modparvec$ contains updated values of both sensitive and insensitive parameters.
We recommend to discard the new values of the sensitive parameters and keep the ones obtained by the LM algorithm.
The reason being that the LM algorithm locates the maximum of the posterior pdf exactly whereas \cref{eq:newtonstep_with_cov} employs an approximation of the posterior covariance matrix.
As the insensitive parameters were excluded from optimization, we have to rely on \cref{eq:newtonstep_with_cov} using the approximative posterior covariance matrix to update the insensitive parameters.


%% file: appendix_download.tex
\section{DOWNLOADING THE PIPELINE AND PACKAGES}
\label{apx:download}
The internet is a fast moving medium.
Domain names appear and vanish after a while, and with them all their content.
Also content evolves and the content accessible under a \textit{uniform resource locator} (URL) or colloquially web address may change within months or years.
Moreover, in many contexts, such as legal documents and scientific publications and data, it is important to have some kind of assurance that the documents have not been tampered with.
An interesting and promising approach to protect against the accidental loss of information and camouflaged modifications of content is the \textit{InterPlanetary File System} (IPFS)~\citep{labs_ipfs_nodate}.
It is a distributed file system build on top of computers accessible over the internet.
Files are not accessed by a name chosen by someone but a unique identifier directly derived from the file content by using a cryptographic hash function.

We anticipate that the pipeline and employed packages will evolve over time.
We want to make sure that the pipeline in the state associated with this version of the paper can be easily located.
Inspired by the idea of content addressable storage, the underlying concept of the IPFS, we follow a similar approach and provide identification strings that allow the localisation of the pipeline and support packages on the internet.
First, we elaborate on the identification strings and how they can be used to locate the pipeline, packages and files on the internet.
Then we provide a catalogue of components including the pipeline itself with a brief description and the identification strings to locate the components.

\subsection{About identification strings}

We use two types of identification strings: (1) git commit hashes and (2) SHA-256 hashes:
\begin{enumerate}
    \item We used \textit{Git} as version control system during the development of the pipeline and all the packages.
    During the development process after a certain number of modifications, these modifications are committed.
    Each \textit{commit} records the exact state of the files in the project at a given time.
    Moreoever, each commit is associated with a unique identifier, which is not only unique within a development project but unique among all the development projects on the planet tracked by git.
    At least, the chance of obtaining identical identifiers of different projects with differing file contents by chance is \textit{extremely} unlikely\footnote{Researchers at CWI Amsterdam and Google have demonstrated a feasible approach to generate two documents with identical SHA-1 hash~\citep{katz_first_2017}. The SHA-1 hash function is also used by Git.
    The amount of compute power required to generate two documents with identical SHA-1 hashes has still to be considered enormous at the time of writing.}.
    The \textit{Git commit hash} is therefore a reasonable identifier for files and projects at a given time if security is not the highest priority.
    \item In the future Git may get upgraded or we may change to another version control system.
    Therefore, there is a need to be able to generate an identifier, which is independent of Git.
    For the files referenced in this paper, we obtain this identifier in the following way:
\begin{lstlisting}
    cat <(find . -type d \
            -not -path '*/\.*' \
            -print0 | \
          LC_ALL=C sort -z) \
        <(find . -type f \
            -not -path '*/\.*' \
            -print0 | \
          LC_ALL=C sort -z | \
          xargs -0 sha256sum) | \
    sha256sum -
\end{lstlisting}
    We refer to this identifier as the \textit{tar hash}.
    If security is a major concern, then the SHA-256 hash function to obtain the \textit{tar hash} offers far greater security than the SHA-1 hash function.
    Please note that hidden files are excluded to avoid the consideration of the \textit{.git} folder and the \textit{.gitignore} file. 
\end{enumerate}

At the time of writing, there are several ways to locate a resource by its identifier:
\begin{itemize}
\item Using either the Git commit hash or the tar hash, the resource can be located via \\
\url{http://nugget.link/<hash>}
It is possible to supply only a part of the full hash instead as long as the sequence is unique among the registered hashes.
\item Using a Git commit hash, it can be located on GitHub via the link: \\
\textit{https://github.com/search?q=<git-commit-id>\&type=Commits} \\
For example, the Dockerfile for setting up the pipeline can be located by following \\
\url{https://github.com/search?q=d16d70579707\&type=Commits}.
In this example we only used the initial part of the identifier, which also works.
\item The tar hash and Git commit hash may have also been added to the webpage associated with a project or package.
A search query with the tar hash or Git commit hash on Google, Bing, DuckDuckGo or any other popular search engine may therefore also return the relevant resource.
At the time of writing, however, we have only tested this with the Google search. 
\end{itemize}

Links to the Dockerfiles of the pipeline and related packages, such as the MongoDB JSON database can also be found on \url{http://www.nucleardata.com}.

\subsection{Catalogue of pipeline modules}

In this section we present the list of packages in~\cref{tbl:package_list} employed in the pipeline along with their identification strings and a brief description.
Also the repository of the pipeline with the sequence of retrieval, preprocessing and fitting scripts is listed.

We decided to assign the version number 0.1.0 to the packages in the state at the time of writing of this article.
Most packages come with a README file with a short example of usage.
Function documentation in most packages is available, created with the help of \textit{Roxygen2}~\citep{wickham_roxygen2}.

Having said that, the documentation of many components does certainly not meet the standards expected in software engineering regarding detail and comprehensiveness at present.
As another note of caution, as the packages have been developed along with the pipeline, the packages have not been thoroughly tested in a context outside of the current implementation of the pipeline.

\cleardoublepage
\onecolumngrid

\begin{table*}[t]
\centering
\newcommand{\cellspacera}[0]{\rule{0pt}{9pt}}
\newcommand{\cellspacerb}[0]{\rule{0pt}{9pt}}
\newcommand{\nuggetlink}[1]{\href{http://nugget.link/#1}{#1}}
\begin{tabular}{l|l}
  \hline
    \cellspacera
    \textbf{eval-fe56-docker} &
    A Dockerfile to install the pipeline and dependencies as one bundle
    \\
    \cellspacerb
    tar hash &
    \nuggetlink{d4b69c92dac3f94af37cb5fe00c60cc8b9ba95ebd5e8f292022d11c69585c1c1} \\
    git commit id & d276ea9ef0efe6d401eee252b7d6990c5fd01d69\\
  \hline
    \cellspacera
    \textbf{eval-fe56} &
    The sequence of script files constituting the pipeline
    \\
    \cellspacerb
    tar hash &
    \nuggetlink{4aa395ec1a7a121cae9fab09e687450bc19034c11c9daba81bcc525acd4887dd} \\
    git commit id & 58a51e018d402362ecdc8d0851d597d9f7890c9b\\
  \hline
    \cellspacera
    \textbf{nucdataBaynet} &
    Management of experimental and model data in the Bayesian context
    \\
    \cellspacerb
    tar hash &
    \nuggetlink{9930f956d4da44bca735672a58464ce844c84f456551c6a7985509c46dbd60a0} \\
    git commit id & 284500f7d09b142ab5a468687612a93d8c51d85b\\
  \hline
    \cellspacera
    \textbf{TALYSeval} &
    Prepare TALYS calculations and retrieve results
    \\
    \cellspacerb
    tar hash &
    \nuggetlink{963bc700027d16f33c06bb9194bf95b1e1512b6eeddfc05190d1a3ce32ae4f88} \\
    git commit id & 857583ba7e259bf5a8bad9fbd558adc77bfe3579\\
  \hline
    \cellspacera
    \textbf{tasmanPatch} &
    Patch to extend functionality of the TASMAN code
    \\
    \cellspacerb
    tar hash &
    \nuggetlink{cfd0a09ecc0b9b135d0f1cd03c1d370bf8ccb55e620cc44282546e4b8a164ed2} \\
    git commit id & 05d243c36708356142dc1828590c8d1f62a9bde4\\
  \hline
    \cellspacera
    \textbf{interactiveSSH} &
    Executing bash commands remotely via SSH from within R
    \\
    \cellspacerb
    tar hash &
    \nuggetlink{4b8af95035326a42542cefb186ab0237c827a31f5286bbbe40c0ad933f0d4e3d} \\
    git commit id & 08adb733de05eb9b4acb76f76e64b0d9cd5a032f\\
  \hline
    \cellspacera
    \textbf{rsyncFacility} &
    Wrapper package around the rsync command line tool
    \\
    \cellspacerb
    tar hash &
    \nuggetlink{5cb3d8e1e9d367eb9d50aea963f9958841eed0987707892314b8549cd899e5fc} \\
    git commit id & fe3825864ee93c0b9fcfd6cf68400051836982bb\\
  \hline
    \cellspacera
    \textbf{remoteFunctionSSH} &
    Execute R functions remotely in the same way as local functions
    \\
    \cellspacerb
    tar hash &
    \nuggetlink{8a8e49572ad0e99e8c56149c63329a951766b7a0c232452f3006b4c39870a42e} \\
    git commit id & 681c4db9f85a712a406b3d49ae47fd235bd9e6b3\\
  \hline
    \cellspacera
    \textbf{clusterSSH} &
    Execute R functions in parallel on a computer cluster
    \\
    \cellspacerb
    tar hash &
    \nuggetlink{78d7ff6205cb96b25335dfb0a96dfc3bcbefed22b10c48809285e7f7ce750f79} \\
    git commit id & c8cc3434aec82f46c003d88c35861757f39dc5b5\\
  \hline
    \cellspacera
    \textbf{clusterTALYS} &
    Launch TALYS calculations in parallel on a computer cluster
    \\
    \cellspacerb
    tar hash &
    \nuggetlink{c8fe3f0abbde33b60017034f980ad0b843afcb200ed86e04c9d51aa9cd4df1ea} \\
    git commit id & 144cecbbcac73f21086020ff3a69958c837b2d74\\
  \hline
    \cellspacera
    \textbf{MongoEXFOR} &
    Convenience interface to the EXFOR JSON MongoDB database
    \\
    \cellspacerb
    tar hash &
    \nuggetlink{cf4c020ea89d3cc2fa10c1298c98c8c4260bdd0708e88501a0d7bcf94b907ed7} \\
    git commit id & 2e2dea55aea90ddca7851f264580bb3ad158c45f\\
  \hline
    \cellspacera
    \textbf{exforUncertainty} &
    Tentative rules for rule-based correction of experimental uncertainties
    \\
    \cellspacerb
    tar hash &
    \nuggetlink{c917f70ac4520a0aef2cca8479e8e28644c8ab3c1062d439a7af9fad2d25a38a} \\
    git commit id & 7b48fa675396e4f2290e6a9361fefc77d66d5ffa\\
  \hline
    \cellspacera
    \textbf{talysExforMapping} &
    Mapping of TALYS predictions to observables recorded in EXFOR subentries
    \\
    \cellspacerb
    tar hash &
    \nuggetlink{5ac8f35314ff5c411cf4409ac48339b35784ac8de250be1de246ed90c0362cef} \\
    git commit id & 2ef3b109b66e0fbe103ec6a6cd2a42b828252926\\
  \hline
    \cellspacera
    \textbf{jsonExforUtils} &
    Utility functions to deal with information in the EXFOR entries
    \\
    \cellspacerb
    tar hash &
    \nuggetlink{e856b7734bd99b81ba02cd002236590b31094ef0d91eb811200c0dc22b2885d7} \\
    git commit id & b464876d312a7316549ed0f3f48dea13217be83d\\
  \hline
    \cellspacera
    \textbf{exforParser} &
    Read EXFOR entries in R and conversion to JSON
    \\
    \cellspacerb
    tar hash &
    \nuggetlink{09104fd60025c4d655d7fad1306a2afc2f049c6293136ed4e4a48652f80eba63} \\
    git commit id & c732c4e05824ac68f9905d603bce10777b2ce9b9\\
  \hline
    \cellspacera
    \textbf{createExforDb} &
    Creation of the EXFOR JSON MongoDB database from EXFOR masterfiles
    \\
    \cellspacerb
    tar hash &
    \nuggetlink{10c618b46ef402be1fc82a0e6583ea09fccae23003c23f99b1448fedc027c29f} \\
    git commit id & af6d2af843cae280f37292858fb2a7c4fb50aeb4\\
  \hline
    \cellspacera
    \textbf{compEXFOR-docker} &
    Dockerfile for a stand-alone installation of the MongoDB EXFOR JSON database
    \\
    \cellspacerb
    tar hash &
    \nuggetlink{e1bc125d69477777a328cf7a20b255cbee29748b187a4c24676e58aa47b6e8dd} \\
    git commit id & 6bd8232157a82e5dce38150e58f257554937d386\\
  \hline
    \cellspacera
    \textbf{exfor-couchdb-docker} &
    Dockerfile for a stand-alone installation of the CouchDB EXFOR JSON database
    \\
    \cellspacerb
    tar hash &
    \nuggetlink{86ac80d845f92bacf0af53c3ec272fef112333c2da5fbea409004d15b03eea31} \\
    git commit id & dd23a4bbea246bafe1e240f7c5dbd5f5cd553cb5\\
\end{tabular}
\caption{List of packages}
\label{tbl:package_list}
\end{table*}

%% file: main.bbl
\begin{thebibliography}{100}

\bibitem{sep-scientific-reproducibility}
F.~Fidler and J.~Wilcox, ``Reproducibility of scientific results,'' in {\em The
  Stanford Encyclopedia of Philosophy} (E.~N. Zalta, ed.), Metaphysics Research
  Lab, Stanford University, winter 2018~ed., 2018.

\bibitem{carlson_evaluation_2018}
A.~Carlson, V.~Pronyaev, R.~Capote, G.~Hale, Z.-P. Chen, I.~Duran, F.-J.
  Hambsch, S.~Kunieda, W.~Mannhart, B.~Marcinkevicius, R.~Nelson, D.~Neudecker,
  G.~Noguere, M.~Paris, S.~Simakov, P.~Schillebeeckx, D.~Smith, X.~Tao,
  A.~Trkov, A.~Wallner, and W.~Wang, ``Evaluation of the {Neutron} {Data}
  {Standards},'' {\em Nuclear Data Sheets}, vol.~148, pp.~143--188, Feb. 2018.

\bibitem{carlson_corrigendum_2020}
A.~Carlson, V.~Pronyaev, R.~Capote, G.~Hale, Z.-P. Chen, I.~Duran, F.-J.
  Hambsch, S.~Kunieda, W.~Mannhart, B.~Marcinkevicius, R.~Nelson, D.~Neudecker,
  G.~Noguere, M.~Paris, S.~Simakov, P.~Schillebeeckx, D.~Smith, X.~Tao,
  A.~Trkov, A.~Wallner, and W.~Wang, ``Corrigendum to “{Evaluation} of the
  {Neutron} {Data} {Standards}” [{Nucl}. {Data} {Sheets} 148, p. 143
  (2018)],'' {\em Nuclear Data Sheets}, vol.~163, pp.~280--281, Jan. 2020.

\bibitem{carlson_international_2009}
A.~Carlson, V.~Pronyaev, D.~Smith, N.~Larson, Z.~Chen, G.~Hale, F.-J. Hambsch,
  E.~Gai, S.-Y. Oh, S.~Badikov, T.~Kawano, H.~Hofmann, H.~Vonach, and
  S.~Tagesen, ``International {Evaluation} of {Neutron} {Cross} {Section}
  {Standards},'' {\em Nuclear Data Sheets}, vol.~110, pp.~3215--3324, Dec.
  2009.

\bibitem{zerkin_experimental_2018}
V.~Zerkin and B.~Pritychenko, ``The experimental nuclear reaction data
  ({EXFOR}): {Extended} computer database and {Web} retrieval system,'' {\em
  Nuclear Instruments and Methods in Physics Research Section A: Accelerators,
  Spectrometers, Detectors and Associated Equipment}, vol.~888, pp.~31--43,
  Apr. 2018.

\bibitem{otuka_towards_2014}
N.~Otuka, E.~Dupont, V.~Semkova, B.~Pritychenko, A.~Blokhin, M.~Aikawa,
  S.~Babykina, M.~Bossant, G.~Chen, S.~Dunaeva, R.~Forrest, T.~Fukahori,
  N.~Furutachi, S.~Ganesan, Z.~Ge, O.~Gritzay, M.~Herman, S.~Hlavač, K.~Katō,
  B.~Lalremruata, Y.~Lee, A.~Makinaga, K.~Matsumoto, M.~Mikhaylyukova,
  G.~Pikulina, V.~Pronyaev, A.~Saxena, O.~Schwerer, S.~Simakov, N.~Soppera,
  R.~Suzuki, S.~Takács, X.~Tao, S.~Taova, F.~Tárkányi, V.~Varlamov, J.~Wang,
  S.~Yang, V.~Zerkin, and Y.~Zhuang, ``Towards a {More} {Complete} and
  {Accurate} {Experimental} {Nuclear} {Reaction} {Data} {Library} ({EXFOR}):
  {International} {Collaboration} {Between} {Nuclear} {Reaction} {Data}
  {Centres} ({NRDC}),'' {\em Nuclear Data Sheets}, vol.~120, pp.~272--276, June
  2014.

\bibitem{schnabel_computational_2020}
G.~Schnabel, ``A computational {EXFOR} database,'' {\em {EPJ} Web of
  Conferences}, vol.~239, p.~16001, 2020.

\bibitem{koning_tendl_2019}
A.~Koning, D.~Rochman, J.-C. Sublet, N.~Dzysiuk, M.~Fleming, and S.~van~der
  Marck, ``{TENDL}: {Complete} {Nuclear} {Data} {Library} for {Innovative}
  {Nuclear} {Science} and {Technology},'' {\em Nuclear Data Sheets}, vol.~155,
  pp.~1--55, Jan. 2019.

\bibitem{helgesson_treating_2018}
P.~Helgesson and H.~Sjöstrand, ``Treating model defects by fitting smoothly
  varying model parameters: {Energy} dependence in nuclear data evaluation,''
  {\em Annals of Nuclear Energy}, vol.~120, pp.~35--47, Oct. 2018.

\bibitem{schnabel_fitting_2018}
G.~Schnabel, ``Fitting and {Analysis} {Technique} for {Inconsistent} {Nuclear}
  {Data},'' {\em arXiv:1803.00960 [nucl-ex, physics:nucl-th, physics:physics]},
  Mar. 2018.
\newblock arXiv: 1803.00960.

\bibitem{helgesson_fitting_2017}
P.~Helgesson and H.~Sjöstrand, ``Fitting a defect non-linear model with or
  without prior, distinguishing nuclear reaction products as an example,'' {\em
  Review of Scientific Instruments}, vol.~88, p.~115114, Nov. 2017.

\bibitem{pearl_causality_2000}
J.~Pearl, {\em Causality: models, reasoning, and inference}.
\newblock Cambridge, U.K. ; New York: Cambridge University Press, 2000.

\bibitem{koning_modern_2012}
A.~Koning and D.~Rochman, ``Modern {Nuclear} {Data} {Evaluation} with the
  {TALYS} {Code} {System},'' {\em Nuclear Data Sheets}, vol.~113,
  pp.~2841--2934, Dec. 2012.

\bibitem{gschnabel-eval-fe56}
{\em {Prototype of evaluation pipeline}}.

\bibitem{gschnabel-compEXFOR-docker}
G.~Schnabel, {\em {Dockerfile to set up EXFOR MongoDB database}}.
\newblock See \cref{apx:download}.

\bibitem{gschnabel-nucdataBaynet}
G.~Schnabel, {\em {R package: nucdataBaynet}}.
\newblock See \cref{apx:download}.

\bibitem{muir_evaluation_1989}
D.~W. Muir, ``Evaluation of correlated data using partitioned least squares: a
  minimum-variance derivation,'' {\em Nuclear Science and Engineering},
  vol.~101, no.~1, pp.~88--93, 1989.

\bibitem{smith_least-squares_1993}
D.~L. Smith, ``A least-squares computational "tool kit",'' Tech. Rep.
  ANL/NDM-128, Argonne National Laboratory, Argonne, Illinois, Apr. 1993.

\bibitem{gluc1980}
D.~M. {Hetrick} and C.~Y. {Fu}, ``{GLUCS: A generalized least-squares program
  for updating cross section evaluations with correlated data sets}.'' Unknown,
  Oct. 1980.

\bibitem{muir_global_2007}
D.~W. Muir, A.~Trkov, I.~Kodeli, R.~Capote, and V.~Zerkin, ``The {Global}
  {Assessment} of {Nuclear} {Data}, {GANDR},'' in {\em Proc. of Int. Conf on
  Nuclear Data for Science and Technology}, EDP Sciences, 2007.

\bibitem{muir_global_2007_website}
D.~Muir, ``Global {Assessment} of {Nuclear} {Data} {Requirements} ({GANDR}
  project),'' 2007.

\bibitem{leeb_consistent_2008}
H.~Leeb, D.~Neudecker, and T.~Srdinko, ``Consistent {Procedure} for {Nuclear}
  {Data} {Evaluation} {Based} on {Modeling},'' {\em Nuclear Data Sheets},
  vol.~109, pp.~2762--2767, Dec. 2008.

\bibitem{neudecker_impact_2013}
D.~Neudecker, R.~Capote, and H.~Leeb, ``Impact of model defect and experimental
  uncertainties on evaluated output,'' {\em Nuclear Instruments and Methods in
  Physics Research Section A: Accelerators, Spectrometers, Detectors and
  Associated Equipment}, vol.~723, pp.~163--172, Sept. 2013.

\bibitem{schnabel_differential_2016}
G.~Schnabel and H.~Leeb, ``Differential {Cross} {Sections} and the {Impact} of
  {Model} {Defects} in {Nuclear} {Data} {Evaluation},'' {\em EPJ Web of
  Conferences}, vol.~111, p.~09001, 2016.

\bibitem{rasmussen_gaussian_2006}
C.~E. Rasmussen and C.~K.~I. Williams, {\em Gaussian {Processes} for {Machine}
  {Learning}}.
\newblock Cambridge, Mass.: MIT Press, 2006.

\bibitem{byrd_limited_1995}
R.~H. Byrd, P.~Lu, J.~Nocedal, and C.~Zhu, ``A {Limited} {Memory} {Algorithm}
  for {Bound} {Constrained} {Optimization},'' {\em SIAM Journal on Scientific
  Computing}, vol.~16, pp.~1190--1208, Sept. 1995.

\bibitem{larson_concise_2008}
N.~M. Larson, ``A {Concise} {Method} for {Storing} and {Communicating} the
  {Data} {Covariance} {Matrix},'' Tech. Rep. ORNL/TM-2008/104, 941045, Oak
  Ridge National Laboratory, Oct. 2008.

\bibitem{harville_matrix_1997}
D.~A. Harville, {\em Matrix {{Algebra From}} a {{Statistician}}'s
  {{Perspective}}}.
\newblock New York, NY: {Springer New York}, 1997.

\bibitem{schnabel_large_2015}
G.~Schnabel, {\em Large scale {Bayesian} nuclear data evaluation with
  consistent model defects}.
\newblock PhD thesis, Technische Universität Wien, Vienna, 2015.

\bibitem{kennedy_bayesian_2001}
M.~C. Kennedy and A.~O'Hagan, ``Bayesian calibration of computer models,'' {\em
  Journal of the Royal Statistical Society: Series B (Statistical
  Methodology)}, vol.~63, pp.~425--464, Aug. 2001.

\bibitem{duvenaud_automatic_2014}
D.~Duvenaud, {\em Automatic model construction with {Gaussian} processes}.
\newblock PhD thesis, University of Cambridge, 2014.

\bibitem{schnabel_first_2018}
G.~Schnabel and H.~Sjöstrand, ``A first sketch: {Construction} of model defect
  priors inspired by dynamic time warping,'' {\em arXiv:1811.03874 [nucl-ex,
  physics:nucl-th, physics:physics]}, Nov. 2018.
\newblock arXiv: 1811.03874.

\bibitem{snelson_sparse_2006}
E.~Snelson and Z.~Ghahramani, ``Sparse {Gaussian} processes using
  pseudo-inputs,'' in {\em Advances in neural information processing systems},
  pp.~1257--1264, 2006.

\bibitem{schnabel_estimating_2018}
G.~Schnabel, ``Estimating model bias over the complete nuclide chart with
  sparse {Gaussian} processes at the example of {INCL}/{ABLA} and
  double-differential neutron spectra,'' {\em EPJ Nuclear Sciences \&
  Technologies}, vol.~4, p.~33, 2018.

\bibitem{quinonero-candela_unifying_2005}
J.~Quiñonero-Candela and C.~E. Rasmussen, ``A unifying view of sparse
  approximate {Gaussian} process regression,'' {\em Journal of Machine Learning
  Research}, vol.~6, no.~Dec, pp.~1939--1959, 2005.

\bibitem{jean_uncertainty_2011}
C.~D.~S. Jean, B.~Habert, P.~Archier, G.~Noguere, D.~Bernard, J.~Tommasi, and
  P.~Blaise, ``Uncertainty {Evaluation} of {Nuclear} {Reaction} {Model}
  {Parameters} using {Integral} and {Microscopic} {Measurements} with the
  {CONRAD} {Code},'' {\em Journal of the Korean Physical Society}, vol.~59,
  pp.~1276--1279, Aug. 2011.

\bibitem{archier_conrad_2014}
P.~Archier, C.~De~Saint~Jean, O.~Litaize, G.~Noguère, L.~Berge, E.~Privas, and
  P.~Tamagno, ``{CONRAD} {Evaluation} {Code}: {Development} {Status} and
  {Perspectives},'' {\em Nuclear Data Sheets}, vol.~118, pp.~488--490, Apr.
  2014.

\bibitem{larson_updated_2008}
N.~M. Larson, ``Updated user’s guide for {SAMMY}: {Multilevel} {R}-{Matrix}
  {Fits} to {Neutron} {Data} {Using} {Bayes}’ {Equation}, {Revision} 4,''
  Tech. Rep. ORNL/TM-9179/R8, Oak Ridge National Laboratory, Oak Ridge, 2008.

\bibitem{levenberg_method_1944}
K.~Levenberg, ``A method for the solution of certain non-linear problems in
  least squares,'' {\em Quarterly of Applied Mathematics}, vol.~2,
  pp.~164--168, July 1944.

\bibitem{marquardt_algorithm_1963}
D.~W. Marquardt, ``An {Algorithm} for {Least}-{Squares} {Estimation} of
  {Nonlinear} {Parameters},'' {\em Journal of the Society for Industrial and
  Applied Mathematics}, vol.~11, pp.~431--441, June 1963.

\bibitem{woodbury_inverting_1950}
M.~A. Woodbury, {\em Inverting Modified Matrices}.
\newblock Statistical Research Group, Memo. Rep. no. 42, Princeton University,
  Princeton, N. J., 1950.

\bibitem{madsen2004}
K.~Madsen, H.~Nielsen, and O.~Tingleff, ``Methods for non-linear least squares
  problems (2nd ed.),'' p.~60, 01 2004.

\bibitem{Rlanguage}
{R Core Team}, {\em R: A Language and Environment for Statistical Computing}.
\newblock R Foundation for Statistical Computing, Vienna, Austria, 2017.

\bibitem{Rdatatable}
M.~Dowle and A.~Srinivasan, {\em data.table: Extension of `data.frame`}, 2019.
\newblock R package version 1.12.8.

\bibitem{RMatrix}
D.~Bates and M.~Maechler, {\em Matrix: Sparse and Dense Matrix Classes and
  Methods}, 2019.
\newblock R package version 1.2-18.

\bibitem{schwerer_exfor_2015}
O.~Schwerer, ``{EXFOR} {Formats} {Manual},'' Tech. Rep. IAEA-NDS-207, IAEA,
  Vienna, Aug. 2015.

\bibitem{cullen_program_2001}
D.~Cullen and A.~Trkov, ``{PROGRAM} {X4TOC4} ({Version} 2001-3) {Translation}
  of {Experimental} {Data} from the {EXFOR} {Format} to a {Computation}
  {Format},'' Tech. Rep. IAEA-NDS-80, IAEA, Vienna, Mar. 2001.

\bibitem{fedynitch_afedynitchx4i3_2020}
A.~Fedynitch, ``afedynitch/x4i3,'' May 2020.

\bibitem{gschnabel-exforParser}
G.~Schnabel, {\em {R package: exfor Parser}}.
\newblock See \cref{apx:download}.

\bibitem{gschnabel-createExforDb}
G.~Schnabel, {\em {Script to add EXFOR masterfiles to MongoDB database}}.
\newblock See \cref{apx:download}.

\bibitem{noauthor_regular_nodate}
``regular expression {\textbar} {Encyclopedia}.com.''

\bibitem{gschnabel-exfor-couchdb-docker}
G.~Schnabel, {\em {Dockerfile to set up EXFOR CouchDB database}}.
\newblock See \cref{apx:download}.

\bibitem{Rmongolite}
J.~Ooms, ``The jsonlite package: A practical and consistent mapping between
  json data and r objects,'' {\em arXiv:1403.2805 [stat.CO]}, 2014.

\bibitem{gschnabel-MongoEXFOR}
G.~Schnabel, {\em {R package: MongoEXFOR}}.
\newblock See \cref{apx:download}.

\bibitem{Rstringr}
H.~Wickham, {\em stringr: Simple, Consistent Wrappers for Common String
  Operations}, 2019.
\newblock R package version 1.4.0.

\bibitem{herman_empire_2007}
M.~Herman, R.~Capote, B.~Carlson, P.~Obložinský, M.~Sin, A.~Trkov, H.~Wienke,
  and V.~Zerkin, ``{EMPIRE}: {Nuclear} {Reaction} {Model} {Code} {System} for
  {Data} {Evaluation},'' {\em Nuclear Data Sheets}, vol.~108, pp.~2655--2715,
  Dec. 2007.

\bibitem{herman_empire-32_2013}
M.~Herman, R.~Capote, M.~Sin, A.~Trkov, B.~V. Carlson, P.~Oblozinsky,
  C.~Mattoon, H.~Wienke, S.~Hoblit, Y.~Cho, G.~Nobre, V.~Plujko, and V.~Zerkin,
  ``{EMPIRE}-3.2 {Malta}-{Modular} system for nuclear reaction calculations and
  nuclear data evaluation,'' Tech. Rep. INDC (NDS)-0603, IAEA, Vienna, Aug.
  2013.

\bibitem{gschnabel-talysExforMapping}
G.~Schnabel, {\em {R package: talysExforMapping}}.
\newblock See \cref{apx:download}.

\bibitem{smith_unified_2008}
D.~L. Smith, ``A unified {Monte} {Carlo} approach to fast neutron cross section
  data evaluation,'' {\em Proc. of the 8 th International Topical Mtg. on Nucl.
  Applics. and Util. of Accelerators, Pocatello, July}, p.~736, 2008.

\bibitem{capote_investigation_2008}
R.~Capote and D.~L. Smith, ``An {Investigation} of the {Performance} of the
  {Unified} {Monte} {Carlo} {Method} of {Neutron} {Cross} {Section} {Data}
  {Evaluation},'' {\em Nuclear Data Sheets}, vol.~109, pp.~2768--2773, Dec.
  2008.

\bibitem{capote_new_2012}
R.~Capote, D.~L. Smith, A.~Trkov, and M.~Meghzifene, ``A {New} {Formulation} of
  the {Unified} {Monte} {Carlo} {Approach} ({UMC}-{B}) and {Cross}-{Section}
  {Evaluation} for the {Dosimetry} {Reaction} $^{\textrm{55}}$ {Mn}(n,g)
  $^{\textrm{56}}$ {Mn},'' {\em Journal of ASTM International}, vol.~9,
  pp.~179--196, Mar. 2012.

\bibitem{bauge_evaluation_2007}
E.~Bauge, S.~Hilaire, and P.~Dossantos-Uzarralde, ``Evaluation of the
  {Covariance} {Matrix} of {Neutronic} {Cross} {Sections} with the
  {Backward}-{Forward} {Monte} {Carlo} {Method},'' in {\em Proc. of Int. Conf.
  on Nuclear Data for Science and Technology"}, EDP Sciences, 2007.

\bibitem{bauge_evaluation_2011}
E.~Bauge and P.~Dossantos-Uzarralde, ``Evaluation of the {Covariance} {Matrix}
  of $^{\textrm{239}}${Pu} {Neutronic} {Cross} {Sections} in the {Continuum}
  {Using} the {Backward}-{Forward} {Monte}-{Carlo} {Method},'' {\em Journal of
  the Korean Physical Society}, vol.~59, p.~1218, Aug. 2011.

\bibitem{koning_bayesian_2015}
A.~J. Koning, ``Bayesian {Monte} {Carlo} method for nuclear data evaluation,''
  {\em The European Physical Journal A}, vol.~51, p.~184, Dec. 2015.

\bibitem{kawano_covariance_1997}
T.~Kawano and K.~Shibata, ``Covariance evaluation system,'' Tech. Rep.
  JAERI-Data/Code 97-037, JAERI, Japan, 1997.

\bibitem{capote_nuclear_2010}
R.~Capote, D.~Smith, and A.~Trkov, ``Nuclear data evaluation methodology
  including estimates of covariances,'' {\em EPJ Web of Conferences}, vol.~8,
  p.~04001, 2010.

\bibitem{herman_covariance_data_2011}
``Covariance data in the fast neutron region,'' Tech. Rep.
  NEA/NSC/WPEC/DOC(2010)427, OECD-NEA, Oct. 2011.

\bibitem{helgesson_comparison_2018}
P.~Helgesson, D.~Neudecker, H.~Sjöstrand, M.~Grosskopf, D.~L. Smith, and
  R.~Capote, ``Assessment of {Novel} {Techniques} for {Nuclear} {Data}
  {Evaluation},'' in {\em Reactor {Dosimetry}: 16th {International}
  {Symposium}} (M.~H. Sparks, K.~R. DePriest, and D.~W. Vehar, eds.),
  pp.~105--116, 100 Barr Harbor Drive, PO Box C700, West Conshohocken, PA
  19428-2959: ASTM International, Nov. 2018.

\bibitem{Gentzsch2001SunGE}
W.~Gentzsch, ``Sun grid engine: towards creating a compute power grid,'' {\em
  Proceedings First IEEE/ACM International Symposium on Cluster Computing and
  the Grid}, pp.~35--36, 2001.

\bibitem{dean_mapreduce_2008}
J.~Dean and S.~Ghemawat, ``{MapReduce}: simplified data processing on large
  clusters,'' {\em Communications of the ACM}, vol.~51, pp.~107--113, Jan.
  2008.

\bibitem{zaharia_apache_2016}
M.~Zaharia, R.~S. Xin, P.~Wendell, T.~Das, M.~Armbrust, A.~Dave, X.~Meng,
  J.~Rosen, S.~Venkataraman, M.~J. Franklin, A.~Ghodsi, J.~Gonzalez,
  S.~Shenker, and I.~Stoica, ``Apache {Spark}: a unified engine for big data
  processing,'' {\em Communications of the ACM}, vol.~59, pp.~56--65, Oct.
  2016.

\bibitem{gschnabel-interactiveSSH}
G.~Schnabel, {\em {R package: interactiveSSH}}.
\newblock See \cref{apx:download}.

\bibitem{gschnabel-rsyncFacility}
G.~Schnabel, {\em {R package: rsyncFacility}}.
\newblock See \cref{apx:download}.

\bibitem{gschnabel-remoteFunctionSSH}
G.~Schnabel, {\em {R package: remoteFunctionSSH}}.
\newblock See \cref{apx:download}.

\bibitem{gschnabel-clusterSSH}
G.~Schnabel, {\em {R package: clusterSSH}}.
\newblock See \cref{apx:download}.

\bibitem{gschnabel-clusterTALYS}
G.~Schnabel, {\em {R package: clusterTALYS}}.
\newblock See \cref{apx:download}.

\bibitem{Rssh}
J.~Ooms, {\em ssh: Secure Shell (SSH) Client for R}, 2019.
\newblock R package version 0.6.

\bibitem{Rstudio}
{RStudio Team}, {\em RStudio: Integrated Development Environment for R}.
\newblock RStudio, PBC., Boston, MA, 2020.

\bibitem{trkov_endf-6_2018}
A.~Trkov, M.~Herman, and D.~Brown, ``{ENDF}-6 {Formats} {Manual},'' Tech. Rep.
  BNL-203218-2018-INRE, Brookhaven National Laboratory, Upton, NY 11973-5000,
  Feb. 2018.

\bibitem{noauthor_specifications_2020}
``Specifications for the {Generalised} {Nuclear} {Database} {Structure},''
  Tech. Rep. NEA No. 7519, OECD Nuclear Energy Agency, 2020.

\bibitem{sjostrand_monte_2019}
H.~Sjöstrand and G.~Schnabel, ``Monte {Carlo} integral adjustment of nuclear
  data libraries – experimental covariances and inconsistent data,'' {\em EPJ
  Web of Conferences}, vol.~211, p.~07007, 2019.

\bibitem{siefman_data_2020}
D.~Siefman, M.~Hursin, H.~Sjostrand, G.~Schnabel, D.~Rochman, and A.~Pautz,
  ``Data assimilation of post-irradiation examination data for fission yields
  from {GEF},'' {\em EPJ Nuclear Sciences \& Technologies}, vol.~6, p.~52,
  2020.

\bibitem{smith_experimental_2012}
D.~Smith and N.~Otuka, ``Experimental {Nuclear} {Reaction} {Data}
  {Uncertainties}: {Basic} {Concepts} and {Documentation},'' {\em Nuclear Data
  Sheets}, vol.~113, pp.~3006--3053, Dec. 2012.

\bibitem{iwamoto_generation_2020}
H.~Iwamoto, ``Generation of nuclear data using {Gaussian} process regression,''
  {\em Journal of Nuclear Science and Technology}, vol.~57, pp.~932--938, Aug.
  2020.

\bibitem{gschnabel-exforUncertainty}
G.~Schnabel, {\em {R package: exforUncertainty}}.
\newblock See \cref{apx:download}.

\bibitem{peelle1987peelle}
R.~Peelle, ``Peelle’s pertinent puzzle,'' {\em Informal memorandum dated
  October}, vol.~13, 1987.

\bibitem{chiba_suggested_1991}
S.~Chiba and D.~Smith, ``A suggested procedure for resolving an anomaly in
  least-squares data analysis known as ``{Peelle}`s {Pertinent} {Puzzle}`` and
  the general implications for nuclear data evaluation,'' Tech. Rep.
  ANL/NDM--121, 10121367, Argonne National Lab., IL (USA), Sept. 1991.

\bibitem{helgesson_combining_2017}
P.~Helgesson, H.~Sjöstrand, A.~Koning, J.~Rydén, D.~Rochman, E.~Alhassan, and
  S.~Pomp, ``Combining {Total} {Monte} {Carlo} and {Unified} {Monte} {Carlo}:
  {Bayesian} nuclear data uncertainty quantification from auto-generated
  experimental covariances,'' {\em Progress in Nuclear Energy}, vol.~96,
  pp.~76--96, Apr. 2017.

\bibitem{neudecker_template_2018}
D.~Neudecker, B.~Hejnal, F.~Tovesson, M.~White, D.~Smith, D.~Vaughan, and
  R.~Capote, ``Template for estimating uncertainties of measured
  neutron-induced fission cross-sections,'' {\em EPJ Nuclear Sciences \&
  Technologies}, vol.~4, p.~21, 01 2018.

\bibitem{neudecker_applyingtemplates_2020}
D.~Neudecker, D.~Smith, F.~Tovesson, R.~Capote, M.~White, N.~Bowden, L.~Snyder,
  A.~Carlson, R.~Casperson, V.~Pronyaev, S.~Sangiorgio, K.~Schmitt, B.~Seilhan,
  N.~Walsh, and W.~Younes, ``Applying a {{Template}} of {{Expected
  Uncertainties}} to {{Updating 239Pu}}(n,f) {{Cross}}-section {{Covariances}}
  in the {{Neutron Data Standards Database}},'' {\em Nuclear Data Sheets},
  vol.~163, pp.~228--248, Jan. 2020.

\bibitem{neudecker_templates_2021}
D.~Neudecker, A.~Lewis, E.~Matthews, and J.~Vanhoy~{\textit{et al.}},
  ``Templates of expected measurement uncertainties,'' Tech. Rep.
  LA-UR-19-13-31156, {Los Alamos National Laboratory}, 2021.

\bibitem{neudecker_validating_2020}
D.~Neudecker, M.~C. White, D.~E. Vaughan, and G.~Srinivasan, ``Validating
  nuclear data uncertainties obtained from a statistical analysis of
  experimental data with the “{Physical} {Uncertainty} {Bounds}” method,''
  {\em EPJ Nuclear Sciences \& Technologies}, vol.~6, p.~19, 2020.

\bibitem{capote_unrecognized_2020}
R.~Capote, S.~Badikov, A.~Carlson, I.~Duran, F.~Gunsing, D.~Neudecker,
  V.~Pronyaev, P.~Schillebeeckx, G.~Schnabel, D.~Smith, and A.~Wallner,
  ``Unrecognized {Sources} of {Uncertainties} ({USU}) in {Experimental}
  {Nuclear} {Data},'' {\em Nuclear Data Sheets}, vol.~163, pp.~191--227, Jan.
  2020.

\bibitem{forrest_statistical_2007}
R.~Forrest and J.~Kopecky, ``Statistical analysis of cross sections—{A} new
  tool for data validation,'' {\em Fusion Engineering and Design}, vol.~82,
  pp.~73--90, Jan. 2007.

\bibitem{forrest_detailed_2008}
R.~Forrest, J.~Kopecky, and A.~Koning, ``Detailed analysis of (n,p) and
  (n,alpha) cross sections in the {EAF}-2007 and {TALYS}-generated libraries,''
  {\em Fusion Engineering and Design}, vol.~83, pp.~634--643, May 2008.

\bibitem{pigni_uncertainty_2003}
M.~T. Pigni and H.~Leeb, ``Uncertainty {Estimates} of {Evaluated}
  $^{\textrm{56}}${Fe} {Cross} {Sections} {Based} on {Extensive} {Modelling} at
  {Energies} {Beyond} 20 {MeV},'' in {\em Proc. {Int}. {Workshop} on {Nuclear}
  {Data} for the {Transmutation} of {Nuclear} {Waste}. {GSI}-{Darmstadt},
  {Germany}}, 2003.

\bibitem{leeb_covariances_2005}
H.~Leeb, ``Covariances for {Evaluations} based on {Extensive} {Modelling},'' in
  {\em {AIP} {Conference} {Proceedings}}, vol.~769, (Santa Fe, New Mexico
  (USA)), pp.~161--164, AIP, 2005.
\newblock ISSN: 0094243X.

\bibitem{blight_bayesian_1975}
B.~J.~N. Blight and L.~Ott, ``A {Bayesian} {Approach} to {Model} {Inadequacy}
  for {Polynomial} {Regression},'' {\em Biometrika}, vol.~62, p.~79, Apr. 1975.

\bibitem{mcbook_owen_2013}
A.~B. Owen, {\em Monte Carlo theory, methods and examples}.
\newblock 2013.

\bibitem{brooks_handbook_2011}
S.~Brooks, A.~Gelman, G.~L. Jones, and X.-L. Meng, eds., {\em Handbook for
  {Markov} chain {Monte} {Carlo}}.
\newblock Boca Raton: Taylor \& Francis, 2011.

\bibitem{koning_towards_2008}
A.~J. Koning and D.~Rochman, ``Towards {Sustainable} {Nuclear} {Energy}:
  {Putting} {Nuclear} {Physics} to {Work},'' {\em Annals of Nuclear Energy},
  vol.~35, pp.~2024--2030, Nov. 2008.

\bibitem{gschnabel-tasmanPatch}
G.~Schnabel, {\em {Patch for TASMAN}}.
\newblock See \cref{apx:download}.

\bibitem{robbins1956}
H.~Robbins, ``An empirical bayes approach to statistics,'' in {\em Proceedings
  of the Third Berkeley Symposium on Mathematical Statistics and Probability,
  Volume 1: Contributions to the Theory of Statistics}, (Berkeley, Calif.),
  pp.~157--163, University of California Press, 1956.

\bibitem{maritz_empirical_2018}
J.~S. Maritz, {\em Empirical {Bayes} methods}.
\newblock Place of publication not identified: Routledge, 2018.
\newblock OCLC: 1036741238.

\bibitem{gschnabel-eval-fe56-docker}
G.~Schnabel, {\em {Dockerfile to set up evaluation pipeline}}.
\newblock See \cref{apx:download}.

\bibitem{hirdt_data_2013}
J.~A. Hirdt and D.~A. Brown, ``Data mining the {EXFOR} database using network
  theory,'' {\em arXiv:1312.6200 [nucl-th, physics:physics]}, Dec. 2013.
\newblock arXiv: 1312.6200.

\bibitem{dwivedi2019trees}
N.~R. Dwivedi, ``Trees and islands -- machine learning approach to nuclear
  physics,'' 2019.

\bibitem{whewell_ml_2020}
B.~Whewell, M.~Grosskopf, and D.~Neudecker, ``Evaluating 239pu(n,f) cross
  sections via machine learning using experimental data, covariances, and
  measurement features,'' {\em Nuclear Instruments and Methods in Physics
  Research Section A: Accelerators, Spectrometers, Detectors and Associated
  Equipment}, vol.~978, p.~164305, 2020.

\bibitem{george_evolution_1989}
A.~George and J.~W. Liu, ``The {Evolution} of the {Minimum} {Degree} {Ordering}
  {Algorithm},'' {\em SIAM Review}, vol.~31, pp.~1--19, Mar. 1989.

\bibitem{davis_direct_2006}
T.~A. Davis, {\em Direct methods for sparse linear systems}.
\newblock Fundamentals of algorithms, Philadelphia: Society for Industrial and
  Applied Mathematics, 2006.

\bibitem{labs_ipfs_nodate}
P.~Labs, ``{IPFS} {Powers} the {Distributed} {Web}.''
\newblock Library Catalog: ipfs.io.

\bibitem{katz_first_2017}
M.~Stevens, E.~Bursztein, P.~Karpman, A.~Albertini, and Y.~Markov, ``The
  {First} {Collision} for {Full} {SHA}-1,'' in {\em Advances in {Cryptology}
  – {CRYPTO} 2017} (J.~Katz and H.~Shacham, eds.), vol.~10401, pp.~570--596,
  Cham: Springer International Publishing, 2017.
\newblock Series Title: Lecture Notes in Computer Science.

\bibitem{wickham_roxygen2}
H.~Wickham, P.~Danenberg, G.~Csárdi, and M.~Eugster, {\em roxygen2: In-Line
  Documentation for R}, 2020.
\newblock R package version 7.1.1.

\end{thebibliography}
